\documentclass{aa}  
\usepackage{ulem}
\usepackage{graphicx}
\usepackage{txfonts}
\usepackage{hyperref}
\usepackage{multirow}
\hypersetup{colorlinks=true, linkcolor=red, citecolor=blue, urlcolor=blue}
\newcommand*\lephare{L\textsc{e} P\textsc{hare}~}

\usepackage{natbib}
\usepackage{color}
\usepackage{amstext}

\begin{document}

\title{ The VIPERS Multi-Lambda Survey. II } 
\subtitle{Diving with massive galaxies in 22 square degrees since $z = 1.5$}


\author{T. Moutard\inst{1} \thanks{thibaud.moutard@lam.fr}
\and S. Arnouts\inst{1}
\and O. Ilbert\inst{1}
\and J. Coupon\inst{2}
\and I. Davidzon\inst{1,3}
\and L. Guzzo\inst{4}
\and P. Hudelot\inst{5}
\and\\ H. J. McCracken\inst{5}
\and L. Van Werbaeke\inst{6}
\and G. E. Morrison\inst{7,8}
\and O. Le F\`evre\inst{1}
\and V. Comte\inst{1}
\and M. Bolzonella\inst{3}
\and\\ A. Fritz\inst{9}
\and B. Garilli\inst{9}
\and M. Scodeggio\inst{9}
}

\institute{
Aix Marseille Universit\'e, CNRS, LAM - Laboratoire d'Astrophysique de Marseille, 38 rue F. Joliot-Curie, F-13388, Marseille, France 
\and
Astronomical Observatory of the University of Geneva, ch. d Ecogia16, 1290 Versoix, Switzerland
\and
INAF - Osservatorio Astronomico di Bologna, via Ranzani 1, 40127 Bologna, Italy
\and
INAF - Osservatorio Astronomico di Brera, via E. Bianchi 46, 23807 Merate/via Brera 28, 20122, Milano, Italy
\and
Sorbonne Universit\'e, UPMC Univ Paris 06, et CNRS, UMR 7095, IAP, 98b bd Arago, 75014 Paris, France
\and
Department of Physics and Astronomy, University of British Columbia, 6224 Agricultural Road, Vancouver, V6T 1Z1, BC, Canada
\and
Institute for Astronomy, University of Hawaii, Honolulu, HI, 96822, USA
\and
Canada-France-Hawaii Telescope, Kamuela, HI, 96743, USA
\and
INAF - Istituto di Astrofisica Spaziale e Fisica Cosmica (IASF)
Milano, via Bassini 15, 20133 Milano, Italy
}

\date{accepted for publication in \aap}
\titlerunning{The VIPERS-MLS -- II: Evolution of massive galaxies at $z < 1.5$}

%
\abstract{We investigate the evolution of the galaxy stellar mass function and stellar mass density from redshift $z=0.2$ to $z=1.5$ of a $K_s < 22$-selected sample with highly reliable photometric redshifts and over an unprecedentedly large area.  Our study is based on near-infrared observations  carried out with the WIRCam instrument at CFHT  over the footprint of the VIPERS spectroscopic survey and benefits from the high-quality optical photometry from the CFHTLS and ultraviolet observations with the GALEX satellite.
The accuracy of our photometric redshifts is $\sigma_{\Delta z/(1+z)} < $0.03 and 0.05 for the bright ($i_{AB}< 22.5$) and faint ($i_{AB} > 22.5$) samples, respectively. 
  The galaxy stellar mass function is measured with $\sim$760,000 galaxies down to $K_s\sim 22$  and over an effective area of $\sim$ 22.4 deg$^2$, the latter of which drastically reduces the statistical uncertainties (i.e. Poissonian error and cosmic variance). We point out the importance of carefully controlling the photometric calibration, whose effect becomes quickly dominant when statistical uncertainties are reduced, which will be a major issue for future cosmological surveys with EUCLID or LSST, for instance.
By exploring the rest-frame $(NUV-r)$ vs $(r-K_s)$ colour-colour diagram with which we separated star-forming and quiescent galaxies, (1)
we find that the density of very massive $\log(M_*/ M_{\odot}) > 11.5$ galaxies is largely dominated by quiescent galaxies and increases by a factor 2 from z$\sim 1$ to $z\sim 0.2$, which allows for additional mass assembly through dry mergers. (2) We also confirm the scenario in which star formation activity is impeded above a stellar mass $\log(\mathcal{M}^\star_{\textsc{sf}} / M_{\odot}) = 10.64 \pm 0.01$. This value is found to be very stable at $0.2 < z < 1.5$. (3) We discuss the existence of a main quenching channel that is followed by massive star-forming galaxies, and we finally (4) characterise another quenching mechanism that is required to explain the clear excess of low-mass quiescent galaxies that is observed at low redshift. }

\maketitle
%
\section{Introduction}
\label{introduction}

The measurement of the stellar mass function (SMF) is a powerful statistical tool for tracing the stellar mass assembly or galaxy growth over cosmic time.  Galaxy formation models rely on the well established $\Lambda$CDM cosmological framework that governs the growth of the dark matter structures and the less well understood baryonic physics at play inside the dark matter haloes (gas accretion, minor or major merging, star formation activity, feedback mechanisms, etc.).  
The shape of the galaxy SMF compared to the expected halo mass function provides valuable information about the physical processes acting at the low- and high-mass ends of the mass function \citep[][]{Silk2012}.

A decade ago, early deep extragalactic surveys have revealed that the average stellar mass density decreased gradually (the integrated form of the SMF) from z$\sim$3 to z$\sim$0 \citep[e.g.][]{Dickinson2003, Fontana2003}. This trend is now confirmed up to redshift $z\sim 8$ \citep[]{Song2015} and is consistent with a hierarchical build-up of the cosmic structures.  Later on, larger surveys have measured the evolution at high redshift of the galaxy bimodality, the well-known separation between star-forming and quiescent galaxies observed in the local Universe \citep[][]{Baldry2004, Moustakas2013}.  They found that the bimodality was already in place at $z\sim 1$ with the quiescent galaxies dominating the massive end of the SMF and the star-forming galaxies dominating its low-mass end \citep[][]{Bundy2006,Borch2006}. This quiescent population had its main build-up epoch between $z=2$ and $z=1$, where the stellar mass density increased by a factor 10 \citep[][]{Cirasuolo2007,Arnouts2007}, while only a factor 2 increase is observed from $z=1$ to $z=0$ \citep[][]{Bell2004,Faber2007}.
 According to the hierarchical scenario, such an early formation epoch of the quiescent population was not a problem as long as the stars formed before this in smaller units and galaxies continued to assemble their masses at a later time \citep[through dry merging phases, e.g.][]{DeLucia2006}. This is a natural support of the \textit{star formation downsizing} picture proposed by \citet[][]{Cowie1996}, where the onset of star formation begins
earlier for most massive galaxies than for lower mass galaxies \citep[see also][]{Gavazzi1996a}. However, the models predict a continuous increase of stellar mass for these massive galaxies with cosmic time \citep[e.g.][]{DeLucia2007}, which is challenged by the last measurements of the SMF, where the massive end does not show significant evolution from $z = 0$ up to redshift $z\sim 1$ \citep[e.g.][]{Marchesini2009,Ilbert2013,Muzzin2013, Moustakas2013,Mortlock2015}, suggesting a mass assembly downsizing.

The predominance of quiescent galaxies at the massive end \citep[e.g.][]{Baldry2012,Moustakas2013,Ilbert2013} supports the idea that the star formation activity is preferentially impeded in galaxies above a given stellar mass or a given dark matter halo mass, if we assume a stellar-to-halo mass relationship \citep[e.g.][]{Coupon2015}.  A wide variety of quenching mechanisms have been proposed to explain the star formation quenching in massive galaxies, such as major mergers \citep[][]{Barnes1992},  virial shock heating \citep[][]{Keres2005}, or radio-AGN feedback \citep{Croton2006,Cattaneo2006} in massive haloes.
  
Several studies have emphasised the role played by the environment for the colour-bimodality and star-formation quenching in the local Universe
\citep{Hogg2003,Kauffmann2004,Baldry2006,Haines2007}. Mechanisms such as \textit{ram-pressure stripping}, in which the gas is expelled from the galaxy \citep[][]{Gunn1972}, or \textit{strangulation}, in which the cold gas supply is heated and then halted \citep[][]{Larson1980}, can be invoked as environmental quenching mechanisms. We emphasise that strangulation processes can either be linked to environment (e.g. when a galaxy enters the hot gas of a cluster) or to peculiar evolution (e.g. when a the radio-AGN feedback stops the cold gas infall). 
The latest  measurements of the quiescent SMFs reveal an upturn at the low-mass end in the local Universe \citep{Baldry2012,Moustakas2013}, whose build-up is observed at higher redshift \citep{Drory2009, Tomczak2014}. This upturn in the low-mass end for quiescent galaxies could be associated to environmental quenching according to \citet{Peng2010, Peng2012}, while \citet{Schawinski2014} suggested a fast process consistent with major merging. Constraining the quenching timescale at different masses might therefore help to highlight the quenching mechanisms.

Until recently, the above conclusions  were mostly based on deep galaxy surveys  such as  GOODS  \citep[][]{Giavalisco2004},  VVDS \citep[][]{LeFevre2005}, COSMOS \citep[][]{Scoville2007}, and DEEP2 \citep[][]{Newman2013}, which are  perfectly suited to provide the global picture of the galaxy stellar mass assembly over a wide range of redshifts. However,  given their small angular coverage (they explore a rather small volume below $z < 1$), they can be particularly sensitive to statistical variance (i.e. \textit{cosmic variance}) at low redshift. 
This is particularly crucial for the very rare galaxies at the high-mass end of the exponentially declining SMF, and it has been claimed that its apparent lack of evolution may be dominated by observational uncertainties  \citep{Fontanot2009,Marchesini2009}. 

A first attempt  to constrain the density evolution of the high-mass galaxies at $z < 1$ has been performed  by \citet{Matsuoka2010}. They combined the SDSS southern strip \citep[][]{York2000} and UKIDSS/LAS survey \citep[][]{Lawrence2007} over a total area of $\sim$ 55 deg$^2$. They observed a  mild-to-high increase of the number density of massive galaxies $10^{11-11.5}M_{\odot}$/$10^{11.5-12.}M_{\odot}$ with a corresponding drop of the fraction of star-forming galaxies in this stellar mass range from $z\sim $1 to $z\sim$0. While subject to large uncertainties in their photometric redshifts, stellar mass estimates, and reliability of the separation into quiescent and star-forming galaxies, this first result suggested that massive galaxies ($M_* > 10^{11} M_{\odot}$) evolve since $z \sim 1$.

 \citet{Moustakas2013} estimated the SMF between $0 < z < 1$  over an area of $\sim$5.5 deg$^{2}$ using PRIMUS, a low-resolution prism survey \citep[for galaxies with $i_{AB}\le23$;][]{Coil2011}.
 The wealth of multi-wavelength information from deep ultraviolet (GALEX satellite) to mid-infrared (Spitzer/IRAC) photometry allowed them to derive accurate stellar masses and a reliable separation between active and quiescent populations.  Their SMF measurements confirmed  the modest change in the number density of the  massive star-forming galaxies ($M_*\ge 10^{11}M_{\odot}$), leaving little room for mergers, but observed a significant drop (50\%) of the fraction of active star-forming galaxies since $z\sim 1$ that is in contrast with the classical picture, in which the star-forming population remains constant across cosmic time.
 Another  major spectroscopic sample is provided by the VIMOS Public Extragalactic Redshift Survey \citep[VIPERS; ][]{Guzzo2014}, whose first $\sim 50,000$ galaxies down to $i_{AB}=22.5$ over an  area of 10.3 deg$^2$ have recently been released \citep[PDR1,][]{Garilli2014}. Using the PDR1 combined with CFHTLS photometry and the same ultraviolet (UV) and near-infrared (NIR) data that we used here, \citet{Davidzon2013} produced the most reliable overall measurement of the high-mass end of the SMF in between $0.5 < z < 1.3$ to date.
 The VIPERS SMF shows to high precision that the most massive galaxies had already assembled most of their stellar mass at $z\sim1$, but that a residual evolution is still present. 
However, as discussed in \citet{Davidzon2013}, although these two studies use spectroscopic redshifts,
 multi-wavelength information, and a large area,  they disagree
slightly concerning the general amplitude of the SMF. These discrepancies might be due~to differences in the stellar mass estimates, for
example, or to selection effects that are not fully accounted for. It highlights how subtle effects become crucial and can introduce significant systematic errors when statistical uncertainties are reduced so drastically.

In this paper we exploit the broad photometric coverage assembled over the footprint of VIPERS  to build  a unique multi-wavelength photometric sample covering more than 22 deg$^2$ down to $K_s < 22$, as part of the VIPERS-Multi Lambda Survey \citep[VIPERS-MLS; see][]{Moutard2016a}.  We benefit of the synergy with the VIPERS spectroscopic survey by using the PDR-1 data to compute reliable photometric redshifts, and we derive stellar masses for 760,000 galaxies out to $z=1.5$.  This allows us to obtain a new estimate of the SMF that
(a) has greater control over the low-mass slope because of the $i < 23.7~/~K_s < 22$ depth of our sample for extended sources (more than 1 mag deeper in the $i$-band than VIPERS), 
(b) extends over a wider redshift range than VIPERS, from $z = 0.2$ out to $z=1.5$, 
(c) is less affected by the cosmic variance because the effective area is doubled with respect to the VIPERS PDR-1 used in \citet{Davidzon2013} (we cover nearly the entire footprint of the final VIPERS survey and avoid the 30\% area loss that is due to the detector gaps in VIPERS), 
(d) suffers a reduced Poisson error because our sample is ten times larger in the common redshift range, and
(e) can be studied separately for star-forming and quiescent objects, which means that the quenching channels that characterise the massive galaxies up to $z = 1.5 $ can be explored, as well as the low-mass galaxies at low redshift.

The paper is organised as follows. In Sect. \ref{data} we describe our photometric and spectroscopic dataset. The photometric redshifts and galaxy classification are presented in Sect. \ref{photoz}, the stellar mass estimates in Sect. \ref{mass}. We detail the measurements of the galaxy SMFs and the associated uncertainties in Sect. \ref{SMF}, where we also point out the effect of the photometric absolute calibration in the new generation of large surveys. We present the evolution of the stellar mass function and stellar mass density in Sect. \ref{evol}. Finally, we discuss our results and their effects on the quenching channels in Sect. \ref{discut}. 

Throughout this paper, we use the standard cosmology ($\Omega_m~=~0.3$, $\Omega_\Lambda~=~0.7$ with $H_{\rm0}~=~70$~km~s$^{-1}$~Mpc$^{-1}$). Magnitudes are given in the $AB$ system \citep{Oke1974}. The galaxy stellar masses are given in units of solar masses ($M_{\sun}$) for a \citet{Chabrier2003} initial mass function (Chabrier IMF).

\begin{figure}[!t]
\center
\includegraphics[width=\hsize, trim = 0.2cm 0cm 0.1cm 0cm, clip]{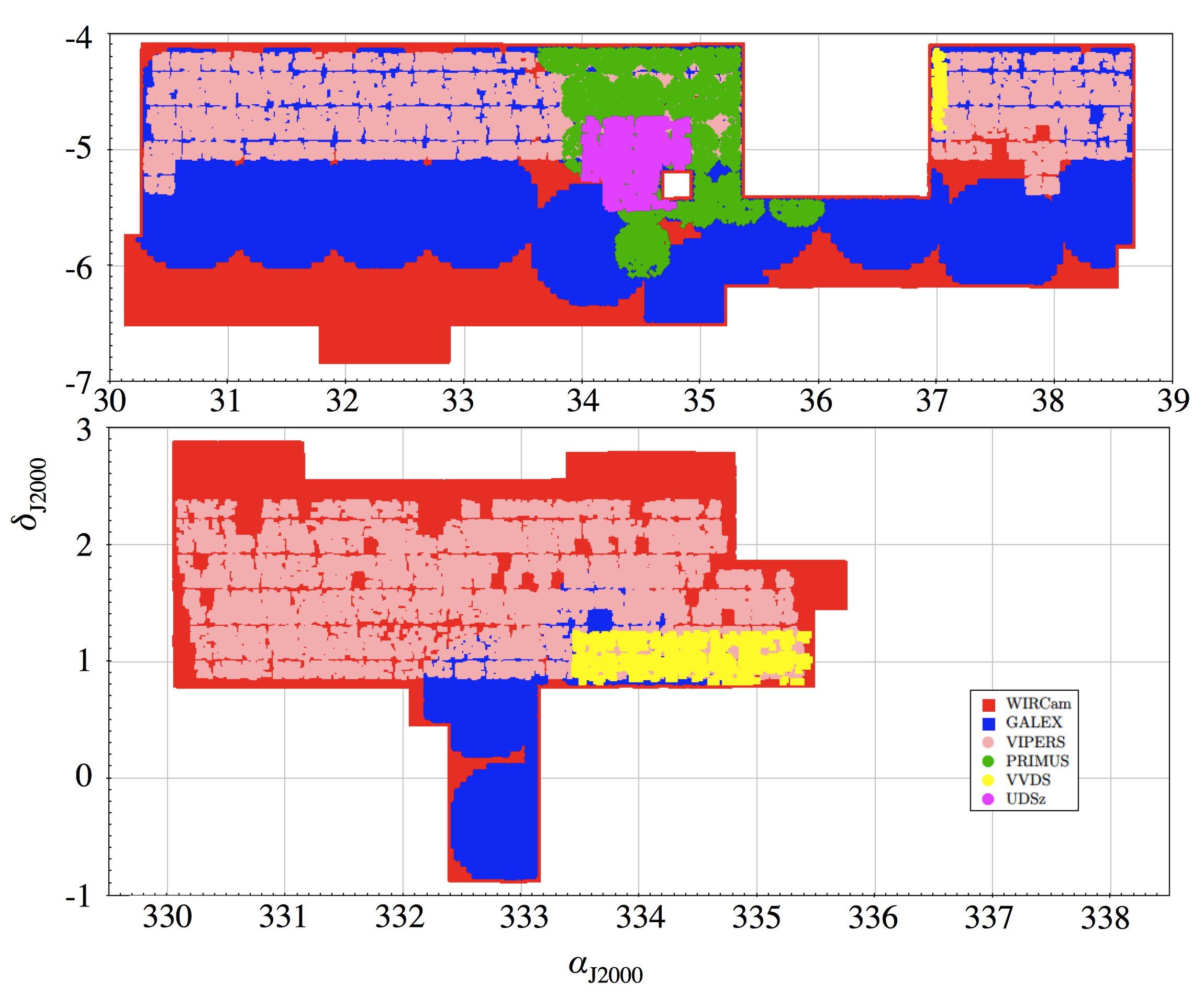}
\caption{Footprints of the WIRCam $K_s$-band (red layout and background) and GALEX $NUV$/$FUV$ (blue circles) observations in the CFHTLS W1 (top) and W4 (bottom) fields. The regions covered by VIPERS (pink), PRIMUS (green), VVDS (yellow) and UDSz (magenta) are over-plotted. The SDSS-BOSS redshifts are distributed over the entire survey.
\label{footprint}}
\end{figure}

\section{Data description}
\label{data}

The observations and the data reduction are described in detail in the companion paper \citep{Moutard2016a} and are briefly summarised below.

\subsection{Optical CFHTLS photometry}
\label{optical}

The CFHTLS\footnote{http://www.cfht.hawaii.edu/Science/CFHTLS/} is an imaging survey performed with the MegaCam\footnote{http://www.cfht.hawaii.edu/Instruments/Imaging/Megacam/} camera in five optical bands, $u, g, r, i,$ and $z$.
It covers $\sim$ 155 deg$^{2}$ over four independent fields with  sub-arcsecond seeing (median $\sim$ 0.8") and reaches a 80\% completeness limit of  $u \sim 24.4$, $g \sim 24.7$, $r \sim 24.0$, $i / y \sim 23.7,$ and $z \sim 22.9$ for extended sources in AB system. We emphasise that the $y$-band refers to the new $i$-band filter, in accordance with the CFHTLS notation. We have used the $y$-band response curve in our analysis when appropriate, but we refer to the "$i$" filter term regardless of whether it was observed with the $i$ or $y$ filter.

In this work we use the W1 ($+02^h18^m00^s$ $-07^{\circ}00^m00^s$) and W4 ($+22^h13^m18s$ $+01^{\circ}19m00^s$) fields. Two independent photometric catalogues have been released to the community: the 7th$^{}$ and final release (noted T0007\footnote{http://terapix.iap.fr/cplt/T0007/doc/T0007-doc.html}) of the CFHTLS produced by Terapix\footnote{http://terapix.iap.fr/} , and the release from the CFHT Lensing Survey team (CFHTLenS\footnote{http://www.cfhtlens.org/}).
Both catalogues are based on the same raw images. The AstrOmatic software suite\footnote{http://www.astromatic.net/} has been used to generate the mosaic images \citep[SWARP,][]{Bertin2002} and to extract the photometric catalogues \citep[SExtractor,][]{Bertin1996}. The two releases differ in several points, however.
\begin{itemize}
\item In T0007, detection is based on $gri-\chi^2$ images, while the galaxies in CFHTLenS are  $i$-detected.
\item A point spread function (PSF) homogenisation is implemented in CFHTLenS \citep[see][]{Hildebrandt2012} to improve the colour estimates. In practice, the PSF is homogenised across the field of view for each filter and degraded in all filters to the one with the highest PSF.
\item A new photometric calibration has been applied to the T0007 release.  While the previous releases and the CFHTLenS release rely on Landolt standard stars \citep[see][]{Erben2013}, T0007 is based on the spectrophotometric tertiary standards from the Super Novae Legacy Survey \citep[SNLS; see the procedure described in][]{Regnault2009}. In brief, each tile of the CFHTLS-Wide is re-observed (with short exposures) during stable photometric conditions and bracketed by two observations of the CFHTLS-Deep field containing the SNLS tertiary standards. 
\end{itemize}
The difference in the calibration scheme of the two releases affects the final photometry.  A comparison of the magnitudes for point-like sources between the T0007 and CFHTLenS releases reveals systematic offsets that are significantly larger than the expected uncertainties. These offsets  are reported in Table \ref{tab_zero_pt} (column $\Delta mag$).
We emphasise that the differences listed in this table are entirely due to the new calibration scheme established for T0007. The procedure used by the T0007 release allows transferring the percent level accuracy of the SNLS photometric calibration to the entire CFHTLS-Wide survey. For this reason, we use the T0007 catalogue as reference in this paper. However, we also perform the complete analysis with the CFHTLenS catalogue to discuss the effect of such differences in the photometric absolute calibration.

\subsection{WIRCam $K_s$ photometry}
\label{wircam_K}

\begin{figure}[!h]
\includegraphics[width=0.95\hsize]{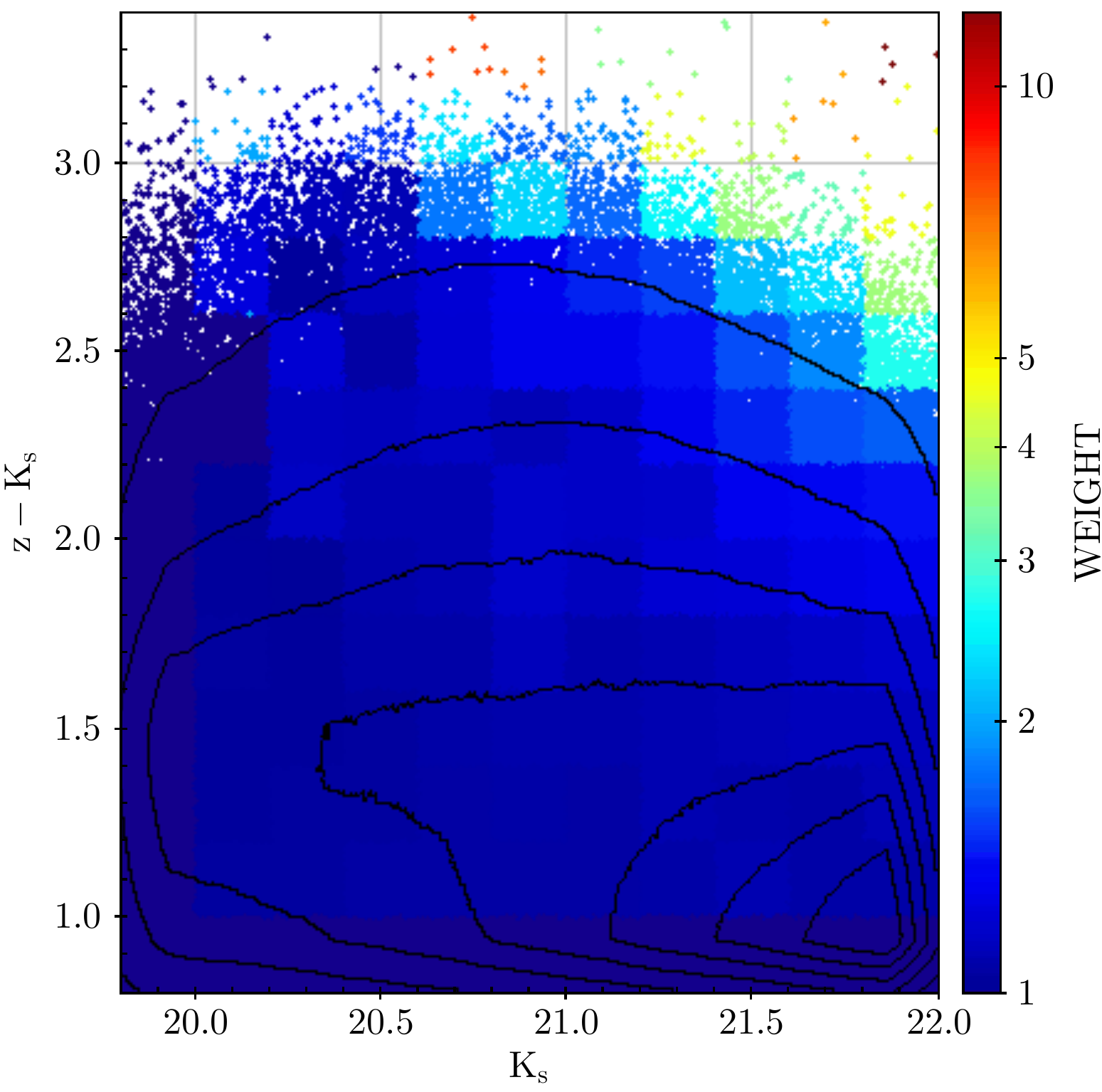}
\caption{Colour$-$magnitude weight map used for our statistical analysis. It takes the missing objects in the $K_s < 22$-limited sample into account. These objects are missed because the $gri$-detection was used to extract the $K_s$ fluxes. 
Weights are multiplicative. This map is restricted to galaxies with redshift $z\le 1.5$ (cf. Sect.~\ref{photoz}) and the contours outline the galaxy density distribution. 
\label{weight_map}}
\end{figure}

We conducted a NIR $K_s$ -band follow-up of the VIPERS fields with the WIRCam instrument at CFHT \citep{Puget2004}.  The layout of the observations is shown in Fig.\ref{footprint} (red background). We covered a total area of $\sim$27 deg$^2$ with an integration time per pixel of 1050 seconds. The image quality is very homogeneous, with an average seeing of all the individual exposures of $<IQ>=0.6\arcsec \pm0.09$. The data have been reduced by the Terapix team\footnote{http://terapix.iap.fr/} and  the individual images were stacked and resampled on the pixel grid of the CFHTLS-T0007 release \citep{Hudelot2012}. The photometry was performed with \texttt{SExtractor} in dual-image mode with a $gri-\chi^2$ image as the detection image and the same settings as those adopted for the T0007 release. The images reach a depth of $K_s= 22$ at $\sim 3\sigma$. 
 The completeness reaches  80\% at $K_s=22;$  this was determined
from a comparison with the deepest surveys UKIDSS Ultra-Deep Survey\citep[UDS, ][]{Lawrence2007}  and VIDEO  \citep{Jarvis2013} in overlapping regions.  Because the primary optical detection is based on the $gri-\chi^2$ image, we may miss the reddest high-redshift galaxies. To account for this possible bias, we measured our source incompleteness as a function of magnitude, $K,$ and colour, $(z-K)$, with all the sources detected in the deep VIDEO survey. We derived a colour-magnitude weight map that we show in Fig.
\ref{weight_map} and use in the remaining paper as multiplicative weighting for our statistical analyses.
We refer to the companion paper, \citet{Moutard2016a}, for a complete description of the method that was used to build this weight map.

\subsection{GALEX photometry}
\label{galex}
When available, we made use of the UV deep-imaging photometry from the
GALEX satellite \citep{Martin2005}. We only considered the observations from the Deep Imaging Survey (DIS), which are shown in Fig.~\ref{footprint} as blue circles ($\varnothing \sim1.1^{\circ}$).  All the GALEX pointings were observed with the near-ultraviolet (NUV) channel with exposure times of $T_{\rm exp} \ge 30$~ksec. Far-ultraviolet (FUV) observations are available for ten pointings in the central part of W1.

The large PSF of GALEX (FWHM$\sim$5\arcsec) means that source confusion becomes a
great problem in the deep survey. To extract the UV photometry, we used a
dedicated photometric code, \texttt{EMphot} \citep{Conseil2011}
that will be described in a separate paper (Vibert et al., in prep.). In
brief, EMphot uses the stamps in the $u$-band (here the T0007 release) as priors, and they are then convolved by the GALEX PSF to produce a simulated image. The scaling factors to be applied to each $u$-band prior is obtained by simultaneously maximising the likelihood between the observed and the predicted fluxes for all the sources in tiles of a few square arc-minutes. The uncertainties on the UV fluxes account for the residuals in the simulated or observed image. A typical depth of $NUV\sim 24.5$ at $\sim 5\sigma$ is observed over the entire survey. The NUV observations cover part of the WIRCam area with $\sim$10.8 and 1.9~deg$^2$ in the W1 and W4 fields, respectively.

\subsection{Final photometric catalogue}
\label{final_cat}

The catalogue of sources comes from the T0007 release and is based on detection in the $gri-\chi^2$ image.  As mentioned above, the same procedure was applied to the $K_s$ images.  Following \citet{Erben2013} and \citet{Hildebrandt2012}, we used the T0007 isophotal apertures for the photometry to estimate the colours. The apertures are smaller than the Kron-like apertures \citep{Kron1980},
which provides less noisy colours and leads to an improved photometric redshift accuracy \citep{Hildebrandt2012}. We also confirmed this with our large spectroscopic dataset (see below), which is especially relevant for faint sources ($i'>23.5$). 

To derive galaxy physical properties, we need to know the total flux in all wavelengths. Therefore, we rescaled the isophotal flux to the Kron-like flux, $m^f_{total} =  m^f_{ISO}+ \delta_m$,  by adopting a unique factor,  $\delta_m$, for each source to preserve the colours. $\delta_m$ is the weighted mean of the individual scaling factors, $\delta_m^f$, and is defined as  $\delta_m= \sum_f  \delta^f_m w^{f} / \sum_f  w^f $, with $f=g,r,i,K_s$ , and $w^f$ its associated errors\footnote{For the CFHTLenS catalogue the scaling factor is computed only in $i$ band. Only the final magnitude is available in the public CFHTLenS catalogue of \citet{Erben2013}.}.
 Finally,  the GALEX photometry,  which corresponds to the total flux measurement (i.e. model PSF photometry) was added in the same way as to the optical and NIR magnitudes.

We here limit the catalogue to galaxies brighter than $K_s<22$. The catalogue includes a total of $\sim$1.3 millions sources over an area of  $\sim$27.1 deg$^2$ , which drops to one million sources over $\sim$22.4 deg$^2$  after applying the masks provided by the CFHTLenS team.

 \subsection{Spectroscopic sample}
\label{zs_sample}

Our WIRCam survey has been designed to cover the VIMOS Public Extragalactic Survey \citep[VIPERS;][]{Guzzo2014} that is carried out with the VIMOS spectrograph and therefore provides many high-quality spectroscopic redshifts. We also added a compilation of the best-quality spectra from the VVDS survey \citep[$I_{AB}\le24$,][]{LeFevre2013},  the $K < 23$ limited UKIDSS spectroscopic Ultra Deep Survey \citep[UDSz,][]{Bradshaw2013,McLure2013}, the low-resolution spectra ($\lambda / \Delta \lambda \sim$40) from the PRIsm MUlti-object Survey \citep[PRIMUS, $i\sim 23$,][]{Coil2011}, and the bright-limited ($i < 19.9$) spectroscopic survey BOSS from the SDSS \citep[Baryon Oscillation Spectroscopic Survey,][]{Dawson2013}. The $K_s < 22$ spectroscopic sample we used is presented in detail in the companion paper \citep{Moutard2016a}.

We selected only the most secure spectroscopic redshifts, which means confidence levels above 95\% for high-resolution surveys and $\sigma < 0.005$ (8\% of outliers with $\delta z/(1+z) > 5 \sigma)$ for PRIMUS best redshifts. When they are available, the redshift measurements from VIPERS were used. Otherwise, the measurements from the deepest high-resolution spectra were favoured. In total, we assembled a $K_s < 22$-limited sample of $45951$ high-quality spectroscopic redshifts to calibrate and measure the accuracy of our photometric redshifts over the unmasked area of the survey (we refer to the companion paper for more details).


\section{Photometric redshifts}
\label{photoz}

\subsection{Photometric redshift measurement}
\label{zp_method}

The photometric redshifts were computed with the spectral energy
distribution (SED) fitting code \lephare \citep{Arnouts2002, Ilbert2006}, using the templates of \citet{Coupon2015}. The new templates are based on the \citet{Ilbert2009} library of 31 empirical templates from \citet{Polletta2007}, complemented by 12 star-forming templates from the Bruzual and Charlot stellar population synthesis models of 2003 \citep[][hereafter BC03]{BC2003}. These templates were optimised to be more representative of the VIPERS spectroscopic sample \citep[for more details we refer to][]{Coupon2015}.

The extinction was added as a free parameter with a reddening excess E(B-V) < 0.3 following different laws: \citet{Prevot1984}, \citet{Calzetti2000}, and a dusty Calzetti curve including a bump at 2175\AA. No extinction was allowed for SEDs redder than Sc. The extinction law of \citet{Prevot1984} was used for templates redder than SB3 templates \citep[see][]{Ilbert2009} and the law
of \citet{Calzetti2000} for bluer templates.

Finally, any possible difference between the photometry and the template library was corrected for by \lephare according to the method described in \citet{Ilbert2006}. In brief, in each band the code tracks a systematic shift between the predicted magnitudes at known redshift and the observed magnitudes. Since our observation area is divided into 47 tiles of $\lesssim 1 deg^2$ with the
relative calibration varying from tile to tile, we performed a tile-by-tile colour optimisation. We used the median offset over all the tiles when there were not enough galaxies with spectroscopic redshift in the tile ($N_{gal}^{spec} \leq 100$) available, which was the case in 12 tiles. 
We stress that the corrections were computed to better fit the colours and are therefore relative. We normalised the median offset on the $K_s$-band because the NIR fluxes are the same (see Sect. \ref{data}). The median relative offsets thus calculated for each photometric band of the T0007 and CFHTLenS catalogues can be found in Table \ref{tab_zero_pt}, with the associated tile-to-tile deviation estimates (namely, the normalised median absolute deviation, NMAD). The difference between the T0007 and CFHTLenS relative offsets is consistent with the difference $\Delta mag$. In other words, we retrieved the shift between the two absolute photometric calibrations through the relative offsets computed with \lephare. This safety check confirms that the colour optimisation achieved with \lephare absorbs the uncertainties that are linked to the photometric calibration.

\begin{table}[h!]
\caption{T0007 - CFHTLenS photometric offsets ($\Delta mag$) obtained by comparing point-like sources in the two catalogues and relative corrections obtained with \lephare to optimise the photometric redshifts. Relative corrections are given using the $K_s$-band as reference (NIR data are identical).  \label{tab_zero_pt}}
\vspace{0.3cm}
\centering
\begin{tabular}{l*{3}{r}}
  \hline
  \noalign{\smallskip}
   &  & \multicolumn{2}{c}{\lephare corrections } \\
  Filter & $\Delta mag^{*}$~~~~~~~ & \begin{tiny}T0007\end{tiny}~~~~~~~ & \begin{tiny}CFHTLenS\end{tiny}~~~~~\\
  
  \hline
  \hline
  \noalign{\medskip}
  $FUV$ & --- ~~~~~~~~  & 0.102 $\pm$ \textit{0.070} & 0.084 $\pm$ \textit{0.079} \\
  $NUV$ & --- ~~~~~~~~  & 0.054 $\pm$ \textit{0.055} & 0.022 $\pm$ \textit{0.065}  \\
  $u$ & -0.013 $\pm$ \textit{0.052} & 0.075 $\pm$ \textit{0.031} & 0.087 $\pm$ \textit{0.042} \\
  $g$ & 0.071 $\pm$ \textit{0.053} & 0.028 $\pm$ \textit{0.019} & -0.053 $\pm$ \textit{0.016} \\
  $r$ & 0.038 $\pm$ \textit{0.052} & 0.022 $\pm$ \textit{0.019} & -0.024 $\pm$ \textit{0.005} \\
  $i$ & 0.066 $\pm$ \textit{0.045}& 0.013 $\pm$ \textit{0.015} & -0.055 $\pm$ \textit{0.009} \\
  $y$ & 0.048 $\pm$ \textit{0.051} & 0.008 $\pm$ \textit{0.009} & -0.042 $\pm$ \textit{0.013} \\
  $z$ & 0.148 $\pm$ \textit{0.054} & 0.087 $\pm$ \textit{0.027} & -0.063 $\pm$ \textit{0.015} \\
  $K_s$ & --- ~~~~~~~~  & 0.0 $\pm$ \textit{0.016} & 0.0 $\pm$ \textit{0.019}   \\
  \noalign{\smallskip}
  \hline
  \noalign{\smallskip}
  \multicolumn{4}{l}{\begin{footnotesize} $^{*}$ $m_{_{T07}} - m_{_{LenS}}$ \end{footnotesize}}  \\
\end{tabular}
\end{table}

\subsection{Accuracy and precision of photometric redshifts}
\label{zp_accu}

\begin{figure*}[!ht]
\centering
\includegraphics[width=0.99\hsize]{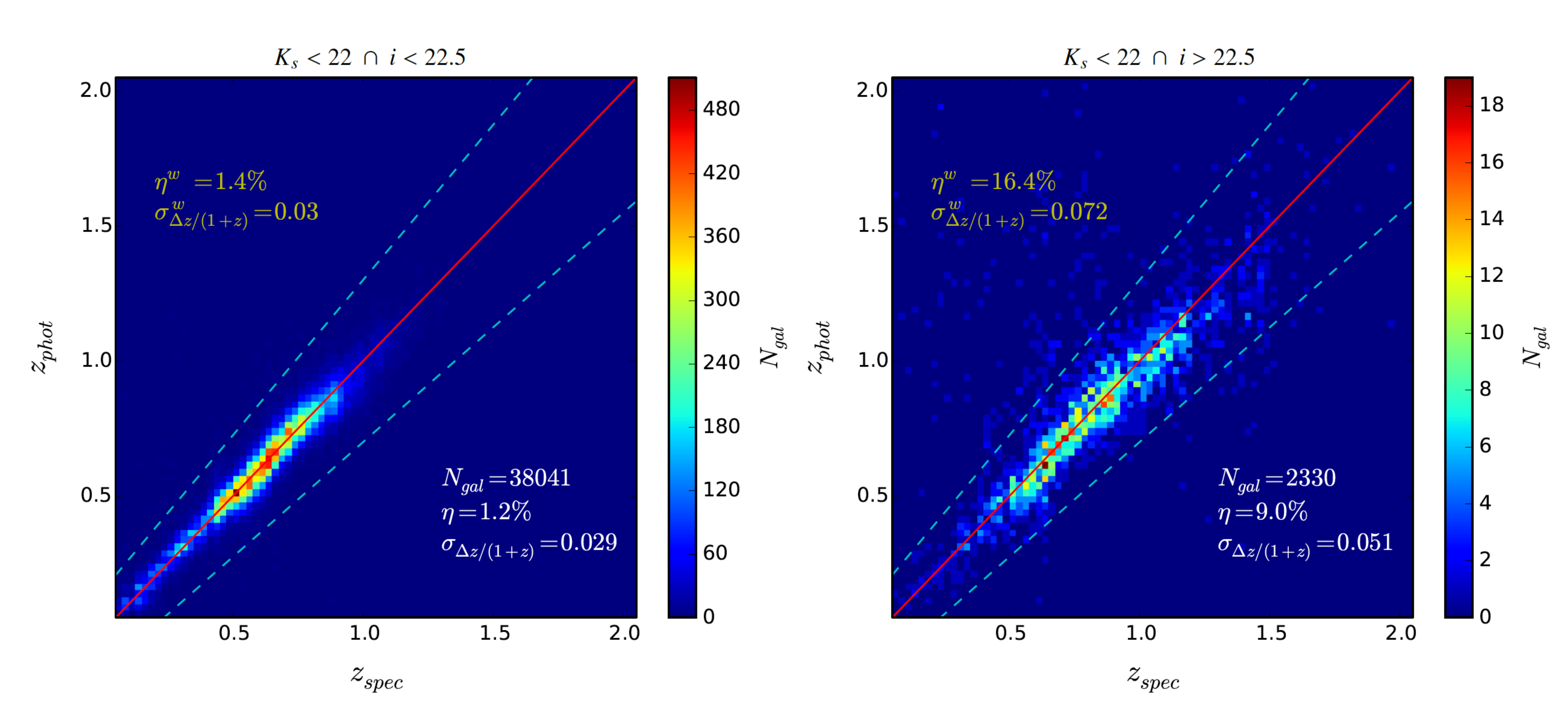}\\

\includegraphics[width=0.49\hsize, trim = 0cm 0cm 0cm 1.5cm, clip]{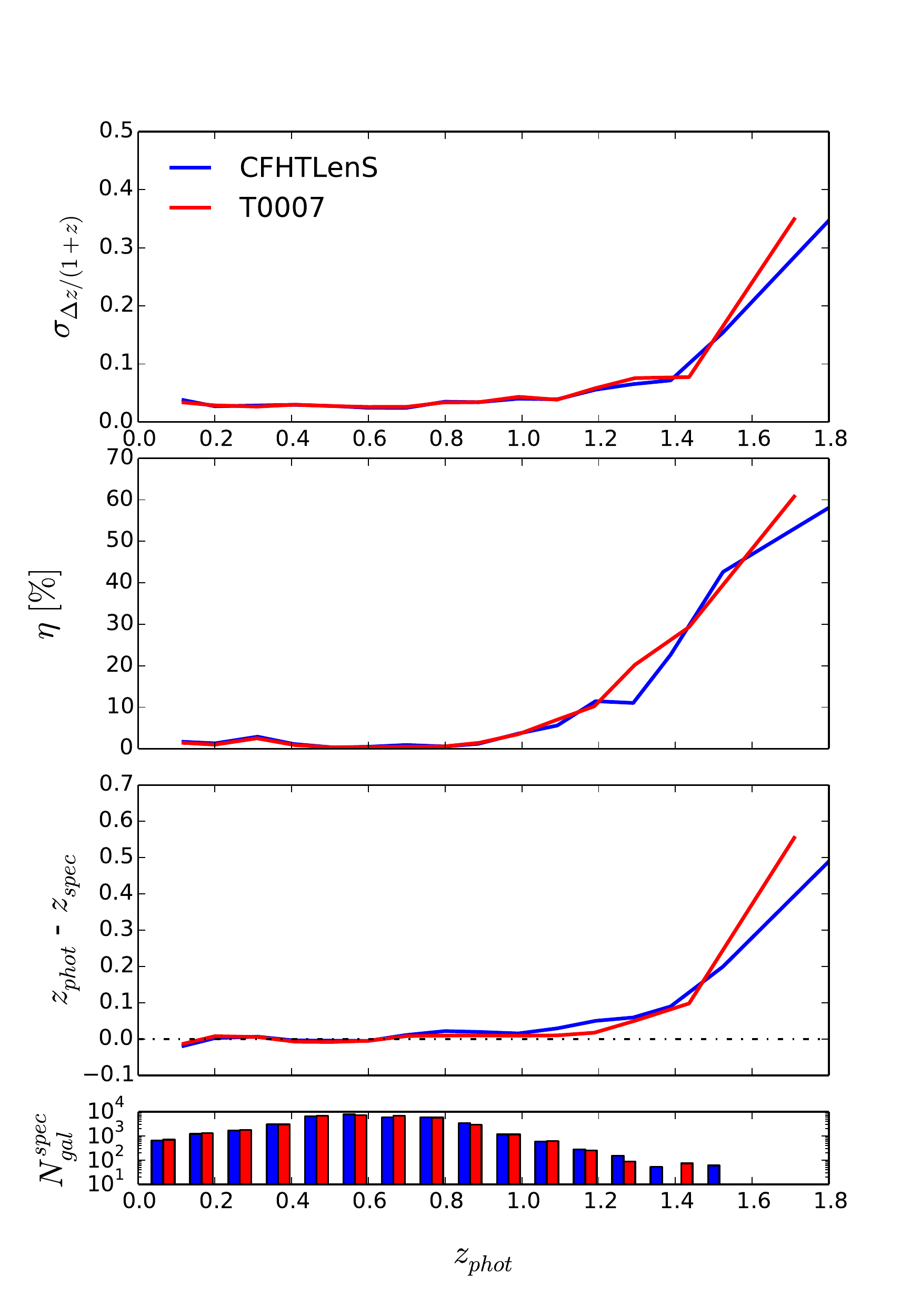}
\includegraphics[width=0.49\hsize, trim = 0cm 0cm 0cm 1.5cm, clip]{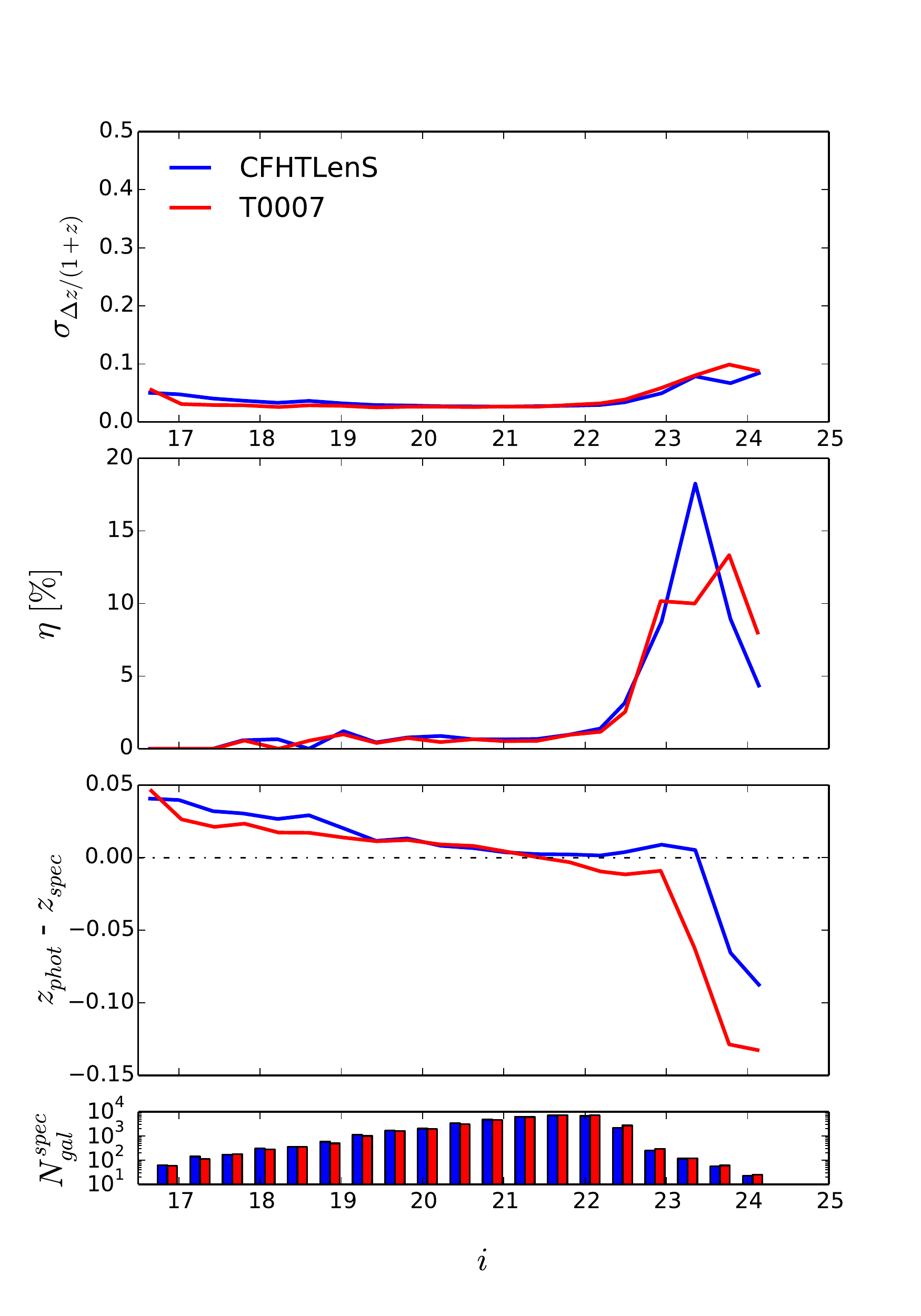}
\caption{Photometric redshift accuracy of our $K_s < 22$-limited sample. \textit{\textbf{Top}} : T0007 photometric redshift as a function of spectroscopic redshift for bright ($i < 22.5 ~\cap~ K_s < 22$) and faint ($i > 22.5 ~\cap~ K_s < 22$) galaxies. The dashed lines delimit the $\sigma_{\Delta z/(1+z)} \leq 0.15$ area, outside which photo-z measurements are considered as outliers. The accuracy estimators written in the upper left corners are weighted with respect to the $i$-band distribution of our photometric sample (see Sect. \ref{zp_accu}).  \textit{\textbf{Bottom}}: Dispersion, outlier rate, bias, and spectroscopic redshift number ($N^{spec}_{gal}$) as a function of photometric redshift (left) and $i$-magnitude (right), using the T0007 (blue) and CFHTLenS (red) optical photometry.  \label{zp_zs}}
\end{figure*}

The comparison between our photometric redshifts and the corresponding spectroscopic redshifts for our $K_s < 22$ -limited sample is shown in Fig. \ref{zp_zs}. Using the NMAD to define the scatter\footnote{$\sigma_z = 1.48~median(~|z_{spec}-z_{phot}| / (1+z_{spec})~)$}, we find  $\sigma_{\Delta z/(1+z)} \sim 0.05$  for faint ($i > 22.5$) galaxies, while the scatter reaches $\sigma_{\Delta z/(1+z)} \sim 0.03$ for the bright ($i < 22.5$) galaxies. Our photo-z outlier rate\footnote{$\eta$ is the percentage of galaxies with $\Delta z/(1+z)>0.15$} is $\eta = 1.2\%$ and $\eta = 9\%$ for corresponding bright and faint samples, respectively (see Fig. \ref{zp_zs}, top panels, lower right corners).

Although the spectroscopic sample has been assembled to be as representative as possible, it is not as deep as the photometric sample. Aiming to correct this effect, we computed estimators that are weighted with respect to the $i$-band distribution of the photometric sample. By using these weighted estimators (marked with an orange $w$  in Fig. \ref{zp_zs}, top panels), we obtained an accuracy $\sigma_{\Delta z/(1+z)}^{w} \sim 0.03$ for bright ($i < 22.5$) galaxies with an outlier rate of $\eta^w = 1.4\%$, and $\sigma_{\Delta z/(1+z)}^{w} \sim 0.07$ and $\eta^w = 16.4\%$ for faint ($i > 22.5$) galaxies in our $K_s < 22$ -limited sample.
 
Even though the T0007 and CFHTLenS calibrations differ, the photometric redshifts obtained in both cases agree well\footnote{$z_{phot}^{ _{ ~T07}}-z_{phot}^{ _{ ~LenS}} = -0.008 \pm 0.048 ~$ for $ ~0.2 < z_{phot}^{ _{ ~T07}} \leq 1.5$} and their accuracies are similar. This is expected from the colour corrections described in Sect. \ref{zp_method}, which absorb the differences between the two calibrations. 

Finally, based on Fig. \ref{zp_zs}, we can define a range of reliable redshifts up to $z = 1.5$, with $\sigma_{\Delta z/(1+z)}(z) < 0.1$. The highest redshift bin that we consider, namely between $z = 1.1$ and $z = 1.5$, is characterised by the weighted accuracy $\sigma_{\Delta z/(1+z)}^w \sim 0.08$ and weighted outlier rate $\eta^w \sim 20\%$.

\subsection{Star and galaxy classification }
\label{star_gal}

Being able to separate galaxies and stars is crucial in our sample, especially for the W4 field, which is close to the Galactic plane and therefore highly populated by stars. \citet{Garilli2008} have found that more than 32\% of the objects in the VVDS-Wide survey are stars. This is a pure $i < 22.5$ -selected spectroscopic survey lying in the CFHTLS W4 field. Aiming to better control the type of the objects that we select as galaxies without compromising the completeness of our sample, we performed a classification based on three different diagnostics. Our classification is presented in detail in the companion paper \citep{Moutard2016a} and is summarised below.
\begin{itemize}
\item[•] First, we used the maximum surface brightness versus magnitude (hereafter $\mu_{max}-m_{obs}$) plane where bright point-like sources are well separated from galaxies \citep[see][]{Bardeau2005, Leauthaud2007}.
\item[•] Secondly, we compared the reduced $\chi^2$ obtained with galaxy templates described in Sect. \ref{zp_method} and a representative stellar library \citep[based on][]{Pickles1998}. An object can be defined as a star when its photometry is better fitted by a stellar spectrum.
\item[•] Finally, we used the $g-z/z-K_s$ plane \citep[equivalent to the $BzK$ plane of][]{Daddi2004} to isolate the stellar sequence and imposed that a star belong  to this colour region.
 This sine qua non condition enabled us to catch faint stars while preventing us from losing faint compact galaxies.
\end{itemize}

We also identified a sample of QSOs (Type-1 AGNs) as point-like sources lying on the galaxy side of the BzK diagram. Dominated by their nucleus, the emission of these AGN galaxies is currently poorly linked to their stellar mass. However, they represent less than 0.5\% of the objects, and we removed them from our sample without compromising its completeness. 

All the objects that were not defined as stars or QSOs were considered as galaxies. We verified on a sample of 1241 spectroscopically confirmed stars that we caught 97\% of them in this way, while we kept more than 99\% of our spectroscopic galaxy sample. With this selection we finally found and removed $\sim$ 8\% and $\sim$ 19\% of objects at $K_s$ < 22 for W1 and W4, respectively, outside the masked area.


\section{Stellar mass estimation}
\label{mass}

\subsection{Method}
\label{mass_method}

Stellar mass, $M_*$, and the other physical parameters were computed by using the stellar population synthesis models of \citet{BC2003} with \lephare. As in \citet{Ilbert2013}, the stellar mass corresponds to the median of the stellar mass probability distribution ($PDF_{M_*}$) marginalised over all other fitted parameters. Two metallicities were considered ($Z=0.008$ and $Z=0.02$ i.e. $Z{\sun}$) and the star formation history declines exponentially following $\tau^{-1} e^{-t/\tau}$ with nine possible $\tau$ values between 0.1 Gyr and 30 Gyr as in \citet{Ilbert2013}. 

The importance of the assumed extinction laws for the physical parameter estimation has been stressed in several recent studies,
for example, by \citet{Ilbert2010} and \citet{Mitchell2013} for the stellar masses, or by \citet{Arnouts2013} for the star formation rate (SFR). We considered three laws and a maximum dust reddening of E(B-V) $\leq$ 0.5: the \citet{Prevot1984}, the \citet{Calzetti2000},
and an intermediate-extinction curve \citep[see][for more details]{Arnouts2013}. As in \citet{Fontana2006}, we imposed a low extinction for low-SFR galaxies (E(B-V) $\leq$ 0.15 if age/$\tau$ > 4). The emission-line contribution was taken into account following an empirical relation between UV and line fluxes \citep{Ilbert2009}.

\begin{figure}[!]
\centering
\includegraphics[width=\hsize]{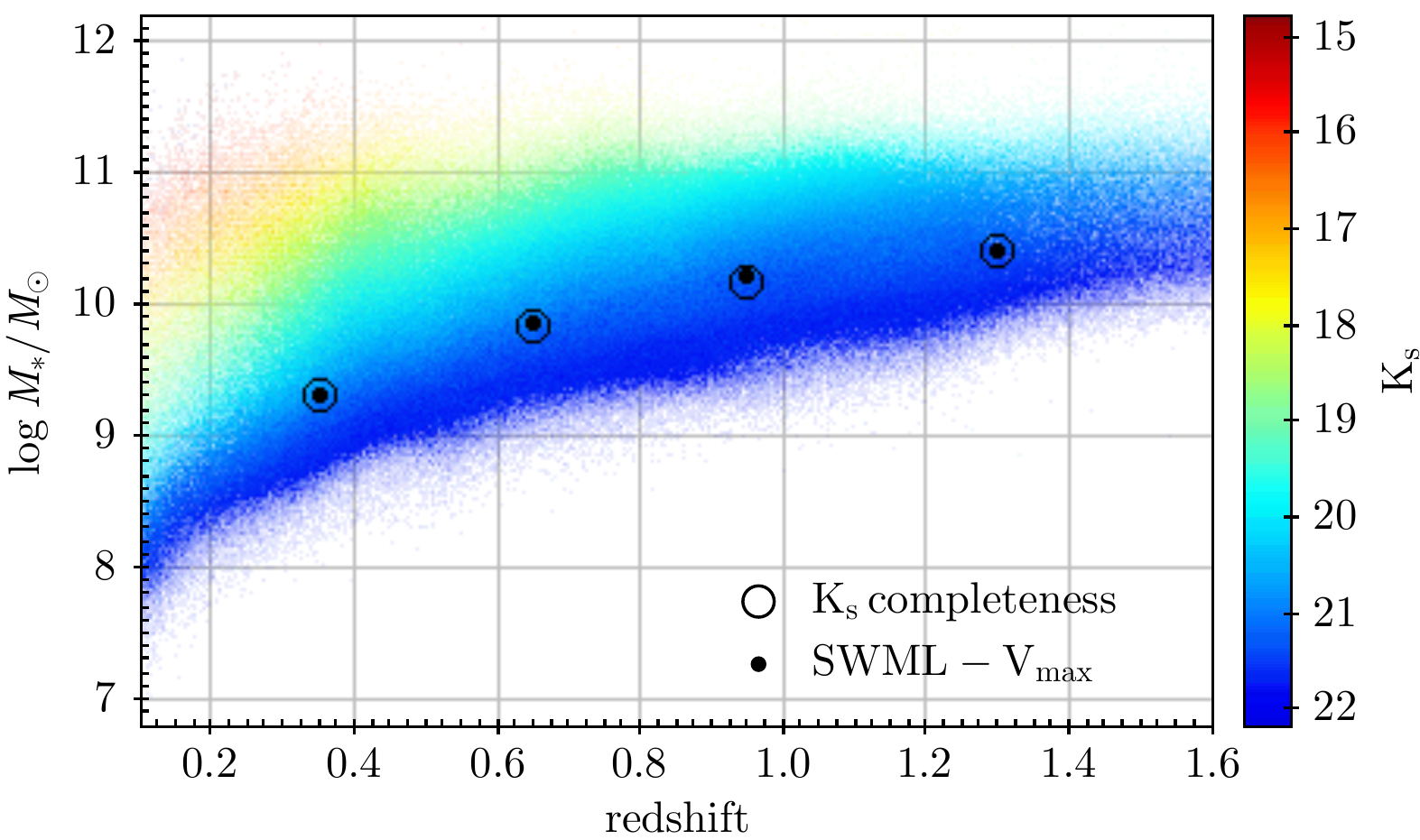}
\caption{Stellar mass versus redshift for our $K_s < 22$-limited sample of galaxies. The black open circles represent the stellar mass completeness limit computed with the $K_s$ completeness limit $M_{lim}$ according to Eq. \ref{mass_lim_eq}. The black dots represent the mass at which the $V_{max}$ and the SWML SMF estimators diverge (see below Sect. \ref{smf_measur}). \label{mass_comp}}
\end{figure}

Using a method similar to \citet{Pozzetti2010}, we based our estimate of the stellar mass completeness limit, $M_{lim}$, on the  distribution of the lowest stellar mass, $M_{min}$, at which a galaxy could have been detected given its redshift. For our sample, which is limited at $K_s < 22$, $M_{min}$ is given by 
\begin{equation}
log(M_{min}) = log(M_*) + 0.4 \ (K_s - 22)
\label{mass_lim_eq}
.\end{equation}
We then considered the upper envelope of the $M_{min}$ distribution. In each redshift bin, $M_{lim}$ is defined by the stellar mass at which 90\% of the population have $M_* > M_{min}$. We show the resulting stellar mass completeness limits (open circles) as a function of redshift in the Fig. \ref{mass_comp} over the $M_*$ distribution for our $K_s < 22$-limited sample.

\subsection{Stellar mass error budget}
\label{mass_errors}

In this section, we quantify the uncertainties associated with the stellar masses, which will be propagated into the error budget of the stellar mass functions. The first to be considered is the uncertainty in the flux measurements. The photon noise is taken into account by the \lephare code during the $\chi^2$ SED fitting procedure  (rescaling and model selection), where it returns the 68\% confidence interval enclosed in  the probability distribution function marginalised on the stellar mass ($PDF_{M_*}$). 

\begin{figure}[!ht]
\centering
\includegraphics[width=0.99\hsize]{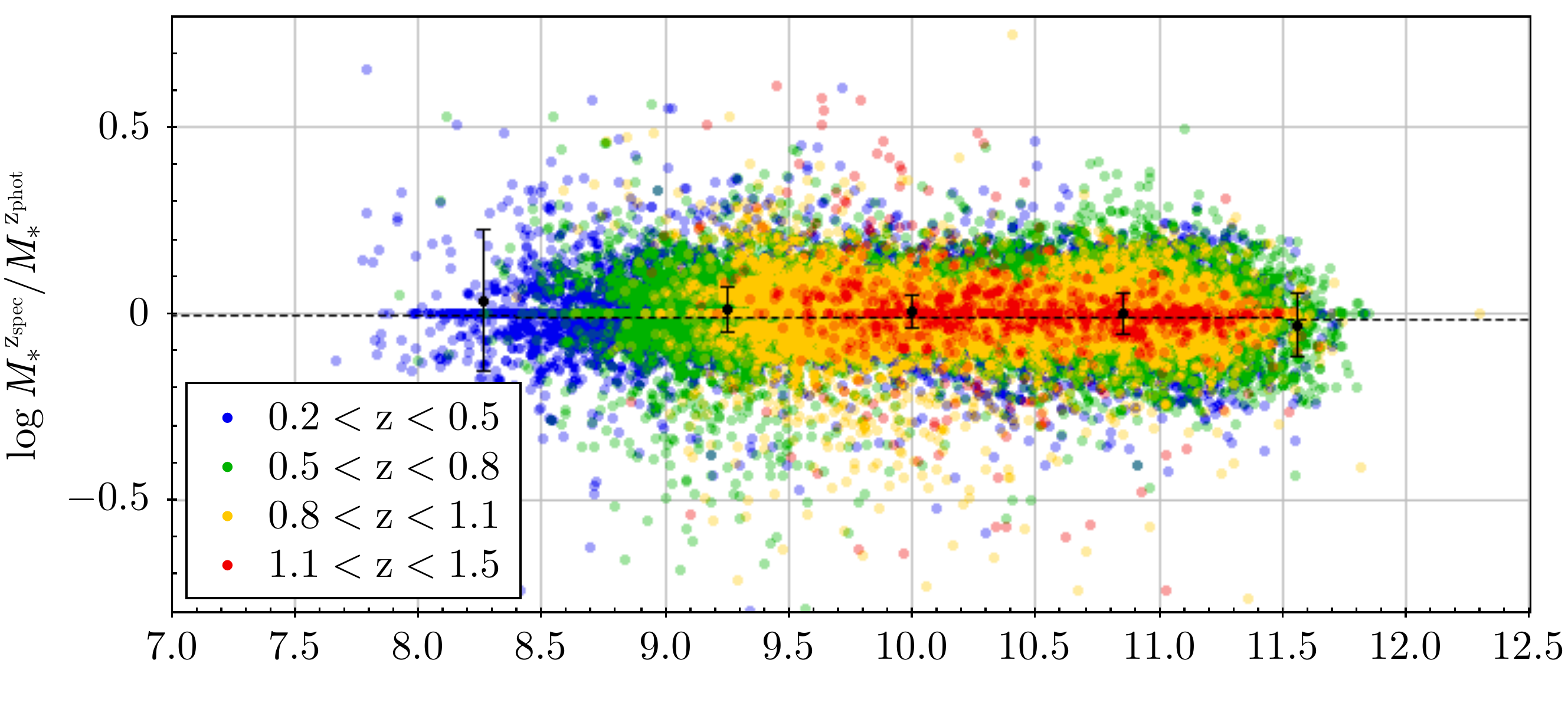}
\includegraphics[width=0.99\hsize]{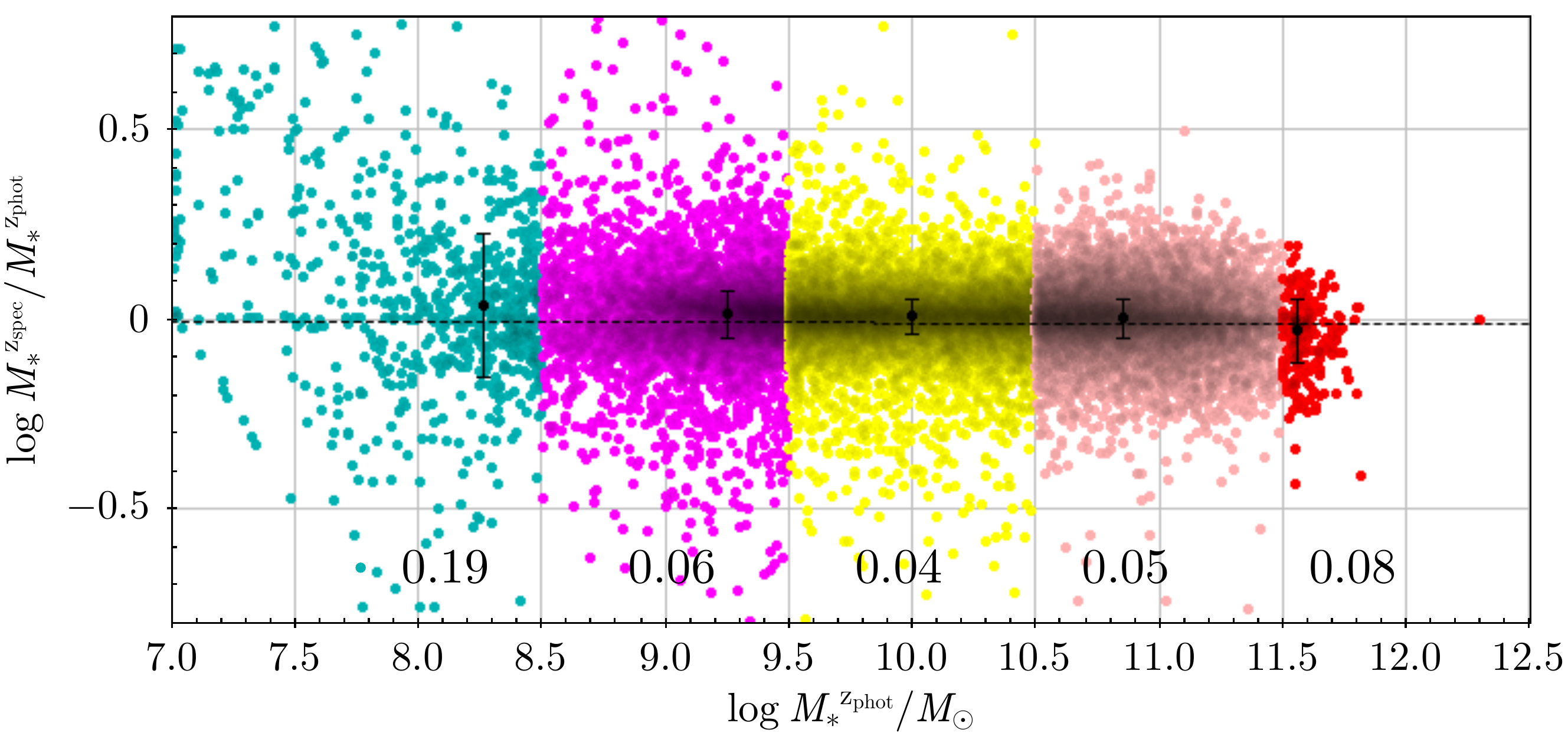}
\caption{Redshift contribution to the stellar mass uncertainty as a function of the stellar mass. The uncertainty is computed from the 1 $\sigma$ dispersion of the ratio $M_*(z_{spec})/M_*(z_{phot})$ in the spectroscopic sample after removing photo-z outliers. In the \textit{\textbf{top panel}}, the distribution is shown in four redshift bins, while in the \textit{\textbf{bottom panel}} it is shown in five stellar mass bins. The error bars correspond to the dispersion reported in each bin of $M_*$, while the dashed line is the linear regression associated with the whole sample.  \label{Mzs_Mzp}}
\end{figure}

The second source of error is introduced by the photometric redshift uncertainty, which is not included in the $PDF_{M_*}$. One way to measure its effect is to compare  the stellar masses derived with the photometric and spectroscopic redshifts. We emphasise that this analysis is probably limited by the completeness of our spectroscopic sample. By contrast, it is powerful when used
to reflect all the photo-z error contributions (quality of the photometry and representativity of the templates).
The difference between the two mass estimates is shown in Fig. \ref{Mzs_Mzp} as a function of stellar mass $M_*^{z_{phot}}$. In the top panel, we show the difference in four redshift bins between $z = 0.2$ and $z = 1.5$. No dependence with redshift is observed.  The linear regression of the whole sample, plotted as a black dashed line, also suggests that it is not mass dependent. The bottom panel shows the $M_*^{z_{phot}} / M_*^{z_{spec}}$ dispersion in five stellar mass bins and reveals a median dispersion of 0.06 dex, with a maximum of 0.19 dex at low mass. 

We then defined the resulting mass uncertainty as the sum in quadrature of all contributions:
\begin{equation}
\sigma_{M} = \sqrt{ ~\sigma_{fit}^2 + \sigma_{z}^2~ }
\label{eq_err_M}
.\end{equation}

However, we have to keep in mind that the stellar mass estimation relies on the numerous assumptions made when we generate our SED templates. For example, \citet{Maraston2005} has shown that a different treatment of the thermally pulsing asymptotic giant branch (TP-AGB) phase in the SSP can lead to a global shift in the stellar mass estimation\footnote{\citet{Pozzetti2007} and \citet{Ilbert2010} estimated on offset of $\sim 0.14$ dex between the stellar masses of BC03 and \citet{Maraston2005}.}. \citet{Ilbert2010} showed that the use of the \citet{Salpeter1955} IMF instead of
the Chabrier IMF decreases the stellar masses by $0.24$ dex. These systematic shifts are therefore not expected to affect the conclusions of our study. \citet{Mitchell2013} also pointed out the potential effect of the assumed dust attenuation on the stellar mass estimation\footnote{\citet{Mitchell2013} estimated that the stellar mass can be underestimated by up to 0.6 dex by assuming the \citet{Calzetti2000} for massive galaxies.}. As presented in the previous section,  we considered three different extinction laws. This allows a higher diversity of possible values for dust attenuation, which is expected to limit the bias that may affect our stellar mass estimation.

\subsection{Effect of the CFHTLS absolute calibration}
\label{systematics}

\begin{figure}[!]
\includegraphics[width=0.499\hsize, trim = 0cm 2.1cm 0cm 0cm, clip]{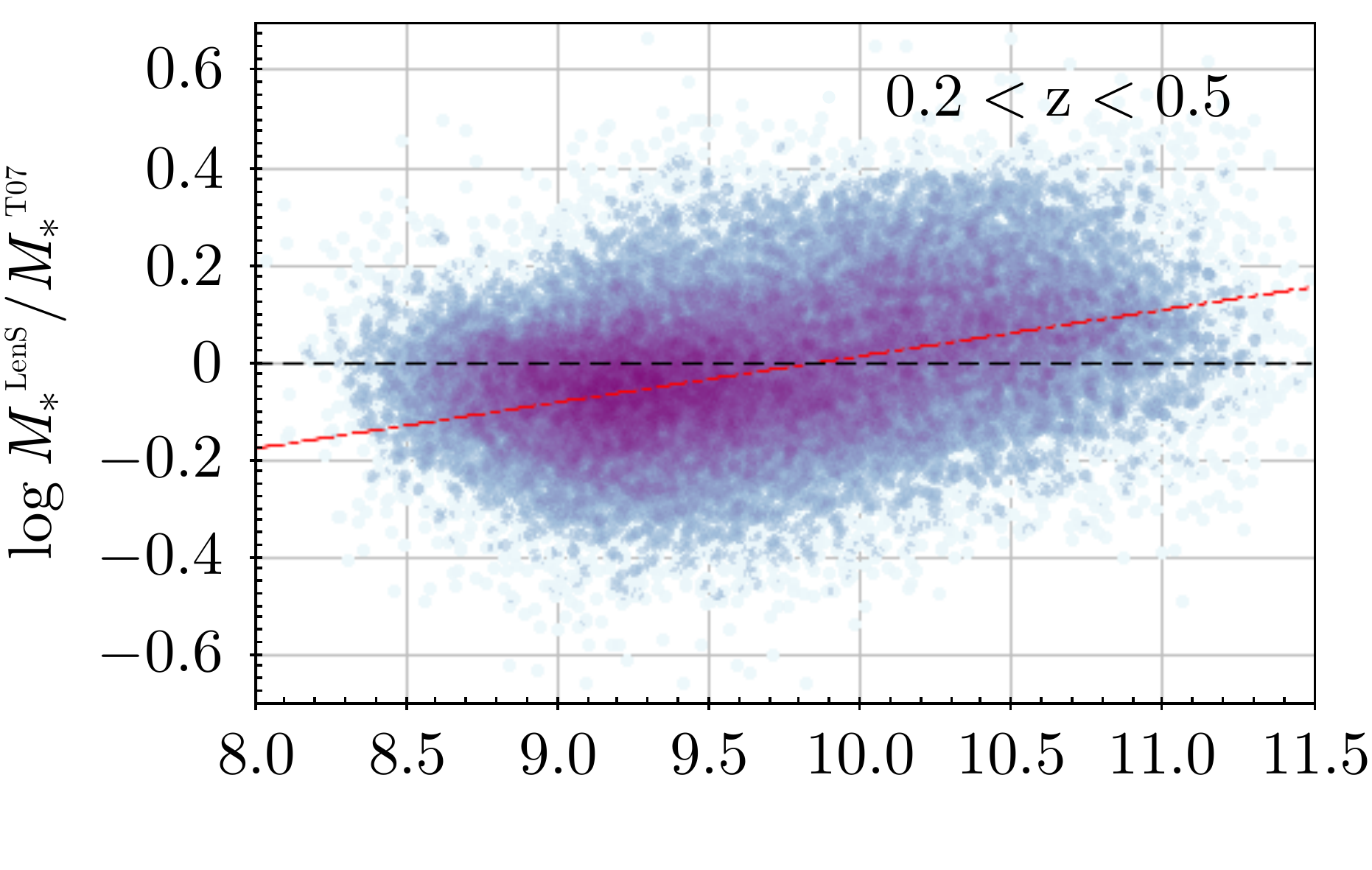}
\hspace{-0.4cm}\includegraphics[width=0.499\columnwidth, trim = 0cm 2.1cm 0cm 0cm, clip]{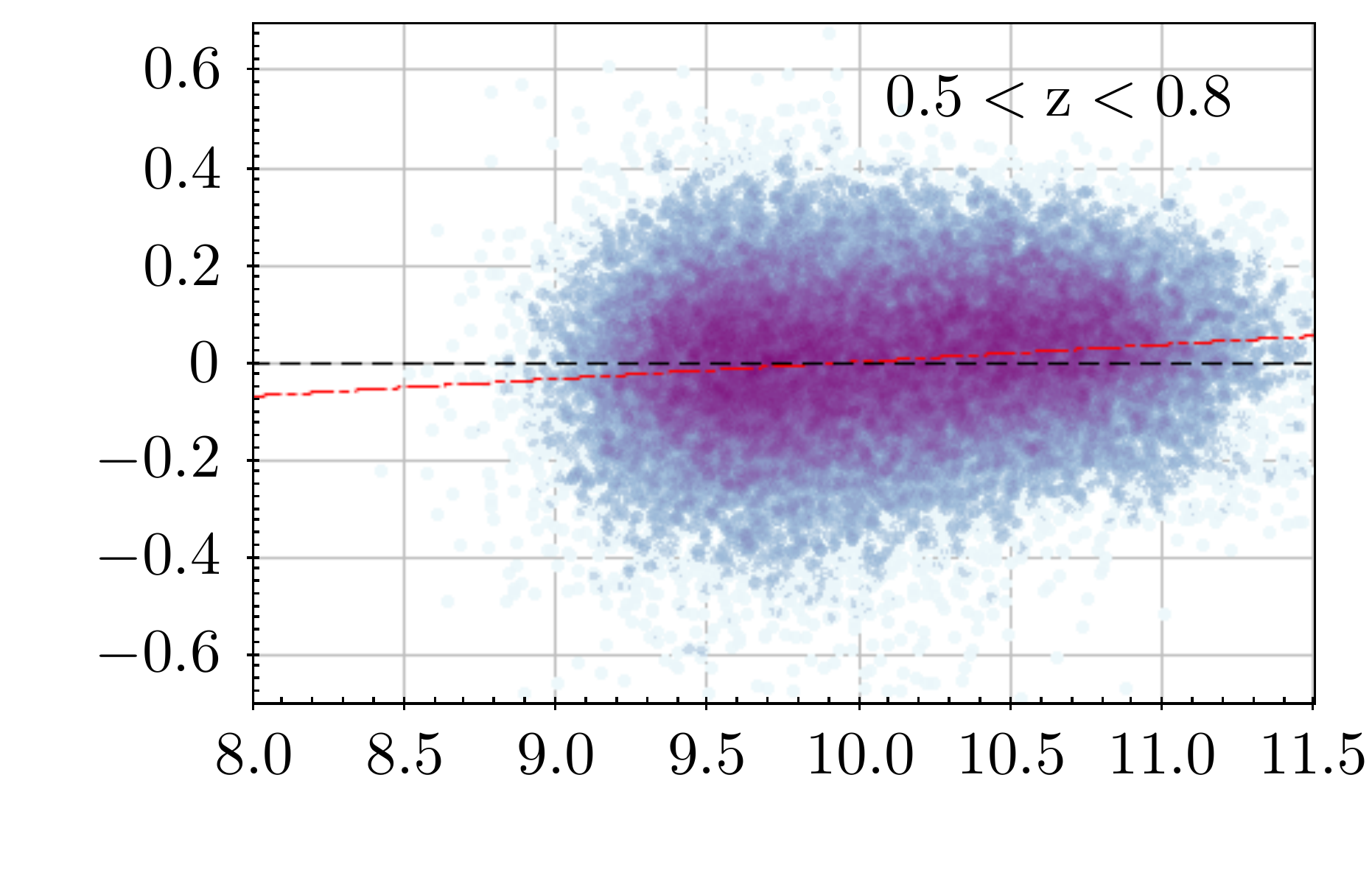}

\hspace{0.01cm}\includegraphics[width=0.497\hsize]{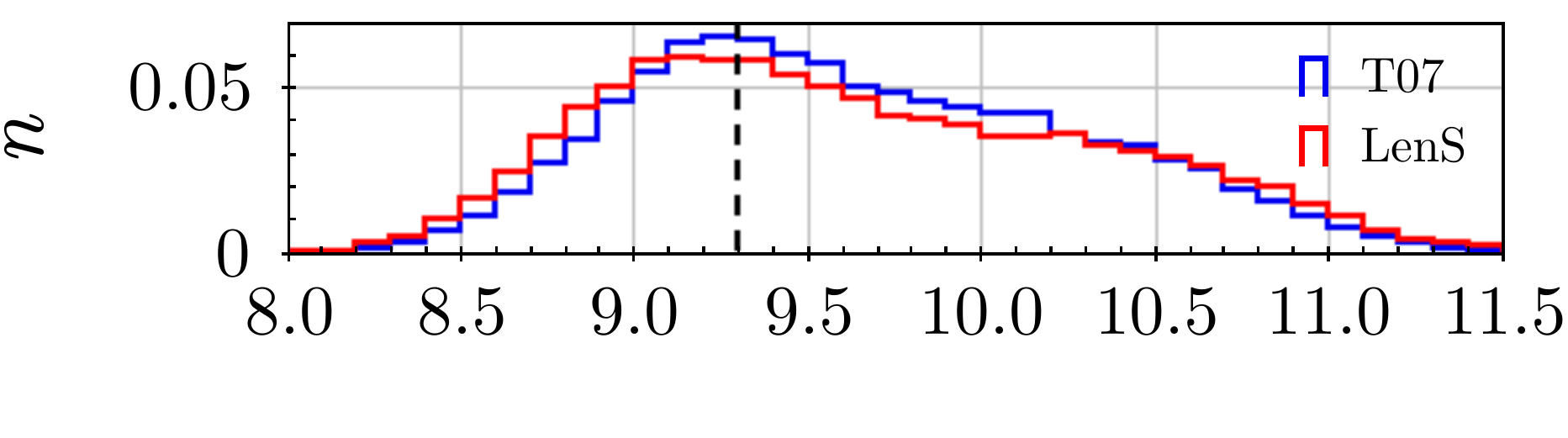}
\hspace{-0.38cm}\includegraphics[width=0.497\columnwidth]{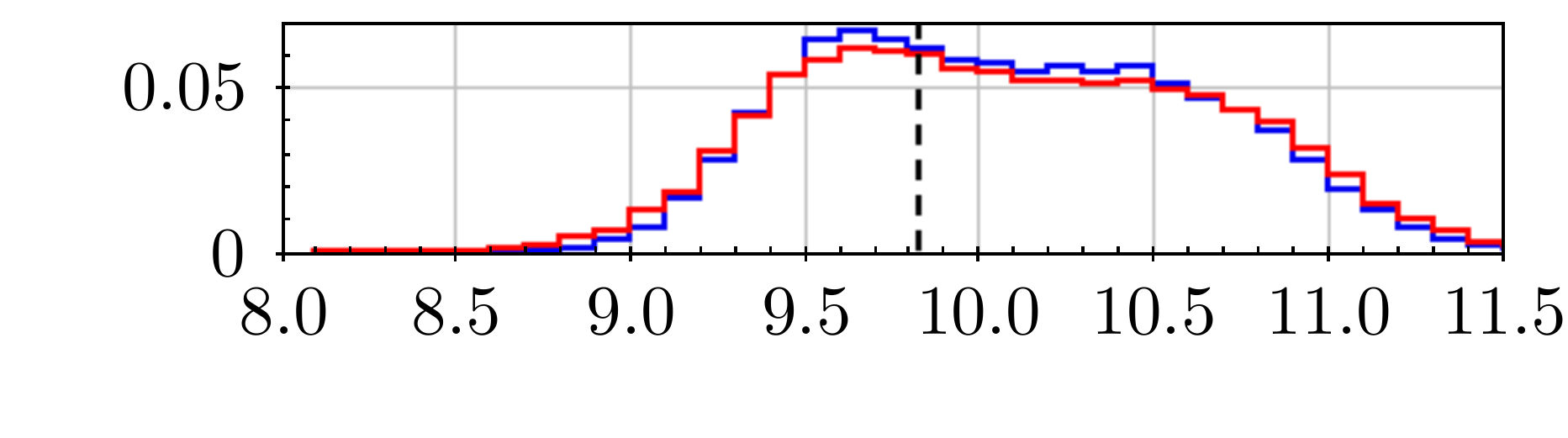}

\includegraphics[width=0.499\hsize, trim = 0cm 2.1cm 0cm 0cm, clip]{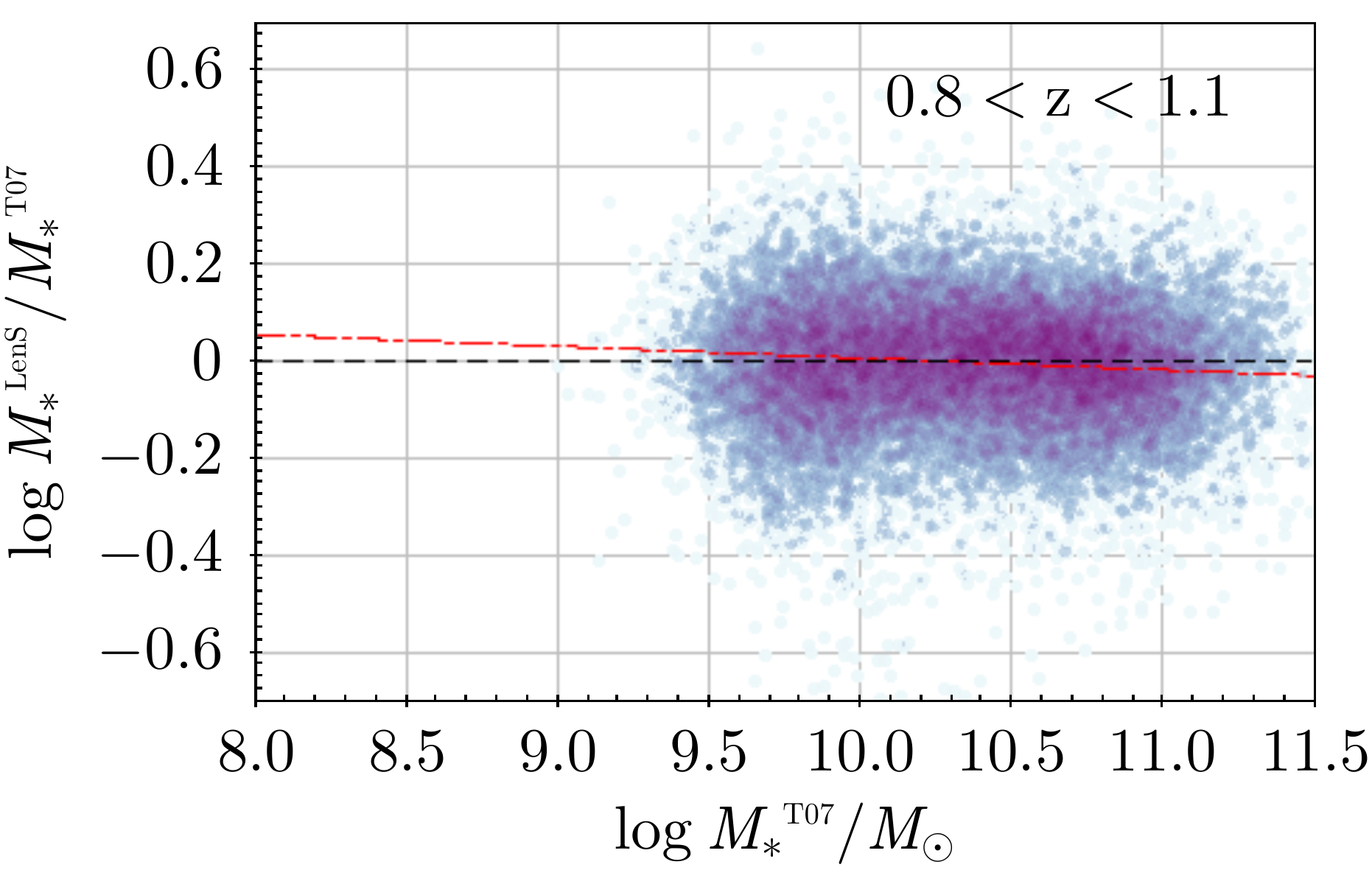}
\hspace{-0.4cm}\includegraphics[width=0.499\columnwidth, trim = 0cm 2.1cm 0cm 0cm, clip]{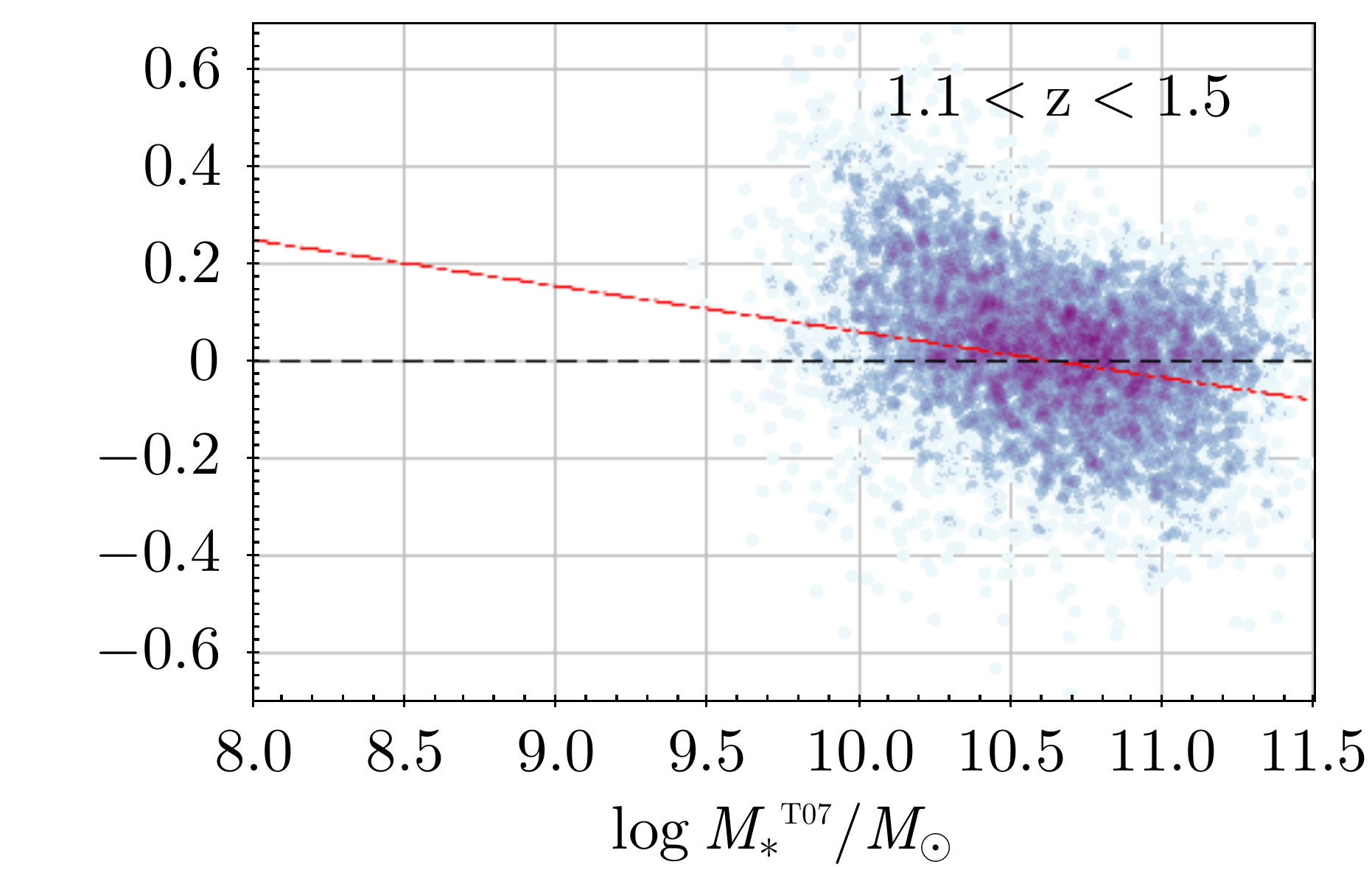}

\hspace{0.01cm}\includegraphics[width=0.497\hsize]{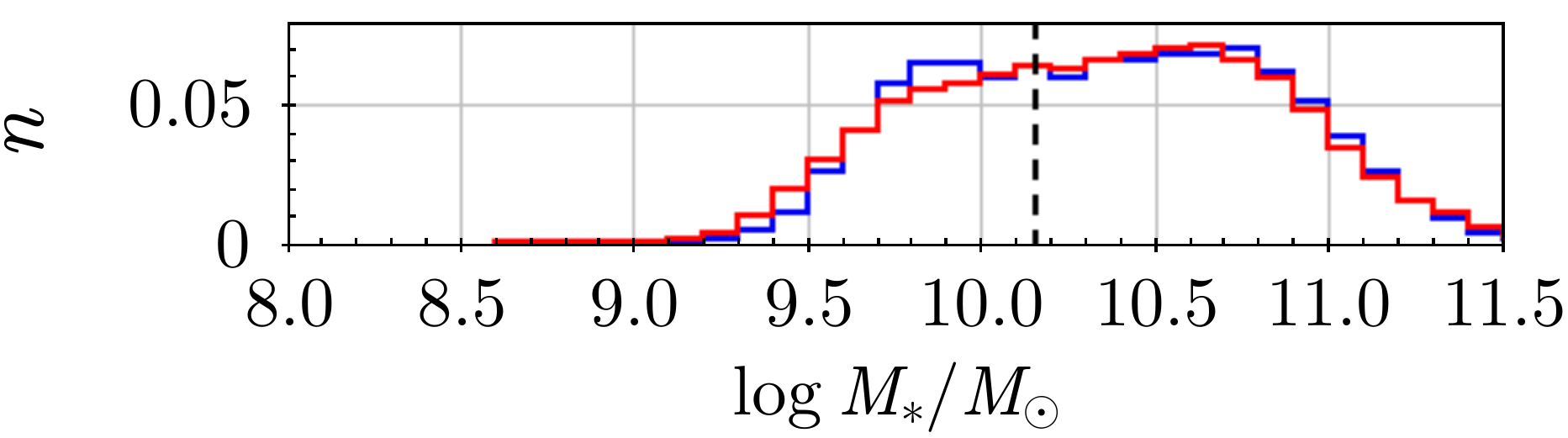}
\hspace{-0.38cm}\includegraphics[width=0.497\columnwidth]{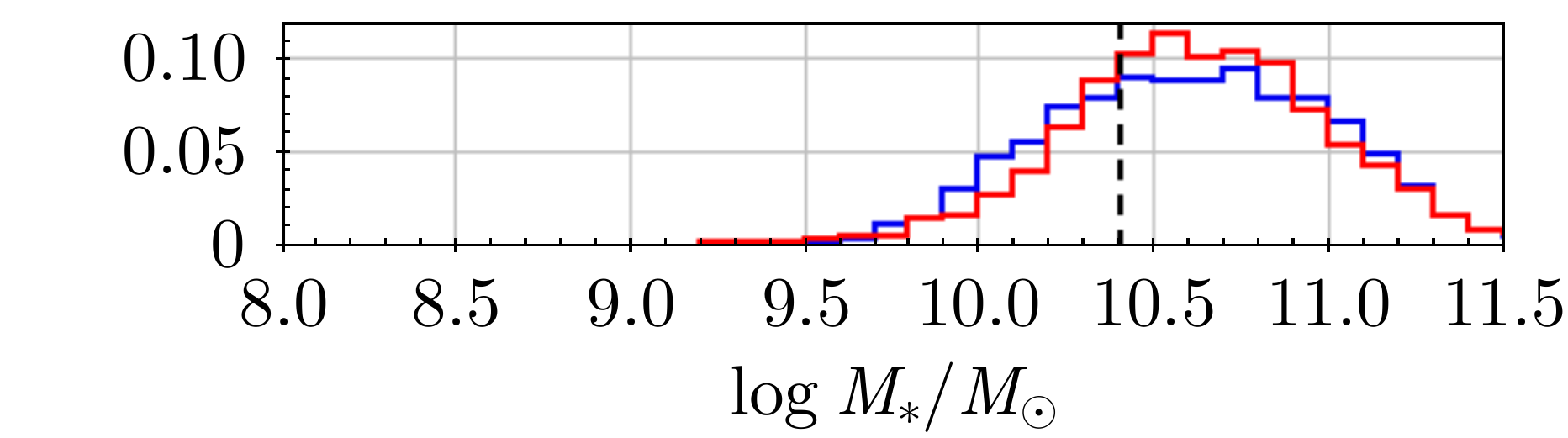}
\caption{Differences between the stellar mass obtained with T0007  and CFHTLenS, $M_*^{ T07}$ and $M_*^{ LenS}$  at different redshifts: the object-by-object $M_*^{ LenS} / M_*^{ T07}$ ratio versus $M_*^{ T07}$ in the upper panel, where the red dashed line is the linear regression, and the $M_*^{ T07}$ (blue) and $M_*^{ LenS}$ (red) normalised number counts in the lower panel, where the vertical black dashed line represents the mass completeness limit.\label{T07_LENS_dMASS}}
\end{figure}

As shown in Sect. \ref{optical}, the absolute photometric calibrations of the T0007 and CFHTLenS magnitudes differ by more than 0.05 mag on average. Even if the T0007 were significantly improved in its calibration, we compare the photometric redshifts and the stellar masses computed with both catalogues blindly to quantify the effect of these offsets.

As seen in Sect. \ref{zp_accu}, the colour corrections applied during the photometric redshift computation allows us to obtain very similar photo-z despite the offset between their calibrations. However, these corrections are a combination of the photometry and the SEDs used to calculate photo-z. As described in Sect. \ref{zp_method}, the templates used for photo-z are different from those used for the masses. Consequently, we did not apply the photo-z colour corrections with the BC03 templates. The differences in the T0007 and CFHTLenS photometries thus directly affect the stellar mass estimation.

Figure \ref{T07_LENS_dMASS} presents these differences in the redshift bins 0.2< z < 0.5, 0.5 < z < 0.8, 0.8 < z < 1.1 and 1.1 < z < 1.5. The difference between the stellar masses obtained with the T0007 ($M_*^{T07}$) and those obtained with the CFHTLenS ($M_*^{LenS}$) is stellar mass dependent. On average, this systematic difference can reach $\pm 0.1$ dex at low redshift ($z < 0.8$). At higher redshift, we do not observe a systematic difference between the two stellar mass catalogues since we used the same $K_s$-band calibration. Even if the object-by-object $M_*^{ LenS}$ to $M_*^{ T07}$ ratio is characterised by a mean offset that never exceeds 0.2 dex, the comparison of the T0007 and CFHTLenS number counts above the mass completeness limit reveals different shapes around $M_* \sim 10^{11} M_{\odot}$, notably at $z < 0.8$. This suggests a significant effect on the SMF massive end at low redshift, as we discussed below.


\section{Measuring the stellar mass functions}
\label{SMF}
 
To compute the galaxy stellar mass function, we selected a sample of $\sim 760,000$ galaxies at $K_s \le 22$ over an effective area of 22.38 deg$^2$. According to what we discussed in Sect. \ref{zp_accu}, we restricted our analysis to the range $0.2\le z \le 1.5$, where we combined reliable redshifts and large volumes.

The galaxy stellar mass function was derived with the tool ALF \citep{Ilbert2005}, which provides three non-parametric estimators: $V_{max}$ \citep{Schmidt1968}, SWML \citep[the step-wise maximum likelihood;][]{Efstathiou1988}, and $C^+$ \citep{Zucca1997}.  The $V_{max}$  estimator is most widely used because of its simplicity. The $1/V_{max}$ is the inverse sum of the volume in which each galaxy was observed. The $V_{max}$  is the only estimator that is directly normalised. The SWML (Efstathiou et al. 1988) determines the SMF by maximising the likelihood of observing a given stellar mass - redshift sample. The $C^+$  method overcomes the assumption of a uniform galaxy distribution, as is the case when using the $V_{max}$\footnote{For more details about these estimators, we refer to \citet{Ilbert2005} and \citet{Johnston2011}.}. As described in \citet{Ilbert2015}, these estimators diverge below a stellar mass limit that should correspond to the limit calculated in Sect. \ref{mass}. In Fig. \ref{mass_comp} we verify that the $V_{max}$ and SWML estimators (black dots) are consistent with our $K_s$-based stellar mass completeness limit (black open circles). We used the colour-magnitude weight map shown in Fig.\ref{weight_map} to correct the SMF for the potential incompleteness described in Sect. \ref{wircam_K}. In the remainder of this study, we work with stellar masses $M_* > M_{min}$ where all the non-parametric estimators agree.

\subsection{Measurements by type and field}
\label{smf_measur}

To separate quiescent and star-forming galaxies, we used the rest frame $(NUV-r)^\textsc{o}$ versus $(r-K)^\textsc{o}$ diagram (hereafter NUVrK) presented by \citet{Arnouts2013}, which is based on the method introduced by \citet{Williams2009}. 
Figure \ref{NUVrK_z} presents the galaxy distribution in the NUVrK diagram for several redshift bins.
This optical-NIR diagram allows us to properly separate red dusty star-forming galaxies from red quiescent ones. Edge-on spirals are clearly identifiable, as is illustrated by the morphological study of the NUVrK diagram at low redshift presented in the companion paper \citep{Moutard2016a}. 

When we computed the rest-frame colours, we adopted the procedure described in Appendix A.1 of \citet{Ilbert2005} to minimise the dependency of the absolute magnitudes to the template library. An absolute magnitude at  $\lambda^0$  was derived from the apparent magnitude in the filter passband that was the closest from $\lambda^0$ $\times (1+z)$ to minimise the k-correction term, except when the apparent magnitude had an error above 0.3 mag, to avoid too noisy colours.
The small break\footnote{It is important to keep in mind that the NUVrK diagram is particularly stretched along the $( r-K_s )^\textsc{o}$ axis.}  in the red clump is artificial and is an effect of the template discretisation, when our procedure used to limit the template dependency fails because of the low signal-to-noise
ratio measurements (here due to the intrinsic low rest-frame NUV emission of quiescent galaxies)\footnote{We note that this effect of discretisation from the templates is smoothed if we use a high number of templates, such as with the BC03 library.  However, this smoothing is somehow artificial since the NUV part is not better constrained in practice. We verified with BC03 that the SMF of quiescent galaxies is not significantly affected by the set of templates we used to compute absolute magnitudes.}.

As shown in Fig. \ref{NUVrK_z}, by following the low-density valley of the NUVrK diagram (the so-called \textit{green valley}), the selection of quiescent galaxies can be defined with the general form
\begin{equation}
[~ ( NUV-r )^\textsc{o} > B_2 ~]~\cap~[~ ( NUV-r )^\textsc{o} > A ~ ( r-K_s )^\textsc{o} + B_1 ~] ~.
\label{eq_NrK}
\end{equation}
$A$, $B_1$ , and $B_2$ are three parameters to be adjusted in each redshift bin,  as suggested by  \citet{Ilbert2015} and \citet{Mortlock2015},
because of the global ageing  of the galaxy population. 

In the four redshift bins, the slope $A$ of Eq. \ref{eq_NrK} seems to be constant, with a typical value of $A = 2.25$.
By projecting the galaxy distribution in a plane perpendicular to the axis of slope $A$\footnote{We selected the galaxies in the range $0.4 < (r - K_s)^\textsc{o} < 0.9$ to avoid the objects in transition.},  we clearly distinguish the red and blue clouds as two normal distributions that we fitted by two Gaussians. We define $B_1$ as the position where the two Gaussians intersect.    

\begin{figure}[t]
\includegraphics[width=\hsize]{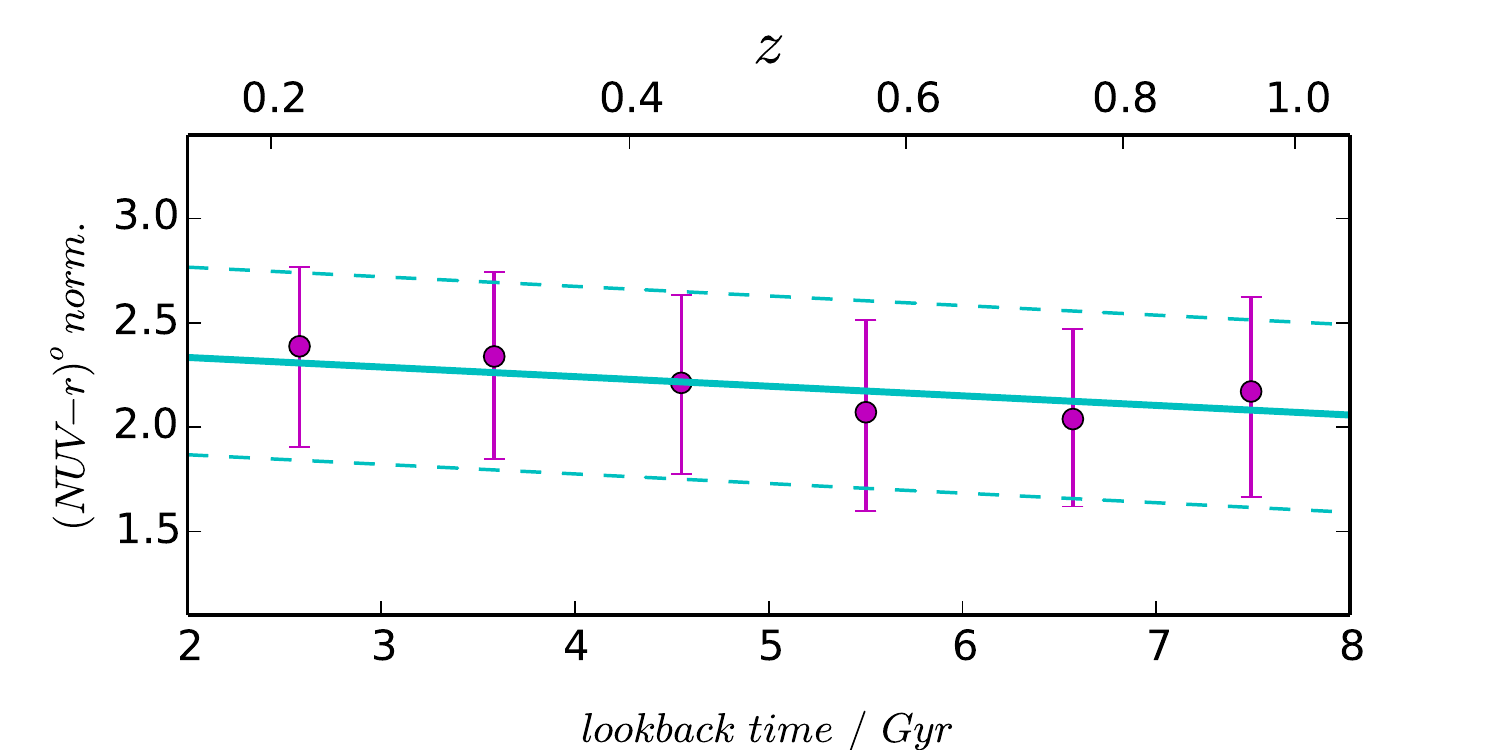}
\caption{Cosmic evolution of the $(NUV-r)^\textsc{o}$ normalisation. The dots represent the position of the minimum density along the $(NUV-r)^\textsc{o}$ axis across cosmic time, while the bars are defined by the extreme values that delimit the NUVrK green valley. The solid line is the linear fit and the dashed lines represent their mean upper and lower envelopes.\label{B_evol}}
\end{figure}

In Fig~\ref{B_evol} we show the evolution of $B_1$ as a function of the look-back time ($t_\textsc{l}$). By assuming a linear relation between $B_1$ and cosmic time, we derive $B_1(t_\textsc{l}) = -0.029$ $t_\textsc{l} +
2.368$ in our highest precision redshift range ($0.2 - 1.0$). Assuming that $B_2$ evolves as $B_1$, we empirically set $B_2(t_\textsc{l} = 2.5 Gyr) = 3.3$, and we find $B_2(t_\textsc{l}) = B_1(t_\textsc{l}) + 1.004$. We can write our selection of quiescent galaxies as
\begin{eqnarray}
\left[ ~( NUV-r )^{\textsc{o}} > 3.372 - 0.029 \ t_{\textsc{l}} ~\right] ~ \cap \nonumber\\
\left[ ~( NUV-r )^{\textsc{o}} > 2.25 \ ( r-K_s )^{\textsc{o}} + 2.368 - 0.029 \ t_\textsc{l} ~\right] ~.
\label{eq_sel}
\end{eqnarray}

All the galaxies that are not selected as quiescent are considered to be star forming. In Fig. \ref{NUVrK_z} the separations between quiescent and star-forming galaxies are shown as white solid line.  We also define the green valley as the region around minimum $B_1$, reaching 10\% of the peak of the red Gaussians, as shown by the white dotted lines. We consider these limits as possible systematic uncertainties when discussing the evolution of the quiescent and star-forming SMFs. 

\begin{figure}[!]
\includegraphics[width=0.532\columnwidth, trim = 0.5cm 1.25cm 1.1cm 0cm, clip]{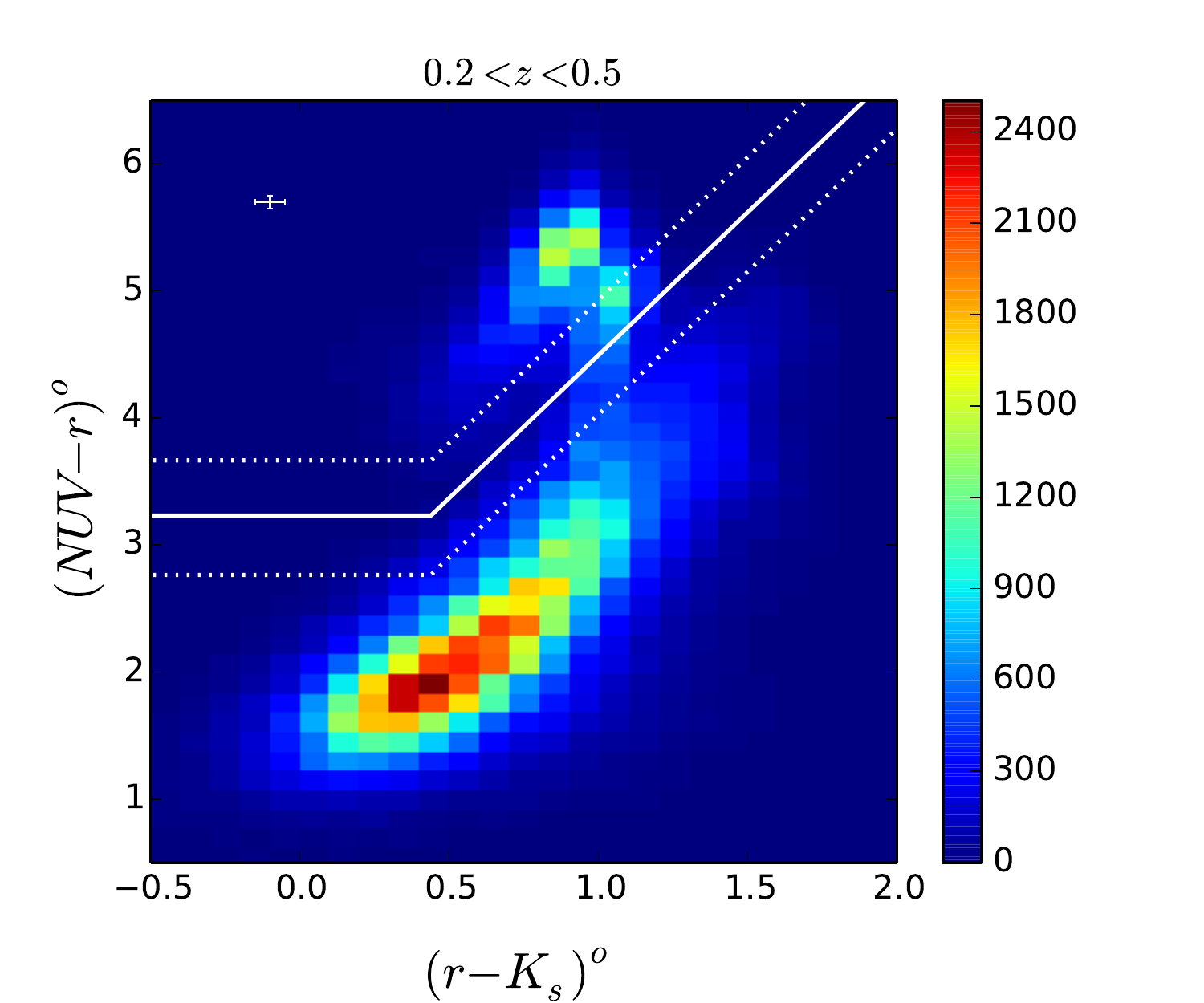}
\hspace{-0.15cm}\includegraphics[width=0.474\columnwidth, trim = 1.5cm 1.25cm 1.6cm 0cm, clip]{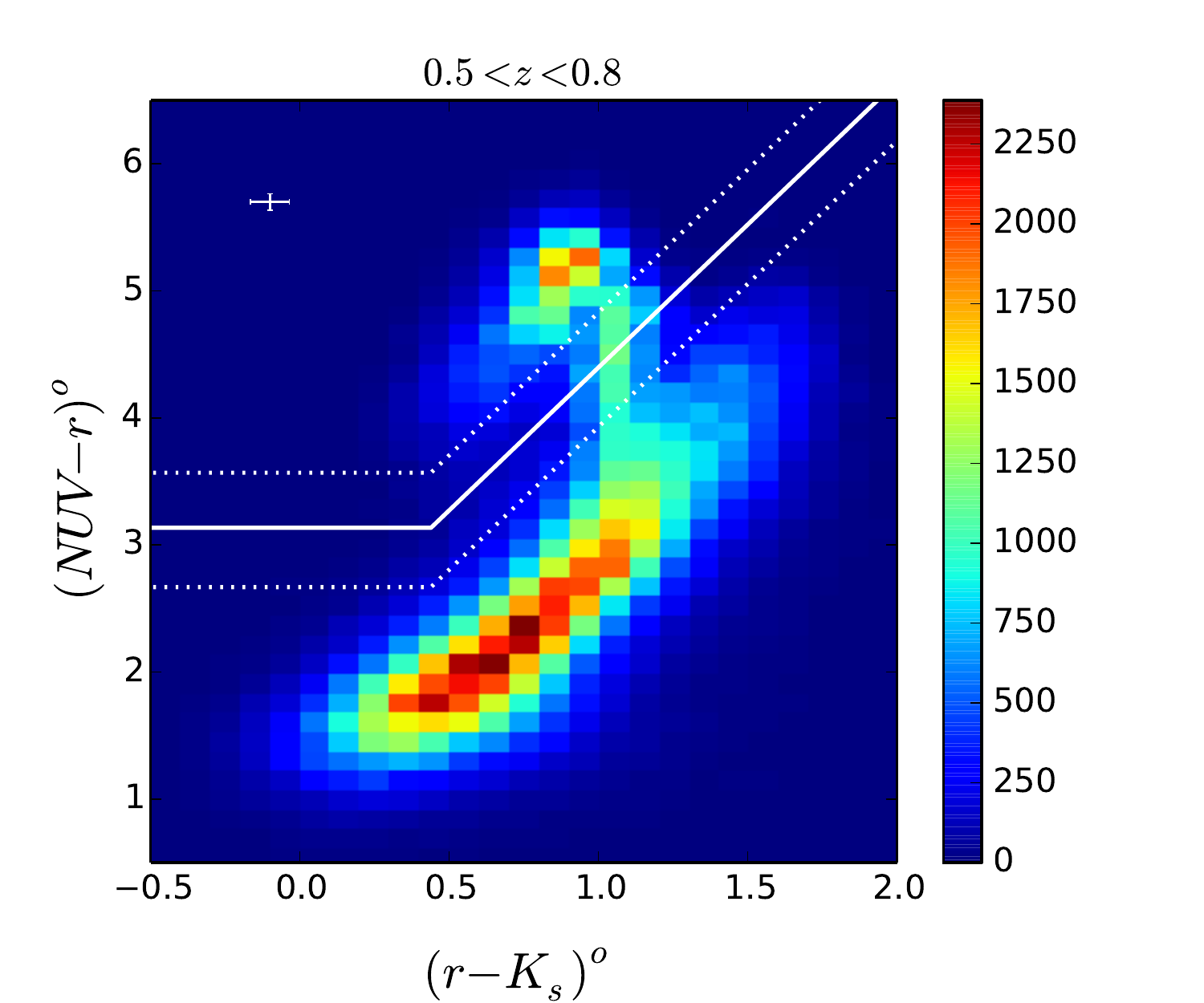}

\includegraphics[width=0.532\columnwidth, trim = 0.5cm 0cm 1.1cm 0cm, clip]{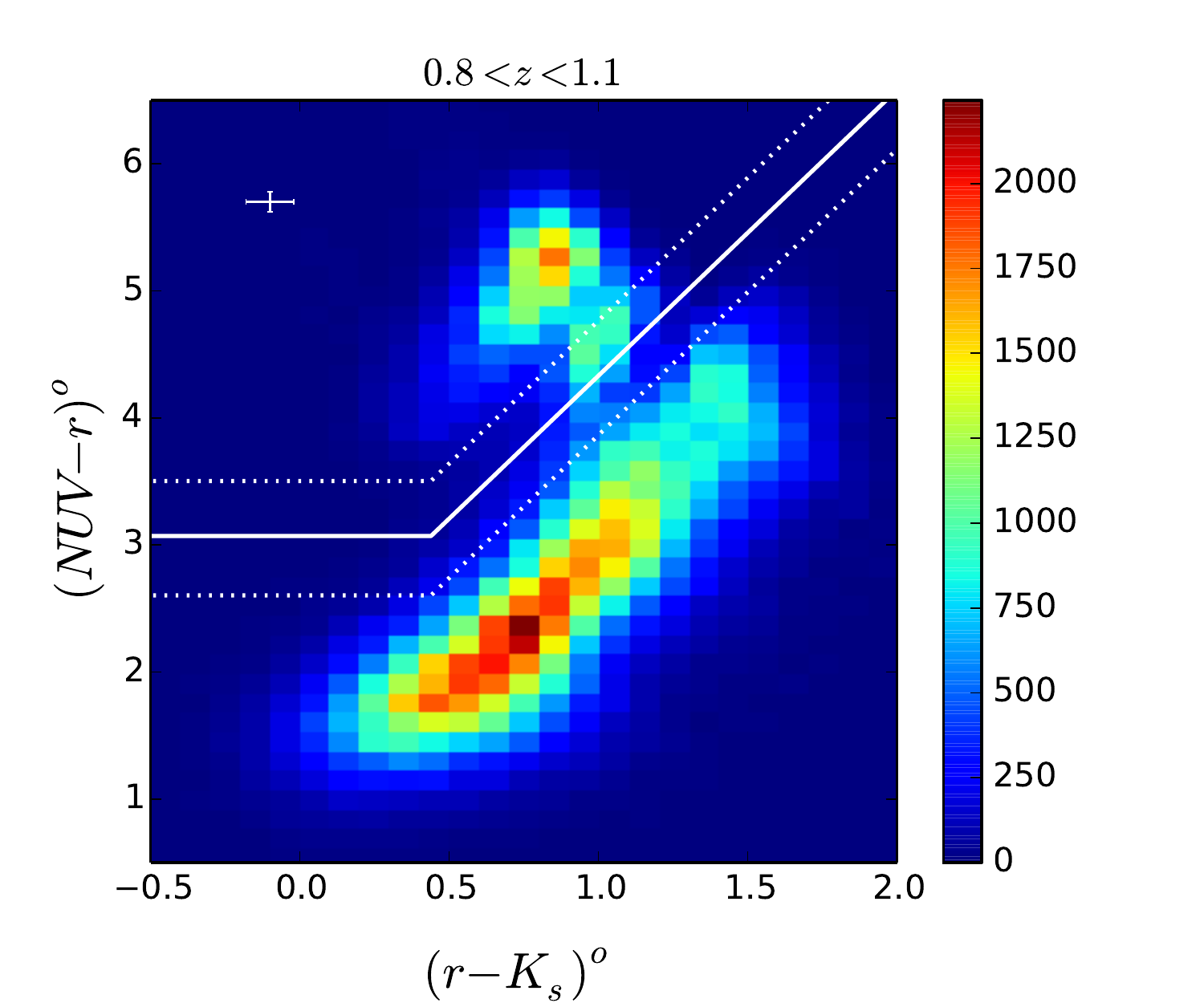}
\hspace{-0.15cm}\includegraphics[width=0.474\columnwidth, trim = 1.5cm 0cm 1.6cm 0cm, clip]{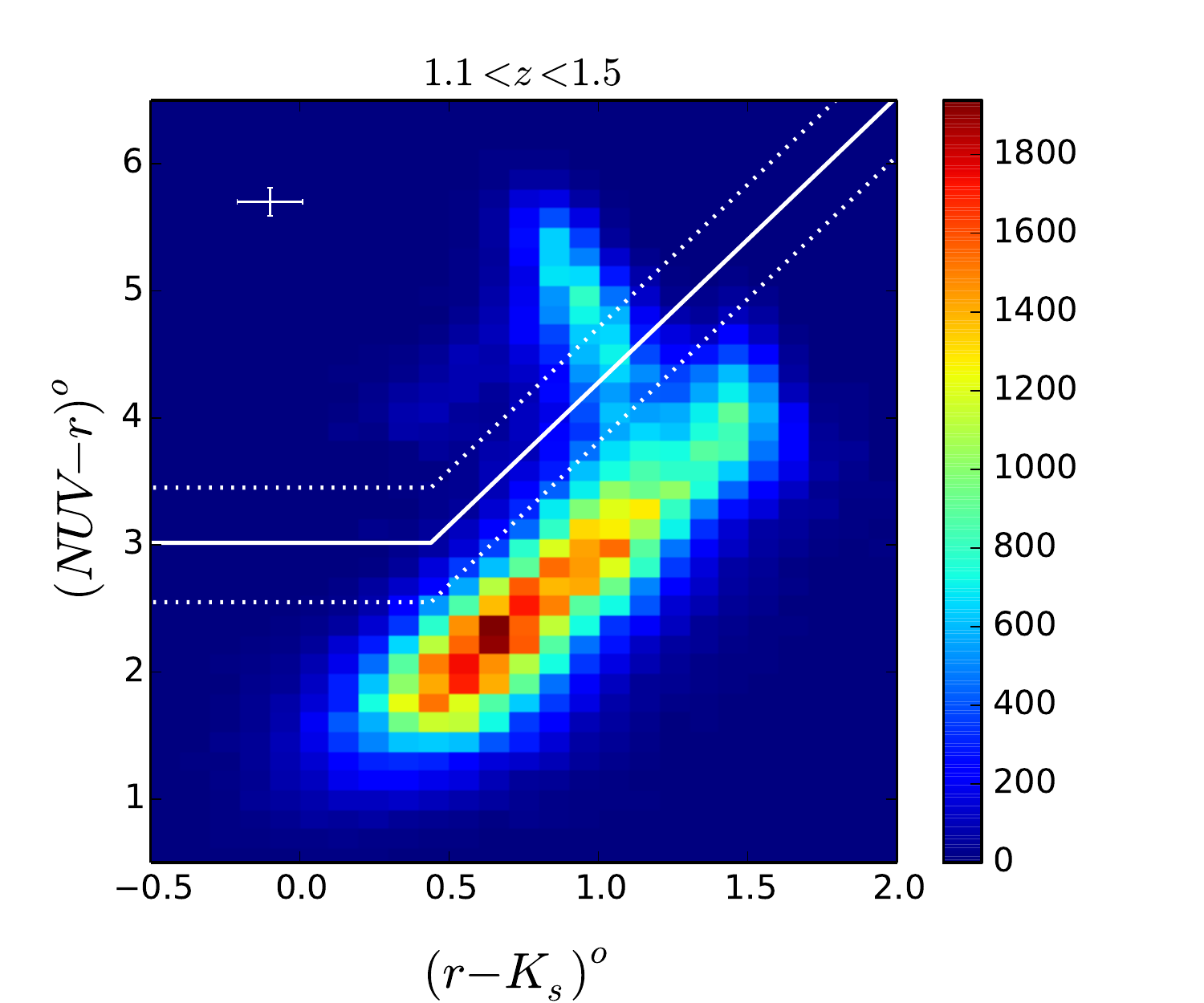}
\caption{Star-forming and quiescent galaxy selection in the NUVrK diagram. The colour code shows the galaxy density. The averaged colour uncertainties (based on the observed photometric errors) are shown in the upper left corner of each panel. The binning used for the density map is tuned to match the typical uncertainties at  $0.2 < z < 0.5$.
The solid line represents the mean selection of quiescent galaxies in a given redshift bin. The dotted lines represent the two extreme selections delimiting the \textit{green valley}.   \label{NUVrK_z}}
\end{figure}

\begin{figure}[!]
\includegraphics[width=0.495\columnwidth, trim = 0cm 1.9cm 2cm 0cm, clip]{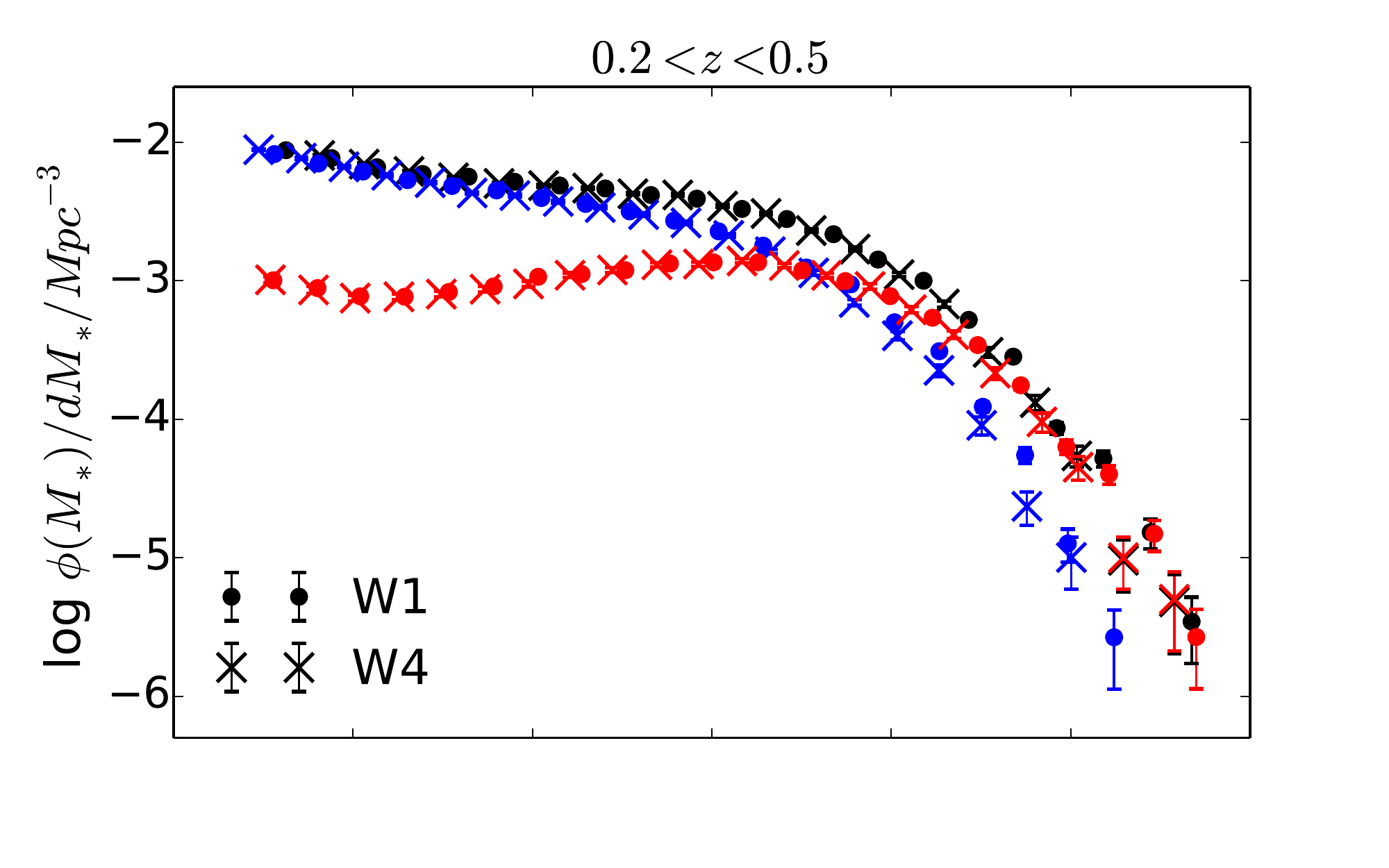}
\includegraphics[width=0.495\columnwidth, trim = 0cm 1.9cm 2cm 0cm, clip]{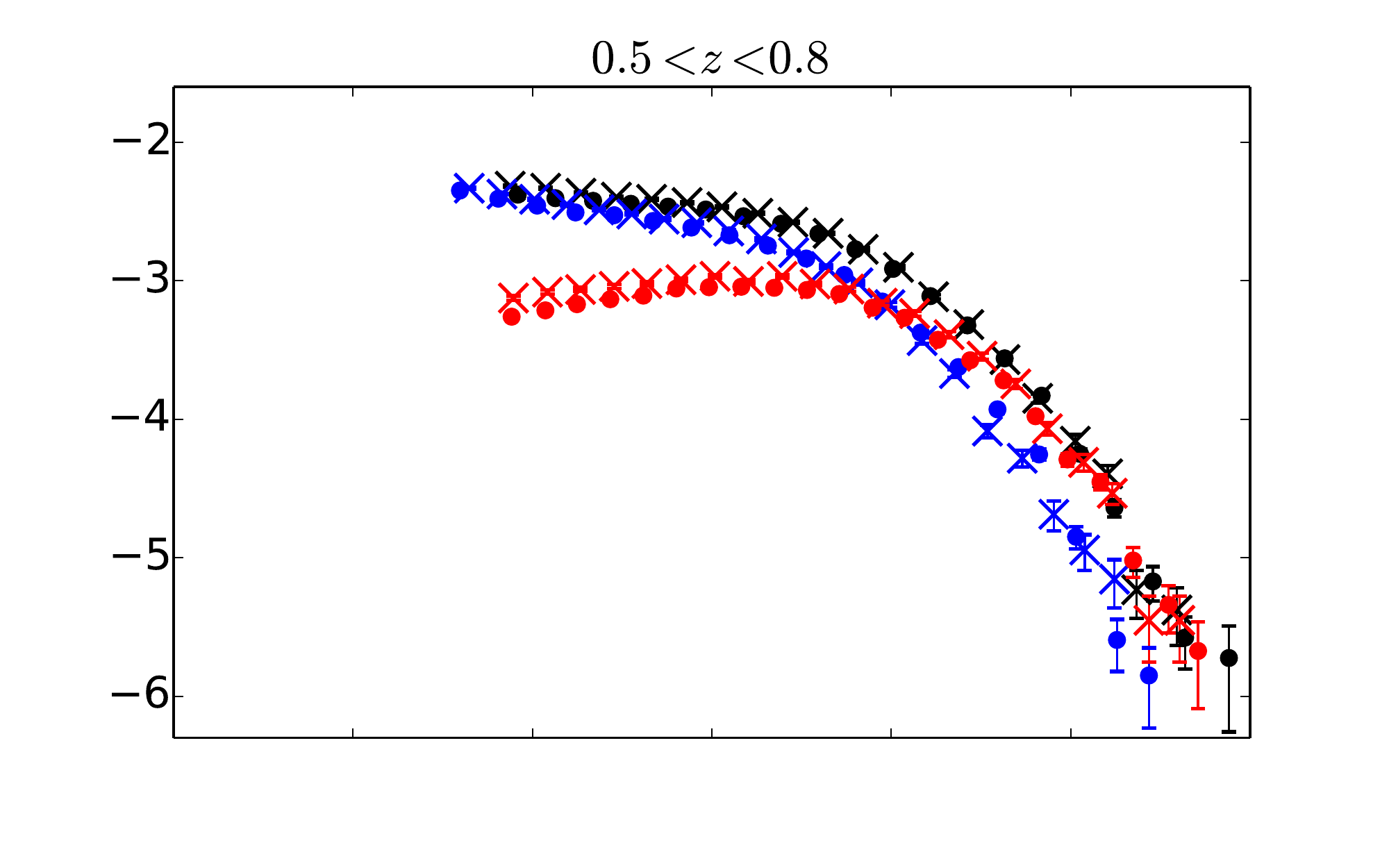}

\includegraphics[width=0.495\columnwidth, trim = 0cm 0cm 2cm 0cm, clip]{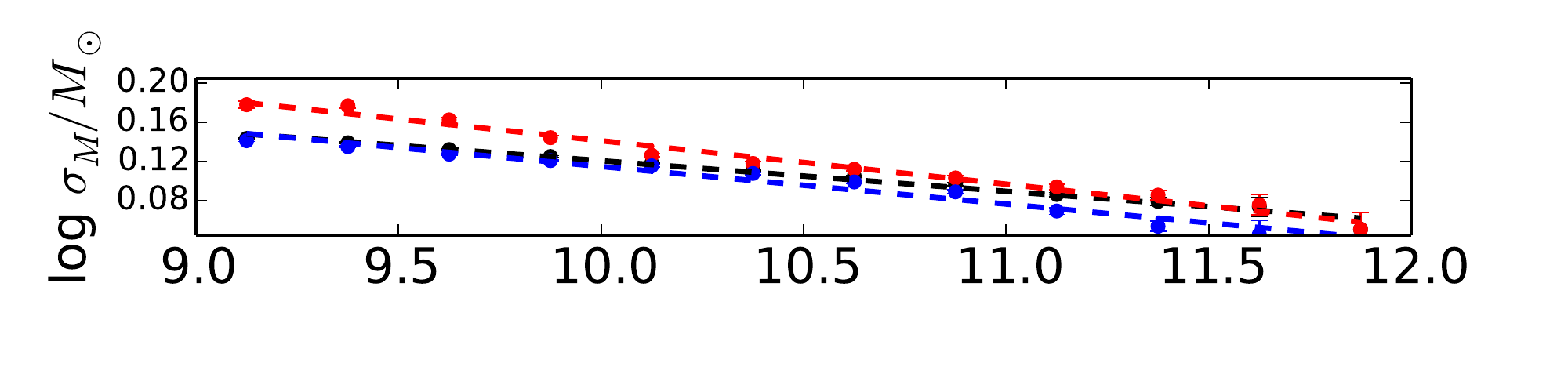}
\includegraphics[width=0.495\columnwidth, trim = 0cm 0cm 2cm 0cm, clip]{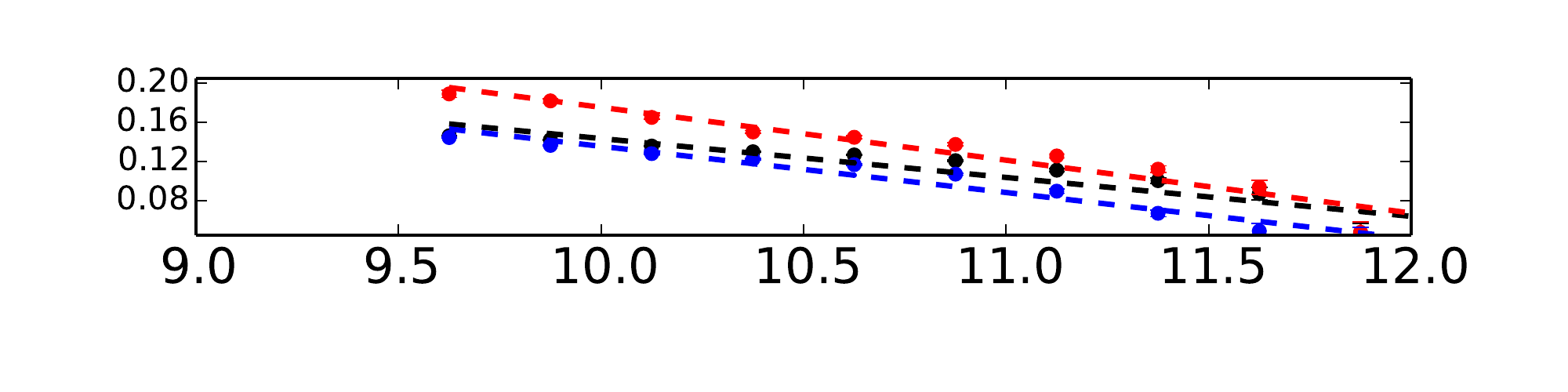}

\includegraphics[width=0.495\columnwidth, trim = 0cm 1.9cm 2cm 0cm, clip]{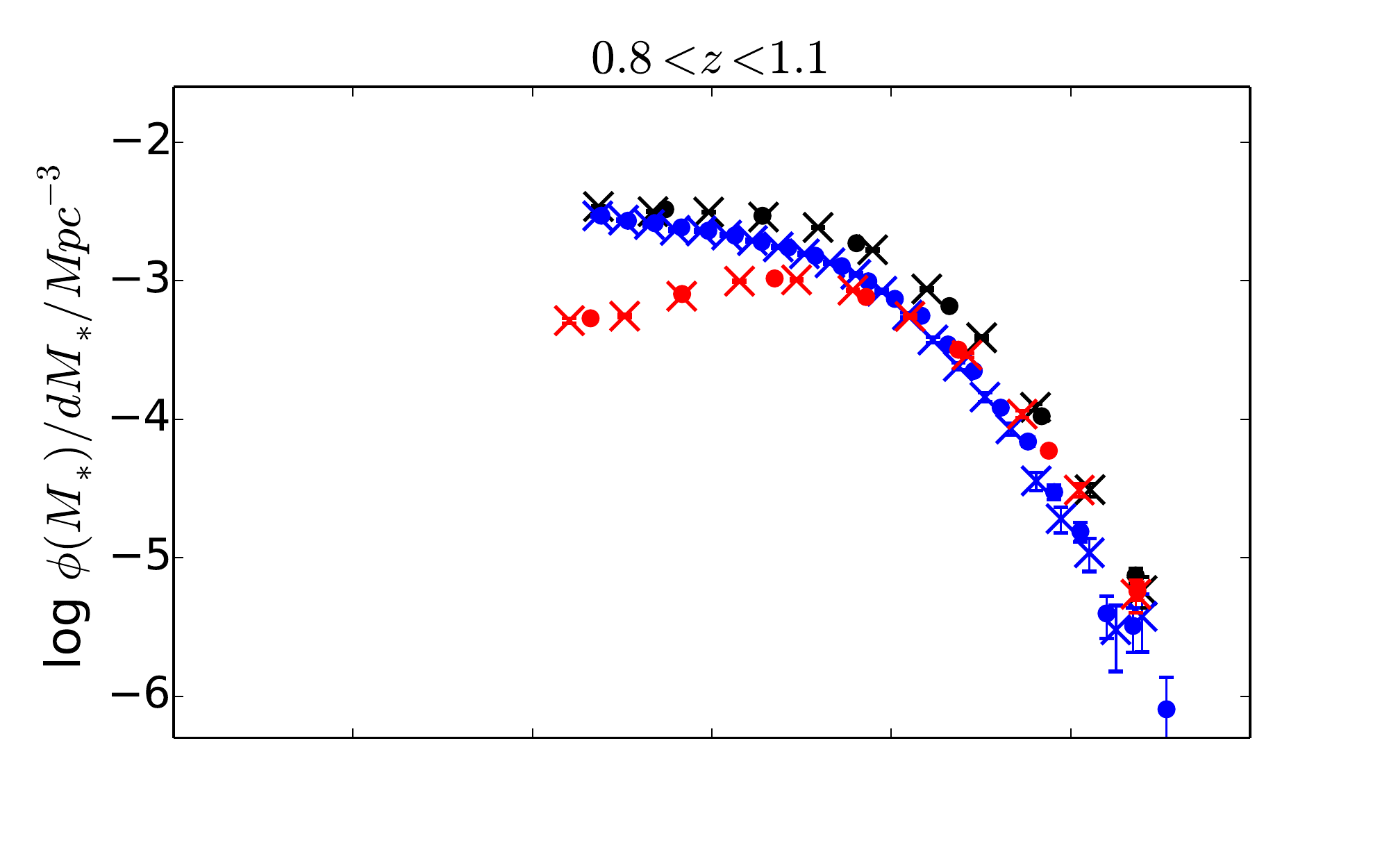}
\includegraphics[width=0.495\columnwidth, trim = 0cm 1.9cm 2cm 0cm, clip]{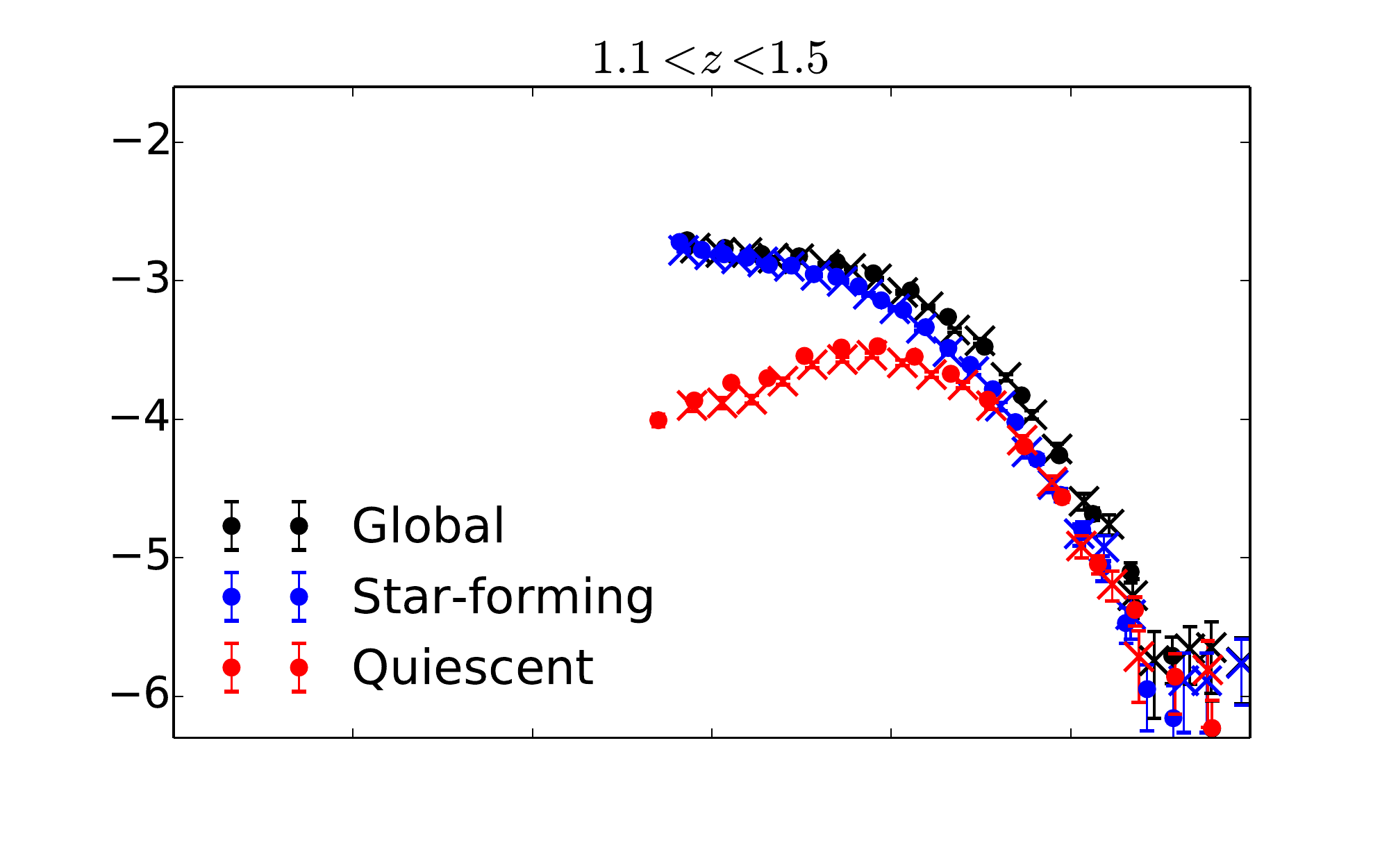}

\includegraphics[width=0.495\columnwidth, trim = 0cm 0cm 2cm 0cm, clip]{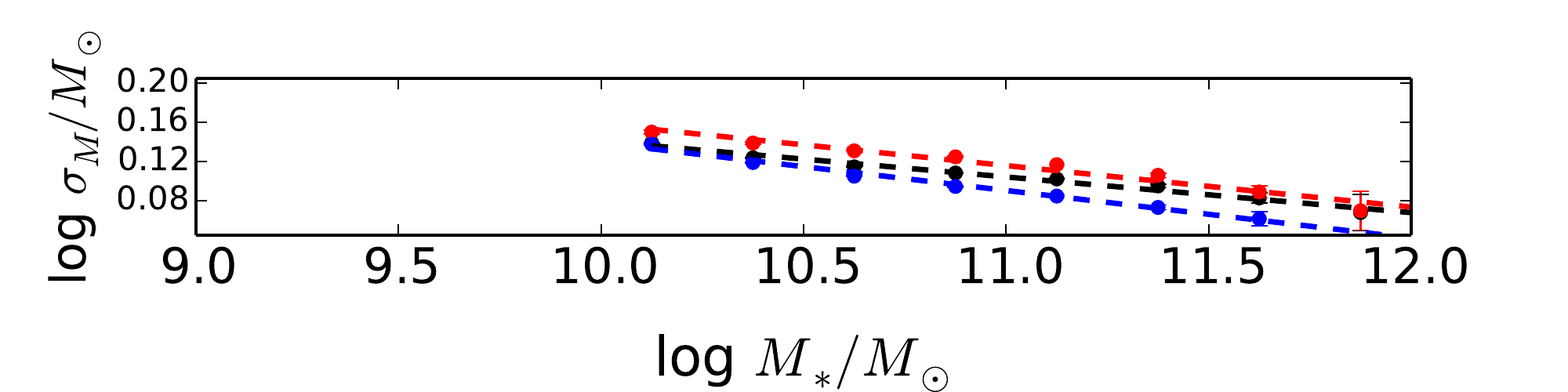}
\includegraphics[width=0.495\columnwidth, trim = 0cm 0cm 2cm 0cm, clip]{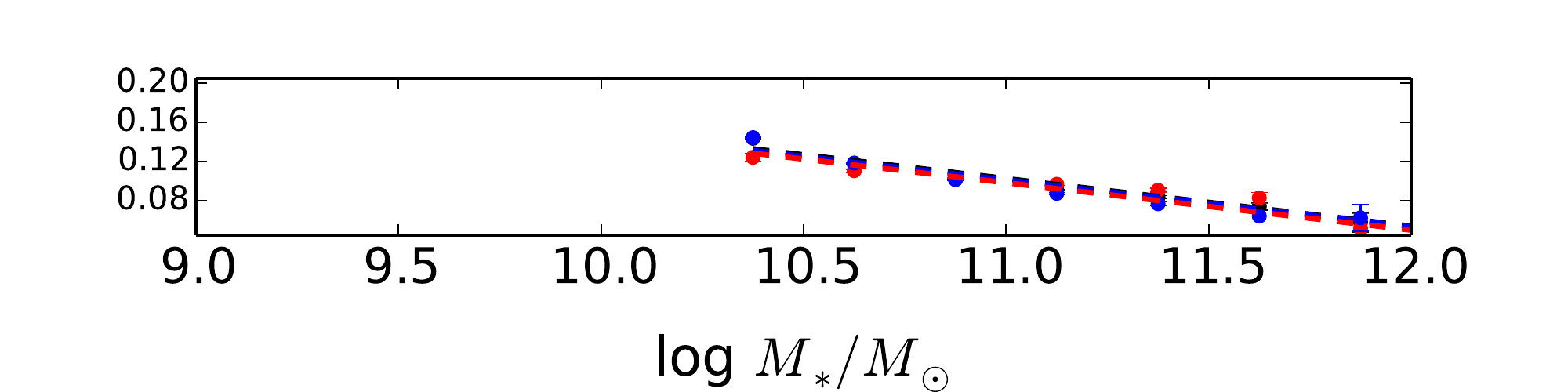}

\caption{Galaxy SMF in the fields W1 (dots) and W4 (crosses) for the global (black), star-forming (blue), and quiescent (red) populations in four redshift bins (\textit{\textbf{upper}} sub-panels). The error bars reflect only the Poissonian contribution, while the corresponding mass uncertainties are shown in the \textit{\textbf{lower}} sub-panels. Only SMF points above the stellar mass completeness are plotted.  \label{SMFs_fig}}
\end{figure}

 Figure  \ref{SMFs_fig} presents the global ($black$), star-forming ($blue$), and quiescent ($red$) galaxy SMFs for the two fields separately (W1: dot and W4: cross)  in four redshift bins.  The sample consists of 481,518 galaxies over 14.43 deg$^2$ in W1 and 268,010 galaxies over 7.96 deg$^2$ in W4.
 The error bars shown in the upper sub-panels reflect only the Poissonian contributions. 
 The SMF comparison between the two fields agrees within the errors. In the lower sub-panels, we plot the stellar mass uncertainty by type, $\sigma_M$, defined in Sect. \ref{mass_errors}, as function of the stellar mass. First, $\sigma_M$ decreases exponentially with stellar mass, as already noted in previous studies \citep[e.g.][]{Grazian2015}. We can then fit the $\sigma_M(M_*)$ relation with a power law (Fig. \ref{SMFs_fig} sub-panels, dashed lines). Secondly, the size of our galaxy sample allows for very small relative Poissonian errors down to densities of around $\sim 10^{-5}$--$10^{-6} Mpc^{-3}$ even if we split by type and field. 
 The cosmic variance contribution in the budget of the errors that affects our SMF measurement is therefore expected to be small, as discussed in the next section.

\subsection{SMF uncertainties}
\label{uncertainties}

In this section, we describe the error budget associated to our SMF measurements. All the contributions to the SMF uncertainties are expressed as a function of the stellar mass and redshift. In addition to the stellar mass and Poissonian errors already mentioned, the  large-scale density inhomogeneities represent a source of uncertainty. This cosmic variance is known to represent a fractional error of 15 -- 25\%  at the high-mass end ($M\ge 10^{11} M_{\odot}$) in the COSMOS survey and of around 20 -- 50\% in narrower pencil-beam surveys, generally dominating the error budget. 

Following the procedure discussed by \citet{Coupon2015}, we investigated the contribution of the cosmic variance in our sample by dividing our survey into N patches of equal areas. Since the effective surface can change from one patch to another, every patch was weighted according to its unmasked area. For a given observed area, we computed the number density dispersion N times over (N-1) patches by discarding a different patch every time. We then considered the mean number density dispersion over the N measurements as our internal estimate of the cosmic variance for a given effective area and the dispersion around the mean as an error estimate of the cosmic variance.

\begin{figure}[!]
\includegraphics[width=\hsize, trim = 0cm 1.2cm 0cm 0cm, clip]{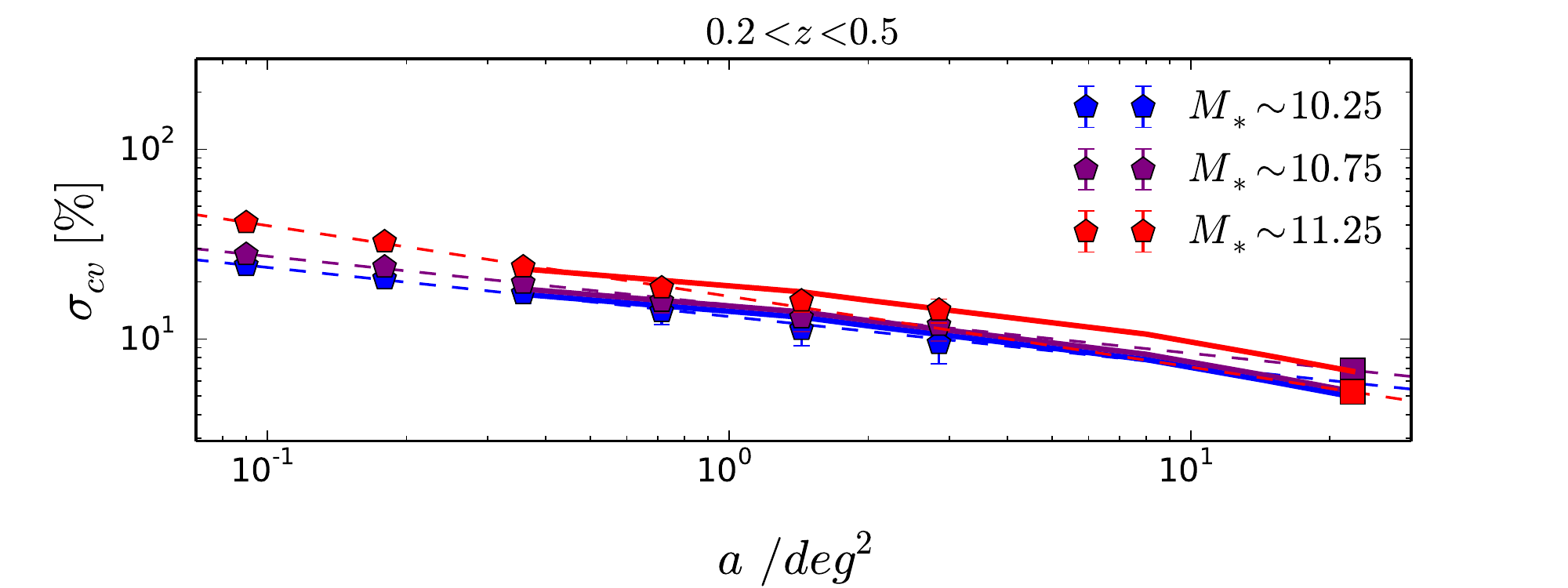}
\includegraphics[width=\hsize]{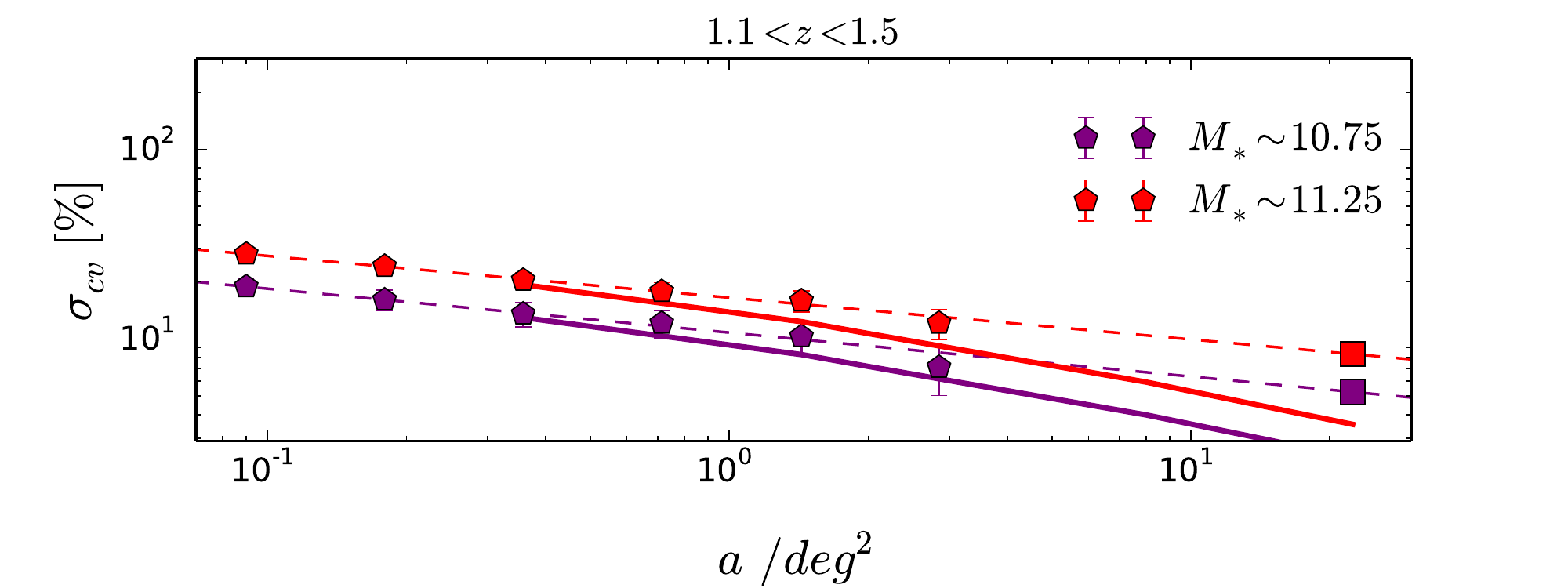}

\caption{Cosmic variance as a function of the effective observed area for three stellar mass bins. The dashed lines correspond to the linear fit of the empirical cosmic variance estimates plotted with pentagons. The squares locate the extrapolated cosmic variance estimate for our entire survey. The solid lines show the corresponding theoretical estimates computed according to \citet{Moster2011}. \label{cv_a}}
\end{figure}

In Fig. \ref{cv_a} we plot our cosmic variance estimate $\sigma_{cv}$ in the redshift bins [0.2, 0.5] and [1.1, 1.5], considering three stellar mass bins from $M_* = 10^{10} M_{\odot}$ up to $M_* = 10^{11.5} M_{\odot}$ (with blue, purple and red dots, respectively) and for mean effective areas ranging from $a \simeq 0.1$ to $a \simeq 2.8$ deg$^2$. 
The relation of cosmic variance -- area  is well fitted by a power-law with $\sigma_{cv}(a)$ = 10 $^{\beta}$ $a$ $^{\alpha}$ (shown as dashed lines).  To estimate the cosmic variance that affects our entire survey, we extrapolated the relations  up to  $a = 22$ deg$^2$, shown as squares.

 For comparison, we also show the cosmic variance predicted for the same redshift and stellar mass bins (triangles) by using the code \texttt{getcv} \citep{Moster2011}. Our internal cosmic variance estimate ($\sigma_{cv}$) and the predicted one agree
remarkably well up to our observed areas of $a = 2$ deg$^2$. For larger areas, the two estimates diverge slightly for high-mass ($M_* > 10^{11} M_{\odot}$) galaxies at $z < 0.5$, where we slightly underestimate $\sigma_{cv}$ with respect to the theoretical prediction. We have to stress that the \citet{Moster2011} procedure is optimised for pencil-beam surveys of areas $a < 1$ deg$^2$. At $z > 0.5$, the theoretical estimators always predict a cosmic variance lower than our own extrapolation. By using our internal estimate, we therefore adopt a conservative approach.

Finally, the last source of error that we need to consider is that of the stellar mass uncertainty  defined in Sect. \ref{mass_errors}.
 To do so, we generated 200 mock catalogues with perturbed stellar masses according to the  expected $\sigma_{M}$ (which includes the photometric redshift uncertainties and the photon noise, Eq. \ref{eq_err_M}) and measured the 
 1$\sigma$ dispersion in the density $\Phi$ of the reconstructed SMFs that we refer to as $\sigma_{\Phi, M}$.
    
 At the end, the error of the stellar mass function is the quadratic sum of all the contributions discussed above and is defined as  
\begin{equation}
\sigma_{tot} = \sqrt{ ~\sigma_{cv}^2 + \sigma_{poi}^2+ \sigma_{\Phi, M}^2~ } ~.
\label{eq_err_tot}
\end{equation}

\subsection{Importance of photometric calibration in large surveys} 

\begin{figure}[!t]
\includegraphics[width=\hsize, trim = 0cm 1.9cm 0cm 0cm, clip]{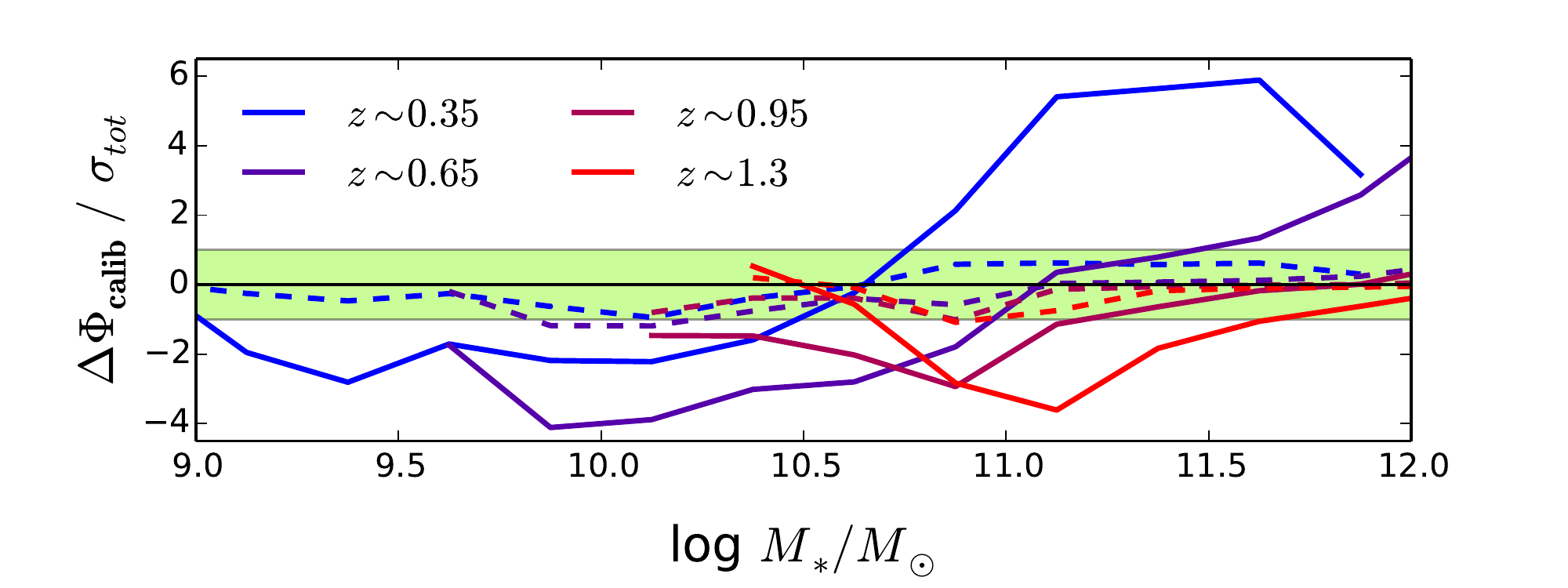}
\includegraphics[width=\hsize]{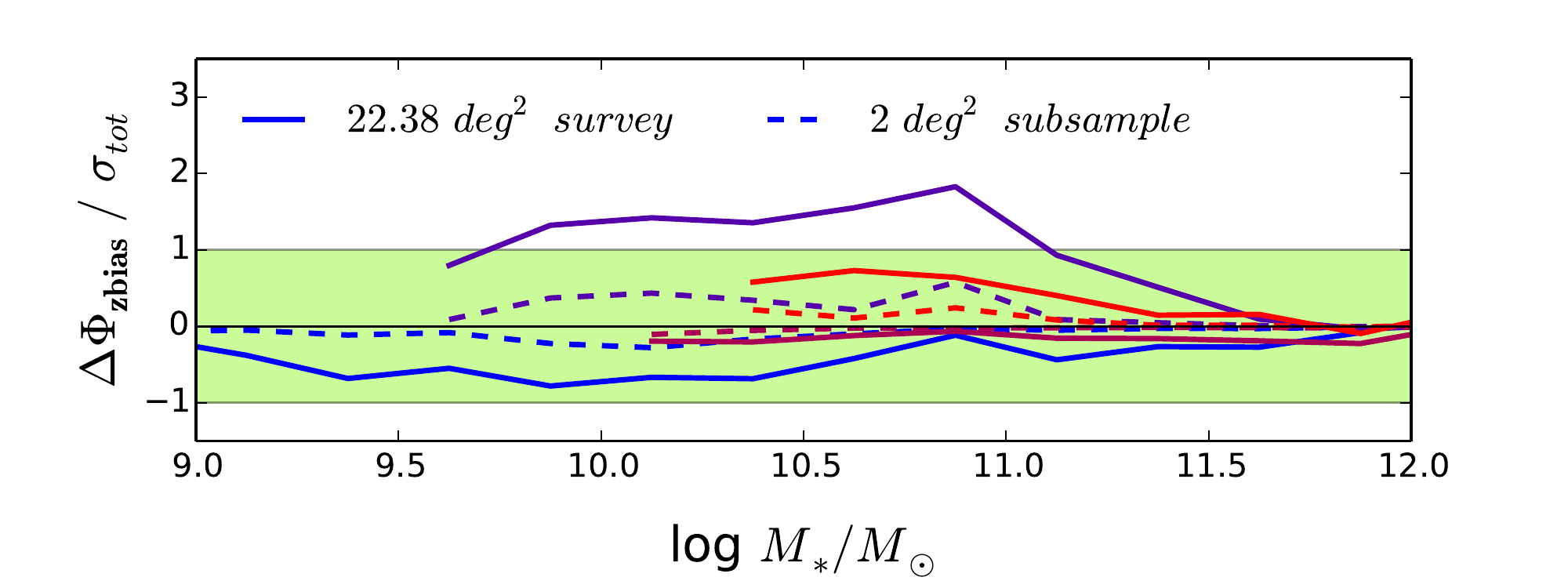}

\caption{Ratio between the systematic stellar mass function difference and the total statistical error, $\Delta \Phi / \sigma_{tot}$, as a function of the stellar mass and in four redshift bins. We consider $\Delta \Phi_{\textbf{calib}}$ (\textit{top} panel) and $\Delta \Phi_{\textbf{zbias}}$ (\textit{bottom} panel), the systematics coming from the absolute photometric calibration and from the photometric redshift bias (cf. Sect. \ref{zp_method}), respectively. The green shaded areas show the region where $| \Delta \Phi | \leq \sigma_{tot}$. \label{sys_t07_lens}}
\end{figure}

As mentioned in Sect. \ref{systematics}, a mean offset of $\sim 0.06$ mag in the optical absolute photometric calibration
 (cf. $\Delta mag$ in Table \ref{tab_zero_pt}) can change the stellar mass estimate by 0.1 dex.  
In the top panel of Fig. \ref{sys_t07_lens}  we show the difference between the two SMFs measured with the T07 and CFHTLens photometries, $\Delta \Phi_{\textbf{calib}} = \Phi_{LenS} - \Phi_{T07}$. This difference is normalised by the total statistical error discussed in the previous section, $\sigma_{tot}$. In general, $\Delta \Phi_{\textbf{calib}} > \sigma_{tot}$ in our survey (solid lines), which means that the SMF variation induced by the calibration offsets is several times larger than the uncertainty of our SMF. It even reaches 5 $\sigma_{tot}$ at low redshift ($0.2 < z < 0.5$; blue solid line), where the stellar mass is essentially driven by the optical photometry.

In contrast, by considering a subsample of 2 deg$^2$ (dashed lines), we find that $| \Delta \Phi_{\textbf{calib}} | \lesssim \sigma_{tot}^{2deg}$ (green shaded area).
This means that the SMF variation driven by the calibration offsets is smaller than the other uncertainties affecting the SMF in a 2 deg$^2$ survey (i.e. the variation is contained within the error bars). In other words, we cannot see the variation that
is due to the calibration because it is hidden by other sources of uncertainties (Poissonian and cosmic variance).
The systematic differences that are due to the T0007-CFHTLenS photometric offsets can therefore be neglected in a 2 deg$^2$ survey, while in surveys of 20 $deg^2$ and more, we reach a regime where the systematic uncertainty that is due to the photometric calibration dominates the error budget. 

For comparison, we also investigated another source of systematic uncertainty: the photometric redshift bias. Using the photo-z bias ($z_{bias} = z_{phot}-z_{spec}$) presented in Sect. \ref{zp_method} (see Fig. \ref{zp_zs}, lower panels), we corrected our photometric redshifts. Instead of using a global correction, we applied a photo-z bias correction for different galaxy types\footnote{We
also checked this by estimating the correction with half of the spectroscopic sample and improving the photo-zs of the other half.}.

Similarly to $\Delta \Phi_{\textbf{calib}}$, the difference between the stellar mass functions computed with the \textit{corrected} photometric redshifts and with the original ones ($\Delta \Phi_{\textbf{zbias}}$) is shown in the lower panel of Fig. \ref{sys_t07_lens}. The effect of the photo-z bias on the SMF measurement is much weaker than the effect of the photometry. The SMF differences induced by the photo-z bias as measured in our sample are largely dominated by the statistical uncertainties in a 2 deg$^2$ subsample. Moreover, in the entire 22 deg$^2$ survey, the difference can only be detected at $z \sim 0.65$, while $|\Delta \Phi_{\textbf{zbias}}| < 2 \ \sigma_{tot}$. 

Given the limited amplitude of its effect on our sample, the photo-z bias can be neglected in our study. By contrast, the SMF variations that are due to the difference in photometry stress the need of carefully controlling the absolute photometric calibration in large surveys. In the present study, the choice of the CFHTLS-T0007 photometry is supported by (1) the SNLS photometric calibration based on a new spectrophotometric standard for high-precision cosmology, and (2) the careful treatment by the Terapix team that enables homogeneously propagating the SNLS photometric calibration over the entire survey.

\section{Evolution of the galaxy stellar mass function and density}
\label{evol}

As shown in Sect. \ref{uncertainties}, the large volume probed by our survey allows us to reduce both the cosmic variance and the Poisson uncertainties. We exploit this  large volume to quantify the evolution of the galaxy SMF, especially at the high-mass end, where it is most relevant.

\subsection{Evolution of the SMF}
\label{SMF_evol}

\subsubsection{Comparison of the global SMF with the literature}
\label{SMF_comparison}

\begin{figure*}[!]
\vspace{-0.05cm}\includegraphics[width=0.499\hsize, trim = 0cm 1.5cm 0cm 0cm, clip]{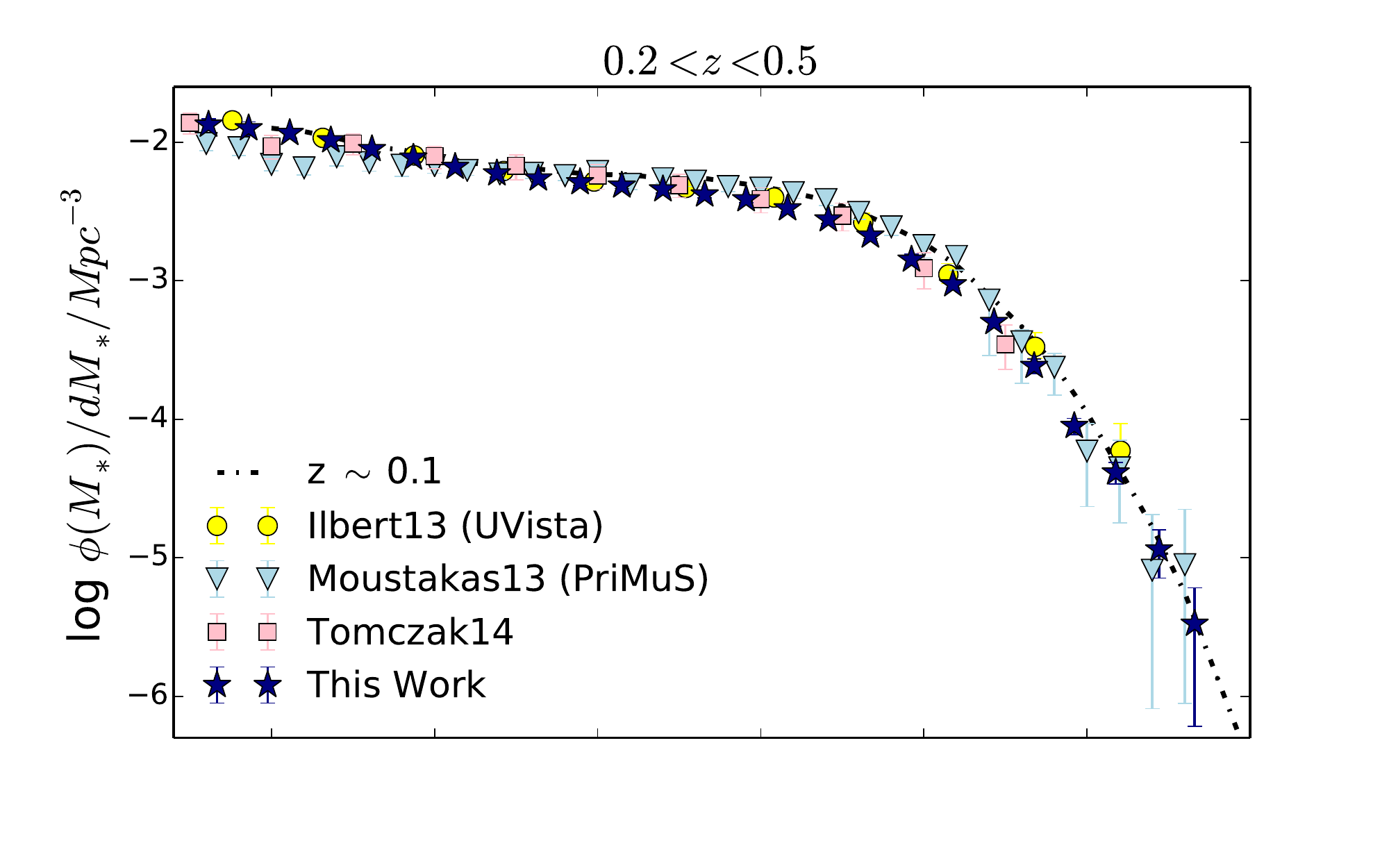}
\vspace{-0.05cm}\includegraphics[width=0.499\hsize, trim = 0cm 1.5cm 0cm 0cm, clip]{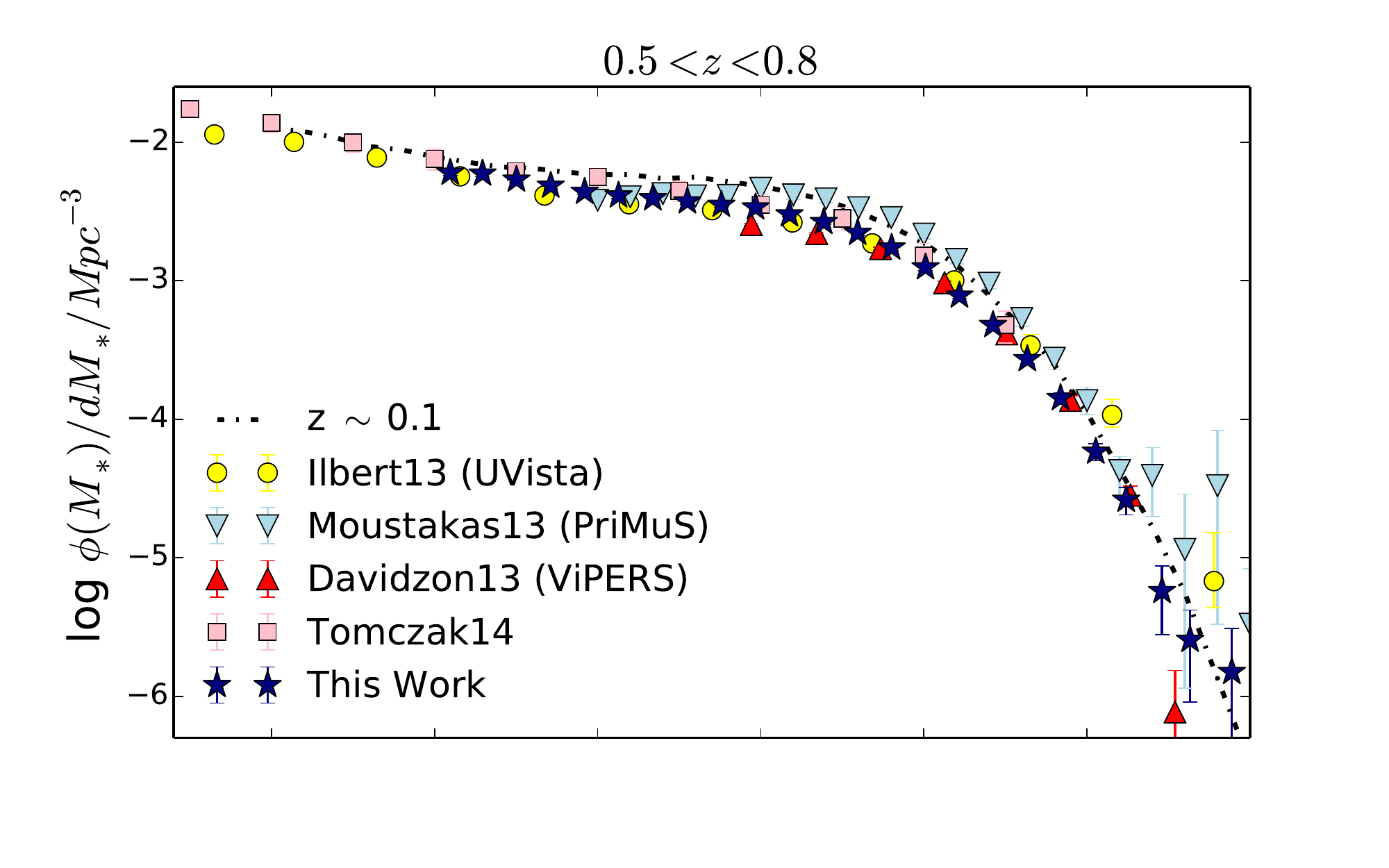}

\includegraphics[width=0.499\hsize, trim = 0cm 0cm 0cm 0.5cm, clip]{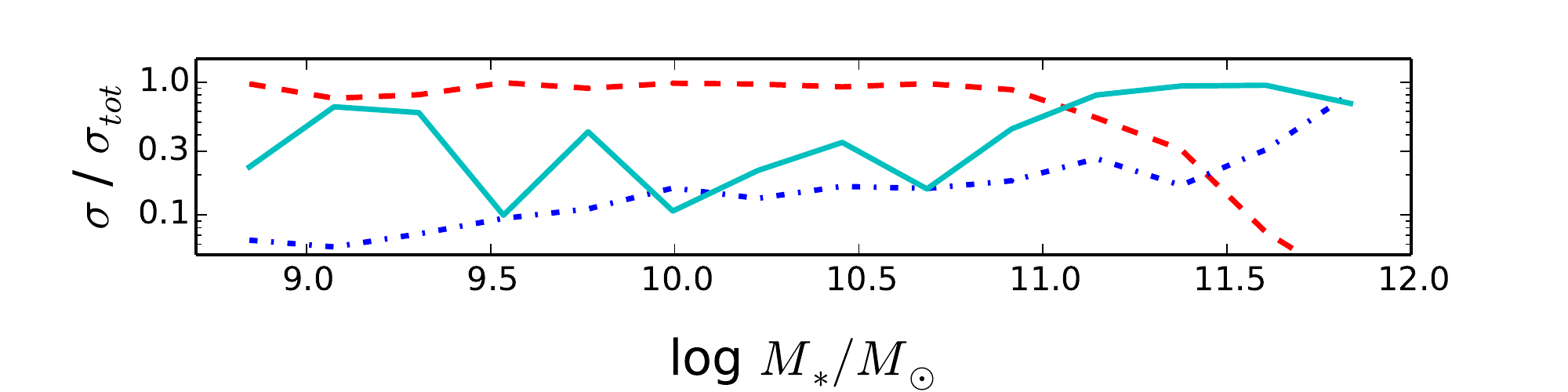}
\includegraphics[width=0.499\hsize, trim = 0cm 0cm 0cm 0.5cm, clip]{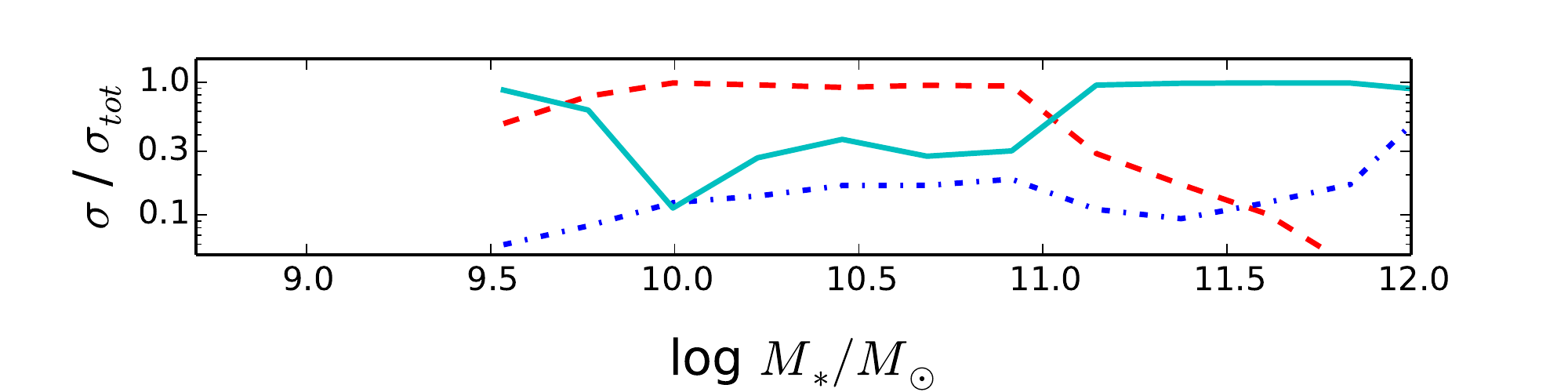}

\vspace{-0.05cm}\includegraphics[width=0.499\hsize, trim = 0cm 1.5cm 0cm 0cm, clip]{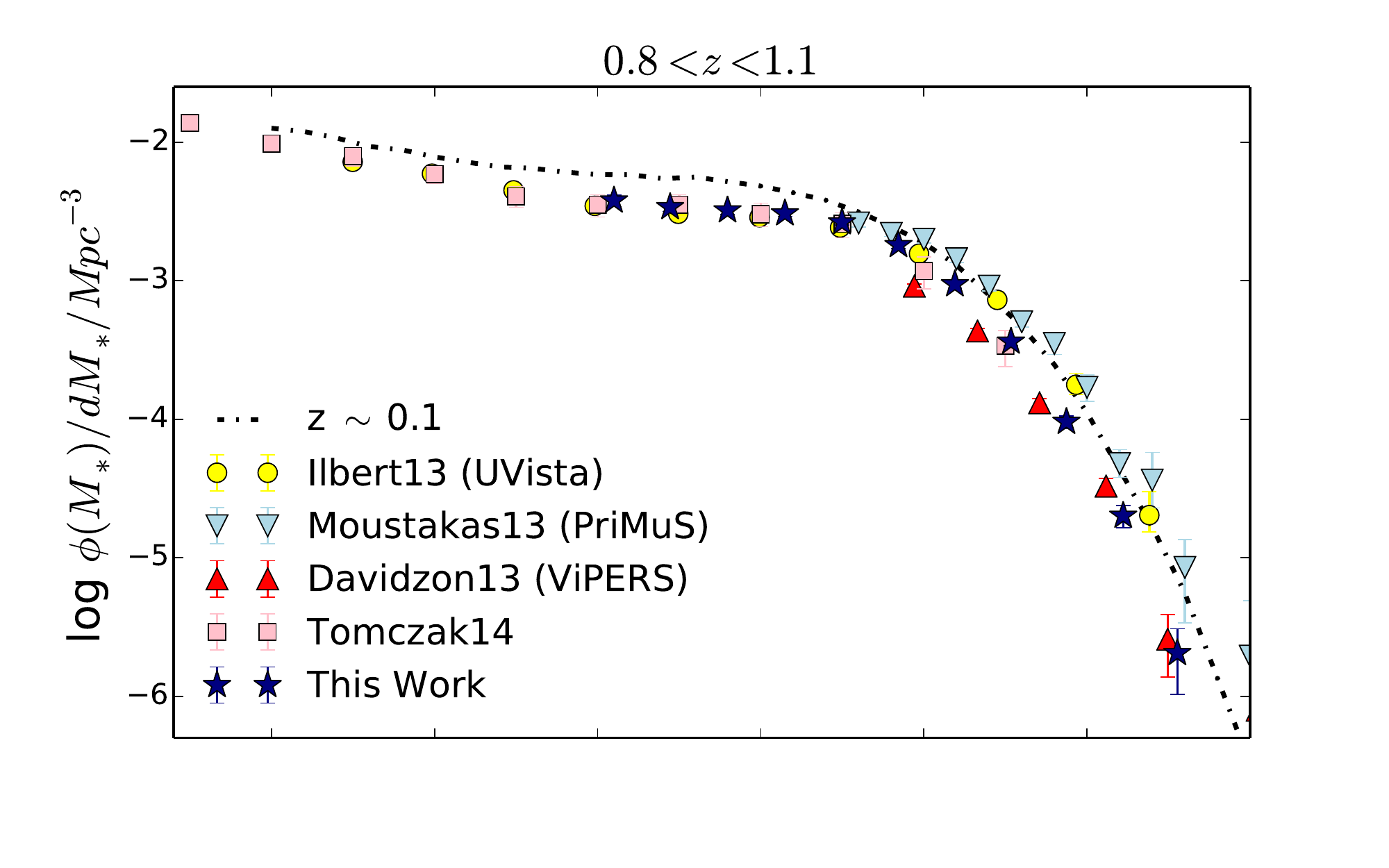}
\vspace{-0.05cm}\includegraphics[width=0.499\hsize, trim = 0cm 1.5cm 0cm 0cm, clip]{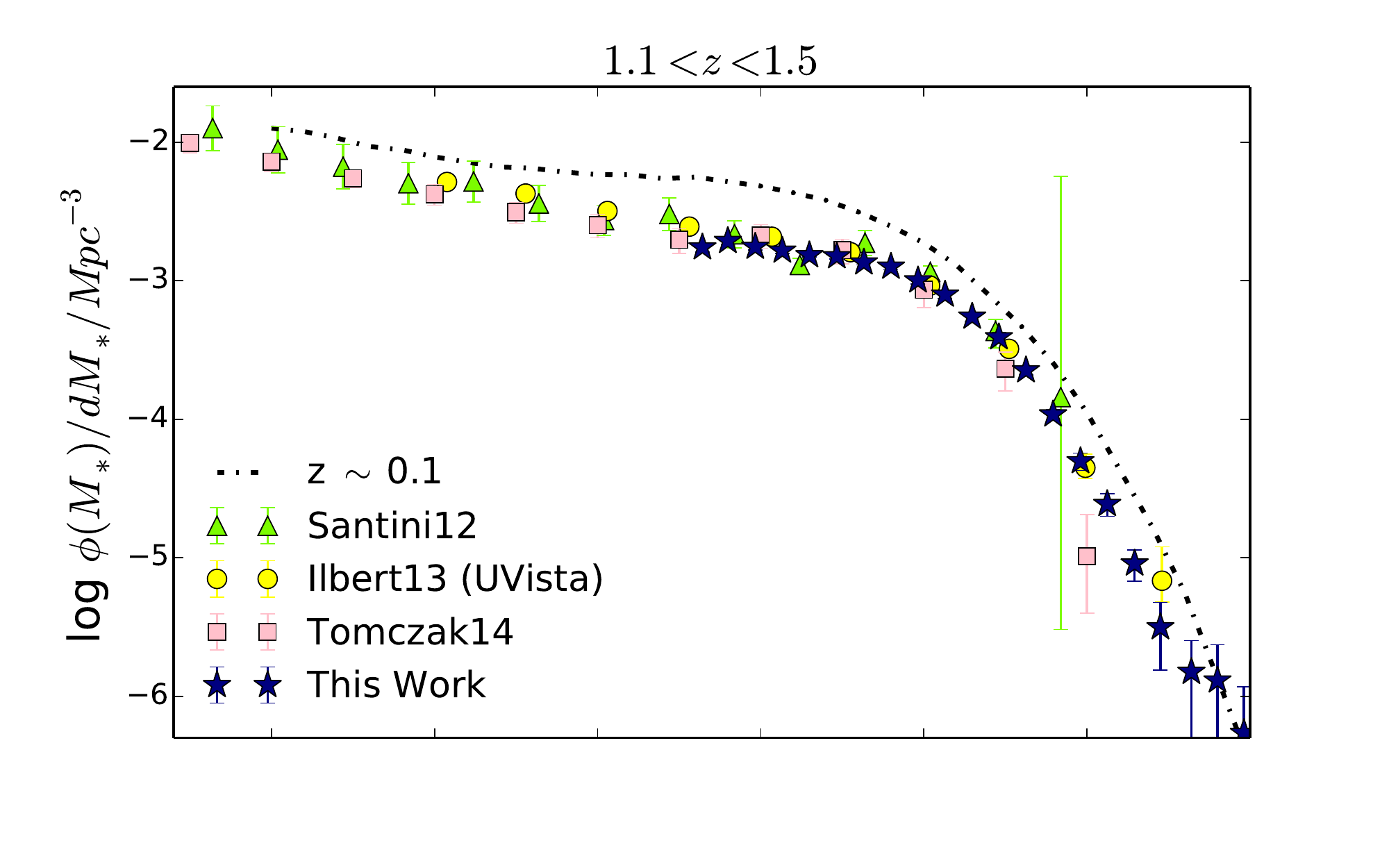}

\includegraphics[width=0.499\hsize, trim = 0cm 0cm 0cm 0.5cm, clip]{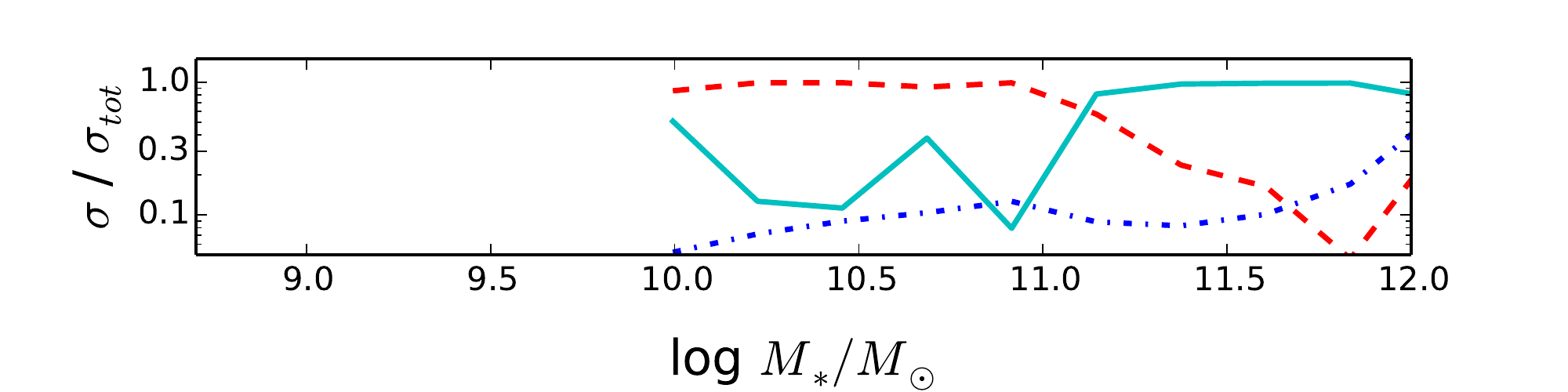}
\includegraphics[width=0.499\hsize, trim = 0cm 0cm 0cm 0.5cm, clip]{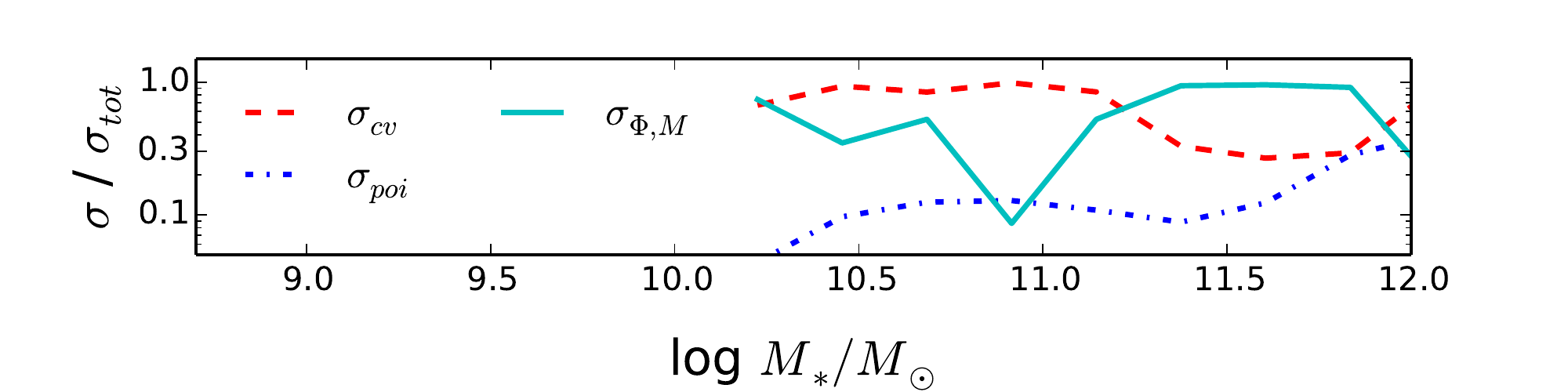}
\caption{Galaxy stellar mass functions (SMF) in four redshift bins. \textit{\textbf{Top sub-panels}}: The SMF measured in the present study (black stars) is compared to previous measurement: \citet{Tomczak2014}, pink squares; \citet{Davidzon2013}, red up triangles; \citet{Moustakas2013}, cyan down triangles; \citet{Ilbert2013}, yellow circles;
and \citet{Santini2012}, green up triangles. The error bars plotted on the measures reflect different contributions to the SMF uncertainty, depending on the considered study: only Poissonian for Ilbert et al., Moustakas et al. and Davidzon et al.; Poissonian and stellar mass for Santini et al.; and Poissonian, stellar mass and cosmic variance for Tomczak et al. and the present study. The dashed line shows the SDSS-Galex local measurement of \citet{Moustakas2013}. \textbf{\textit{Lower sub-panels}}: The corresponding SMF error contributions normalised by the total SMF uncertainty (see Eq. \ref{eq_err_tot}). The blue dash-dotted line represents the Poissonian contribution. The red dashed line and the cyan solid line represent the cosmic variance and the mass uncertainty contributions, respectively.  \label{SMF_litt}}
\end{figure*}

In the upper sub-panels of the Fig. \ref{SMF_litt}, we compare our global SMF measurements with the literature. Our results
agree well overall with many previous studies, although some differences exist. We discuss these in this section. The error bars corresponding to our measurement (black) reflect the total error $\sigma_{tot}$ defined in Eq. \ref{eq_err_tot}. In the lower sub-panels, we show the contribution of each error to the error budget. We show each contribution normalised by the total error $\sigma_{tot}$ as a function of the stellar mass. First, we note that the Poissonian error (blue dash-dotted line) represents a minor contribution to the total SMF uncertainty up to the very high-mass end (i.e. above $10^{11.5} M_{\odot}$). Secondly, the contribution of the cosmic variance ($\sigma_{ cv}$; red dashed line) is dominant up to stellar masses around the SMF knee ($M_* < 10^{11} M_{\odot}$). We  finally note that the contribution of the stellar mass uncertainty  ($\sigma_{\Phi, M}$; cyan solid line) drives the total uncertainty of the SMF high-mass end. We recall that while the Poissonian uncertainty is always taken into account in the literature, the error bars may reflect different contributions to the SMF uncertainty depending on the study considered in Fig. \ref{SMF_litt} (as specified in the caption).

The comparison with \citet{Davidzon2013} is straightforward since their observations were taken in the same two fields of the  CFHTLS survey, covering an effective area of 5.34 and 4.97 deg$^2$ in W1 and W4, respectively.  
The authors derived the SMF between $z = 0.5$ and $z = 1.3$ using the VIPERS-PDR1 dataset ($\sim$ 50,000 galaxies), that is,~the main spectroscopic sample used to calibrate our redshifts (Sect.~\ref{zp_accu}). The work of \citet{Davidzon2013} clearly illustrates the advantages of  using spectroscopic redshifts (e.g.~the easier removal of stellar interlopers and QSO). However, to estimate the SMF through spectroscopic data, some difficulties need to be solved, such
as the statistical weighting to account for the spectroscopic sampling rate \citep[see][for more details about how these weights are computed in VIPERS]{Garilli2014}. We observe a good agreement between the two SMF estimates, especially at $M_* > 10^{11} M_{\odot}$. 

The statistical uncertainties are very low in both  VIPERS and our analysis, and the two surveys are additionally collected almost in the same area. Any difference is likely due to some combination of the photometric redshift uncertainty of our sample, the spectroscopic incompleteness affecting VIPERS, the adopted SED fitting method, or the photometric calibration used in VIPERS (T0005) and in our survey (T0007). However, the only significant discrepancy is observed close to the stellar mass completeness limit of VIPERS, where the measurements of \citet{Davidzon2013} are $\lesssim 0.2$ dex lower. This difference at low masses could be due to some incompleteness correction that is due to the $i$-band selection, while our sample is $K_s$-band selected.

\citet{Moustakas2013} also measured the SMF by relying on the spectroscopic redshift sample of PRIMUS ($\sim$ 40,000 galaxies between $z=0.2$ and 1 and cover $\sim5.5$ deg$^2$ over five fields). In general, we observe that the SMF measurements from PRIMUS form the upper limit of the literature. Their SMF estimate is significantly above the others at $0.5<z<0.8$.
In the range $10^{10.5}<M_* < 10^{11}\,M_{\odot}$, the difference reaches 0.2 dex, while the authors predict that the cosmic variance should not affect the measurement by more than 10\%; a larger offset is observed at $M_*>10^{11.5} M_{\odot}$, which could be mainly explained by the cosmic variance affecting their measurement, which is estimated to be very strong at high mass\footnote{\citet{Moustakas2013} estimated $\sigma_\mathrm{cv} = 0.1-1.4$ for $\log M_* > 11.5 M_{\odot}$ at $0.5<z<0.8$.}. In the next redshift bin ($0.8<z<1.1$), the SMF of \citet{Moustakas2013} is also significantly higher than ours. The reason for the discrepancy may be linked to the different recipe (dust models, template libraries, etc.) adopted by \citet{Moustakas2013} in their SED fitting procedure \citep[see also][for a discussion about the effect of different SED fitting methods on the SMF]{Davidzon2013}. We compared their stellar masses in the XMM-LSS field, which overlaps the W1 field. We found that the PRIMUS masses are higher than ours by 
$0.17 \pm 0.09$ dex at $0.2 < z < 0.5$, $0.15 \pm 0.08$ dex at $0.5 < z < 0.8$ and $0.12 \pm 0.1$ dex at $0.8 < z < 1$\footnote{Even by using the same photometry as \citet{Moustakas2013}, i.e.~including GALEX, CFHTLS, and SWIRE (3.6 and 4.5$\mu$m), similar differences in the stellar masses are observed.}. This could explain part of the observed shift in the SMFs. It is worth noting that the two largest spectroscopic surveys so far, VIPERS and PRIMUS, lead to the largest difference in the SMF measurements. This highlights the great effect of systematic uncertainties in the latest large surveys \citep[see also][]{Coupon2015}.

The comparison of our measurements with deep photometric surveys
shows that our results agree well with those of \citet{Tomczak2014} and \citet{Ilbert2013}, down to the lowest stellar masses we can explore. Their analysis was based on much deeper data, which confirms the estimate of our lower mass limits. Only in the redshift bin $0.8 < z < 1.1$,  we note  a significant difference with \citet{Ilbert2013} in the high-mass end of the SMF.  This can be explained by the well-known over-density in the  COSMOS field \citep{Kovac2010a, Bielby2012}\footnote{The SMFs at $0.8 < z < 1.1$ are consistent with each other if $\sigma_\mathrm{cv}$ computed by \citet{Ilbert2013} is included in the error budget of their SMF ($\sigma_\mathrm{cv} = 0.1-0.25$ for $\log M_*/M_{\odot} = 11-12$).}. 

Finally, we show the local, $z\sim 0.1$,  GALEX-SDSS SMF from \citet{Moustakas2013} in all panels of Fig.~\ref{SMF_litt} (as a dash-dotted line). A small but clear and progressive deviation of the SMF with redshift is obvious, in comparison to the local SMF. Far from evident in the previous studies, the trend observed at high mass is confirmed and quantified in Sects. \ref{quant_SMF_evol} and \ref{dens_evol}, and is discussed in Sect. \ref{high_mass_evol}.

\subsubsection{Fitting the global, star-forming, and quiescent SMF}
\label{fitting}

\begin{figure*}[!]
\vspace{-0.05cm}\includegraphics[width=0.499\hsize, trim = 0.5cm 1.5cm 1.1cm 0cm, clip]{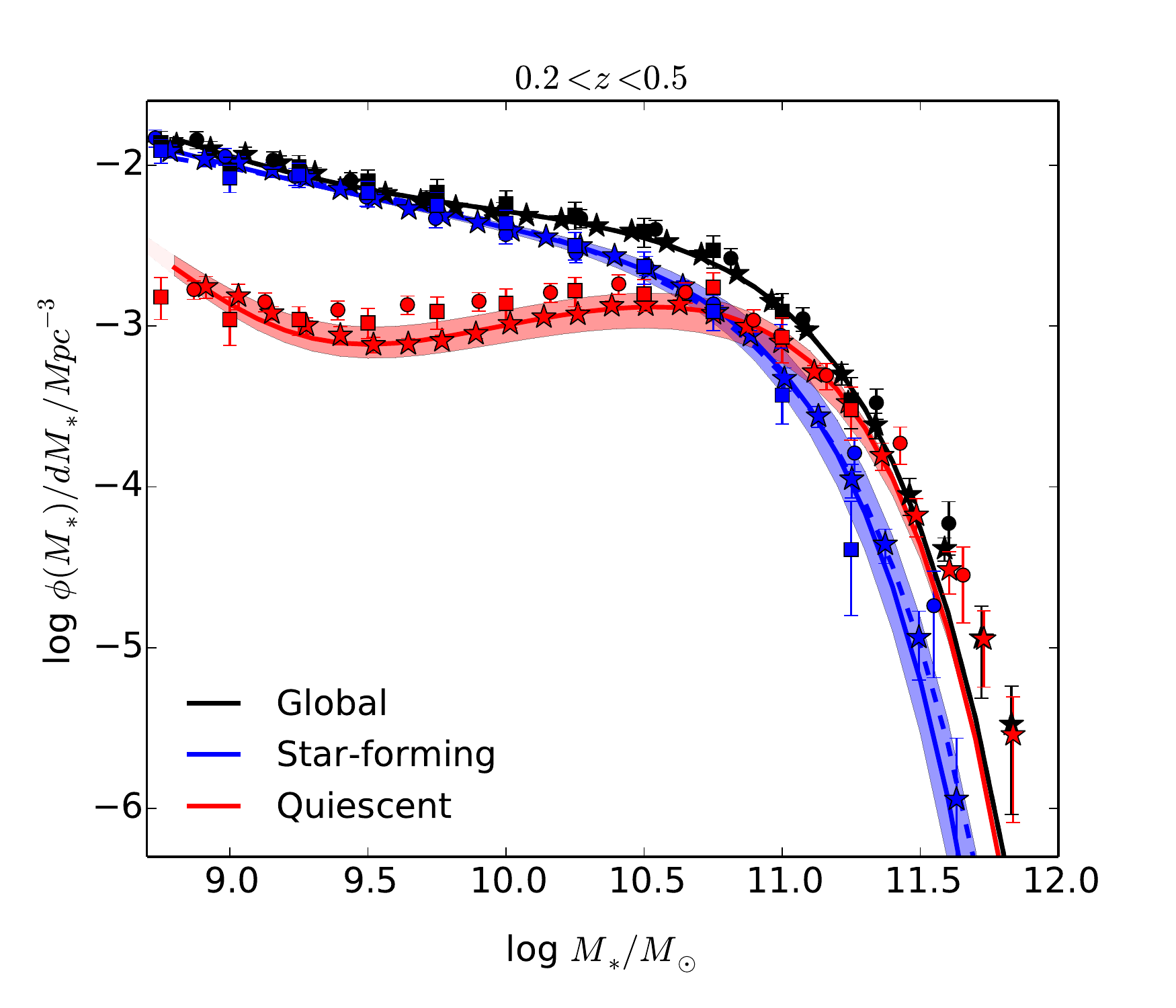}
\vspace{-0.05cm}\includegraphics[width=0.499\hsize, trim = 0.5cm 1.5cm 1.1cm 0cm, clip]{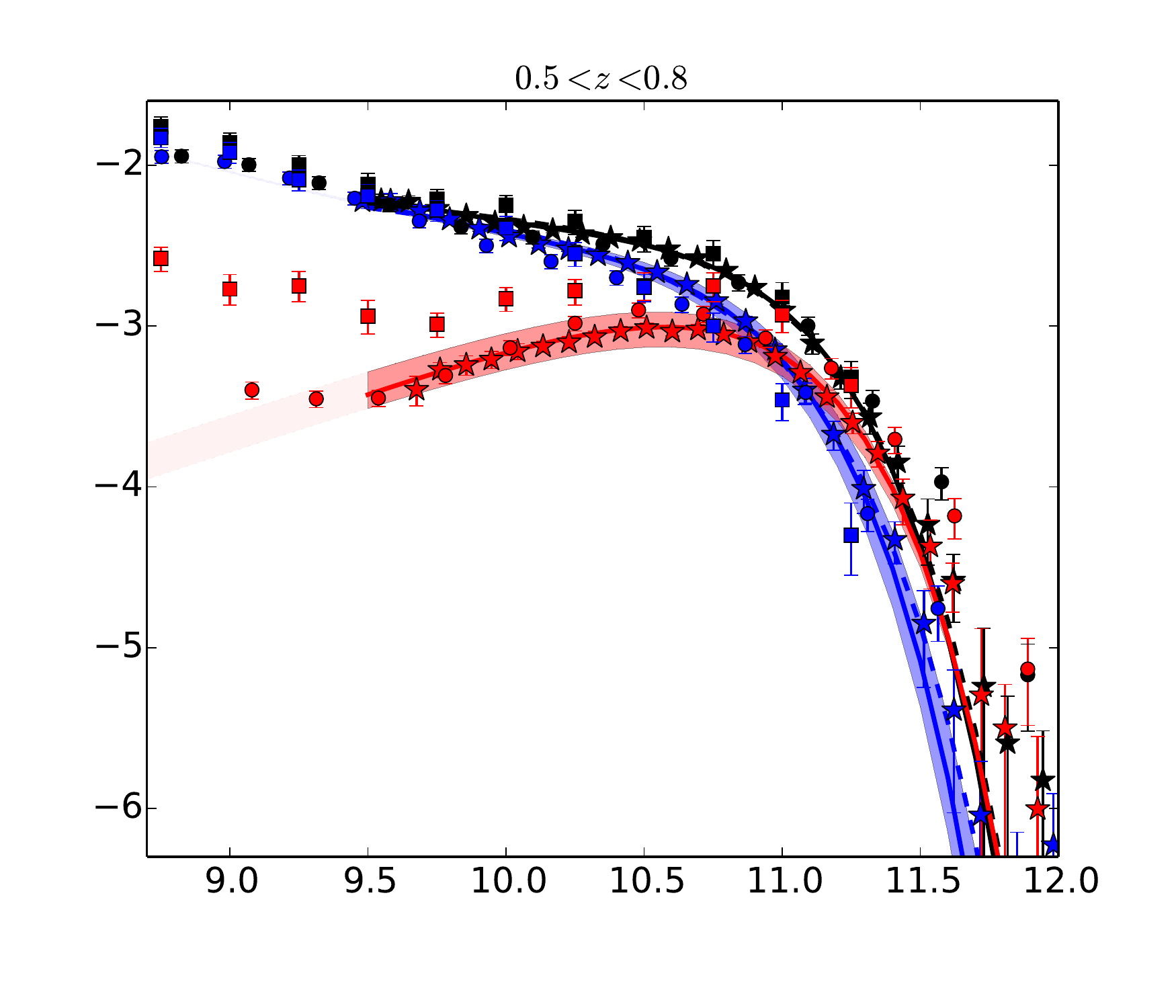}
\vspace{-0.05cm}\includegraphics[width=0.499\hsize, trim = 0.5cm 0cm 1.1cm 0cm, clip]{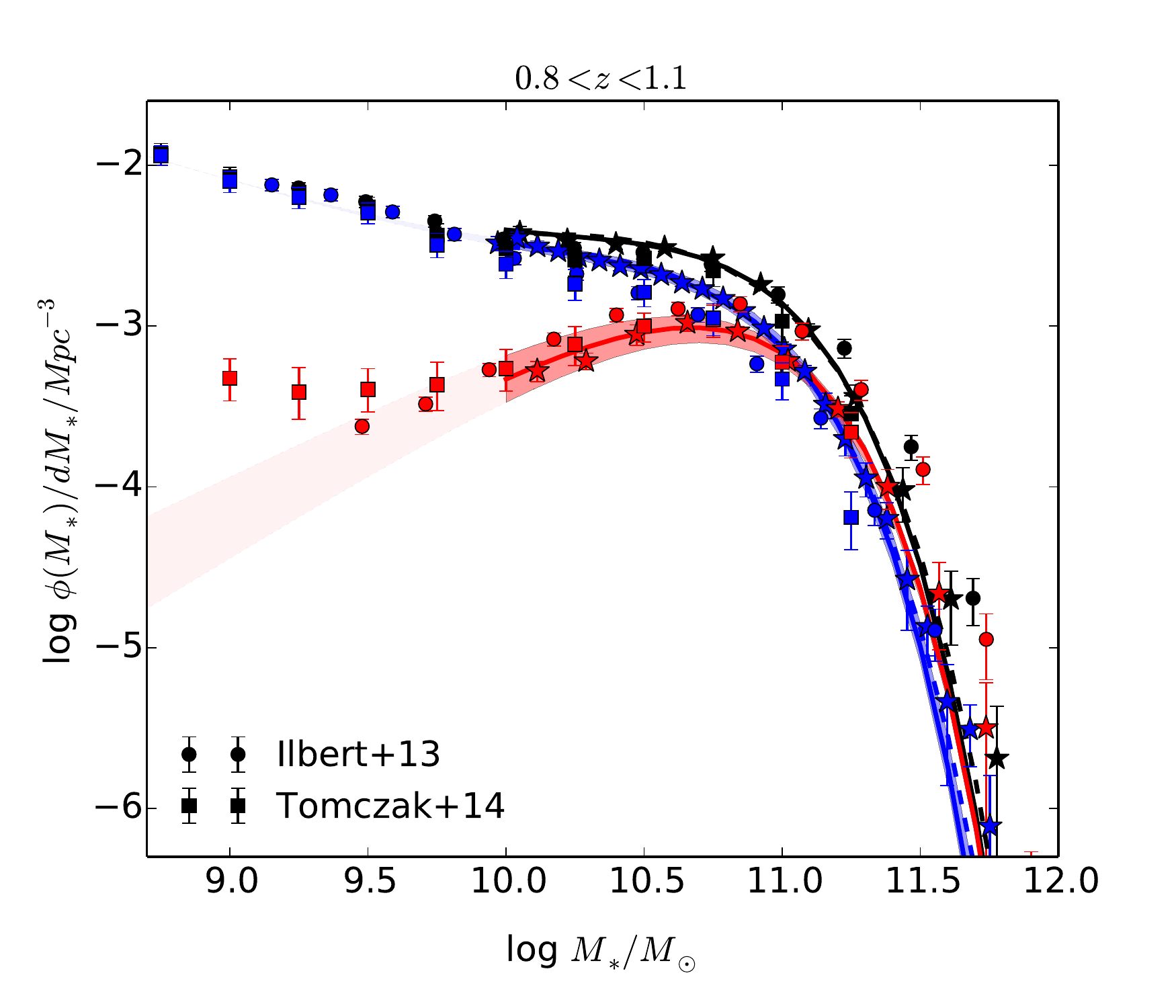}
\vspace{-0.05cm}\includegraphics[width=0.499\hsize, trim = 0.5cm 0cm 1.1cm 0cm, clip]{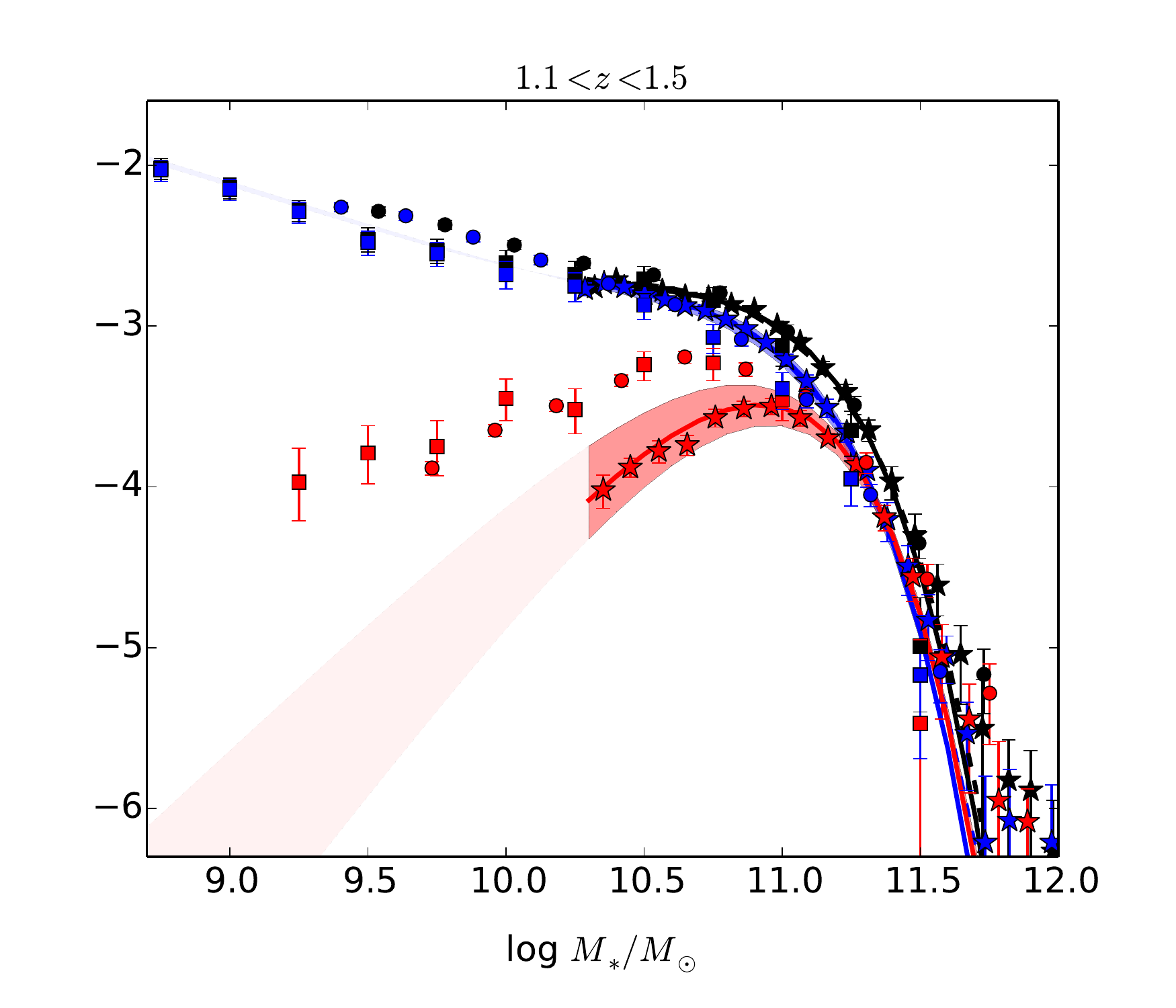}
\caption{Stellar mass function for all (black), star-forming (blue), and quiescent (red) galaxies in four redshift bins. The solid lines show the best parametric form of our SMF measurements (stars), while the shaded areas represent the systematic uncertainty due to the SF/Q separation (cf. Sect. \ref{smf_measur}). The dashed lines show the parametric forms obtained if a single-Schechter function is assumed to fit the SF population. The measurements of \citet[][squares]{Tomczak2014} and \citet[][circles]{Ilbert2013} are plotted for comparison. \label{SMF_fitt}}
\end{figure*}

To quantify the evolution of the SMF, we adopted the parametrisation proposed by   \citet{Schechter1976}. As already noted, the total stellar mass function is better fitted with a double Schechter function \citep{Baldry2008,Pozzetti2010,Ilbert2013,Tomczak2014}, defined as
\begin{equation}
\Phi(M_*) \ dM_* = e^{-\frac{M_*}{\mathcal{M}^\star}} \ \left[ \Phi^\star_1 \left(\dfrac{M_*}{\mathcal{M}^\star} \right) ^{\alpha_1} + \Phi^\star_2 \left(\dfrac{M_*}{\mathcal{M}^\star} \right) ^{\alpha_2} \right] \ \dfrac{dM_*}{\mathcal{M}^\star} ~,
\label{eq_double_sch}
\end{equation}
where $\mathcal{M}^\star$ is the characteristic stellar mass, $\Phi^\star_1$ and $\Phi^\star_2$ are the normalisation factors, and $\alpha_1$ and $\alpha_2$ are the power-law slopes satisfying $\alpha_2 < \alpha_1$.

It has been shown that the massive end of the stellar mass function can be significantly affected by the stellar mass uncertainty \citep{Caputi2011} through the so-called Eddington bias \citep{Eddington1913}. We corrected the SMF for the Eddington bias during the fitting process by convolving the SMF parametric form by the stellar mass uncertainty $\sigma_M$\footnote{Only statistical uncertainties (Poisson and cosmic variance) are considered during the fitting process, while the mass uncertainty is already taken into account in the convolution with the SMF.} following the procedure described in \citet{Ilbert2013}. The authors estimate d$\sigma_M$ for each redshift bin, but \citet{Grazian2015} have pointed out the importance of using an estimate of $\sigma_M$ that depends on the stellar mass in addition to the redshift dependence\footnote{By considering the mass dependency of $\sigma_M$ , we find that the deconvolution has a weaker effect than if we use the mean estimate of $\sigma_M$ at a given redshift.}. We used the $\sigma_M(M_*, z)$ estimate described in Sect. \ref{smf_measur} (cf. Fig. \ref{SMFs_fig}, sub-panels).

Figure \ref{SMF_fitt} shows the SMF of the global (black stars), star-forming (blue stars), and quiescent (red stars) populations. We included the SMFs measured by \citet{Tomczak2014}
and \citet{Ilbert2013}, who probed the very low-mass populations for SF and Q galaxies. A simple Schechter function (i.e. $\Phi^\star_2 = 0$ in Eq. \ref{eq_double_sch}) seems to be sufficient to fit the star-forming contribution above the stellar mass completeness limit (blue dashed lines). However, as already shown by several studies working with deeper surveys \citep[see e.g.][]{Drory2009,Ilbert2013,Tomczak2014}, the SMF of star-forming galaxies reveals an upturn at low mass and is better fitted with a double-Schechter function \citep[  ][Sect. 3.2]{Tomczak2014}. Given our stellar mass completeness limit, we can only constrain the low-mass end of the star-forming SMF at $z < 0.5$. Therefore we set the low-mass components of the double-Schechter function to the values found at $0.2 < z < 0.5$, $\alpha_{2\ \textsc{sf}} = -1.49$ and $\Phi^\star_{2\ \textsc{sf}} = -3.24$. Our choice is supported the lack of evolution that is observed for $\alpha_2$ and $\Phi^\star_2$ by \citet{Ilbert2013} and \citet{Tomczak2014}. In addition, our values agree quite well with \citet{Tomczak2014}, who probed the SMF at lower stellar mass. The resulting double-Schechter function is plotted in Fig. \ref{SMF_fitt} (blue solid line). 

For the quiescent galaxies, we clearly need a double-Schechter function to fit the SMF at low redshift (red stars in the Fig. \ref{SMF_fitt} upper left panel). The upturn at low mass is slightly more pronounced in our measurement than in the literature, regardless of the position of the quiescent galaxy selection in the NUVrK green valley (cf. Sect. \ref{smf_measur}). In other words, the low-mass slope that we measure does not depend significantly on our selection of quiescent galaxies. We also verified that we find the same shape when we select the quiescent galaxies based on their specific star formation rate (sSFR), using $sSFR < 10^{-11} Gyr^{-1}$ \citep[see][for more details on this threshold]{Ilbert2010}. We find the upturn position around $M_* \simeq 10^{9.5} M_{\odot}$ , in good agreement with previous measurements, that is, $M_* \simeq 10^{9.2} M_{\odot}$, $M_* \simeq 10^{9.4} M_{\odot}$ and $M_* \simeq 10^{9.6} M_{\odot}$ for \citet{Tomczak2014}, \citet{Ilbert2013} and \citet{Drory2009}, respectively.
Even though several deep surveys show that the low-mass upturn of the quiescent SMF is still present at $z > 0.5$, using a single-Schechter
function is sufficient given our survey stellar mass limit. The discrepancies between our star-forming and quiescent SMF and the literature are mainly explained by the different criteria used to separate quiescent and star-forming galaxies. If we include the galaxies lying in the green valley in the quiescent sample (i.e. by considering the upper or lower envelopes of the quiescent
or star-forming SMF), our measurements of the SMF agree with those of \citet{Tomczak2014} and \citet{Ilbert2013} at $z < 1.1$. At higher redshift, including the green valley in the quiescent \textit{locus} of the NUVrK diagram is not enough to reconcile the estimates. 

We cannot exclude the possibility that we may have missed some fainter red galaxies as a result of the $gri$-detection described in Sect. \ref{final_cat}. However, this effect should be limited since we corrected for this incompleteness according to the weight colour map shown in Fig. \ref{weight_map}, as previously explained. 
To add an independent validation of our procedure of correcting for this incompleteness, we used the CFHTLS-Deep/WIRcam Deep Survey \citep[WIRDS;][]{Bielby2012}, which overlaps our CFHTLS-Wide/$K_s<22$ survey. We estimated the completeness of the $gri$ selection as a function of redshift and stellar masses separately for quiescent and star-forming galaxies. Below $z<1.1$, we did not find any completeness problems, regardless of galaxy type or stellar mass range. At $1.1 < z < 1.5$, the quiescent sample is $>85\%$ complete after applying our weighting scheme, and applying a correction based on the WIRDS sample would shift the density by less than 0.1 dex, which is well inside our uncertainties. It appears that star-forming galaxies can also be affected by incompleteness around $M_* \sim 10^{10.9} M_{\odot}$ (probably because of dust extinction in massive galaxies at high redshift)\footnote{A similar trend for extremely dusty star-forming galaxies is visible in Fig. 8 of \citet{Ilbert2010}.}. However, comparison with the literature suggests that our SF sample does not significantly suffer from this incompleteness (Fig. \ref{SMF_fitt}). 

Moreover, we have to highlight that our total SMF agrees with \citet{Tomczak2014} at $M_* > M_{lim}$, while our SMF estimate for SF galaxies is continuously higher (by 0.02 dex at $M_* = 10^{10.5} M_{\odot}$ and 0.08 dex at $M_* = 10^{10.75} M_{\odot}$). This SMF difference for SF galaxies would allow a transfer (between the SF and Q populations) that is sufficient to reconcile the our SMF estimate for quiescent galaxies with the estimate of these authors. This stresses the sensitivity of the SMF to the Q/SF selection\footnote{We recall that \citet{Ilbert2013} and \citet{Tomczak2014} used a constant selection of quiescent galaxies at $z < 1.5$, while we used a time-dependent selection (cf. Sect. \ref{smf_measur}, Eq. \ref{eq_sel}).}.

Since the low-mass end of the global SMF is strongly dominated by the star-forming population at $z > 0.5$, we assumed the same parametrisation of $\alpha^*_2$ and $\Phi^\star_2$ (i.e. $\alpha^*_2 = \alpha^*_{2\ \textsc{sf}}$ and $\Phi^\star_2 = \Phi^\star_{2\ \textsc{sf}}$). We derived two parametric forms of the global SMF, depending on whether the double or the simple Schechter form of the star-forming SMF is considered, as shown in Fig. \ref{SMF_fitt} (with dashed and solid black lines, respectively). 
The corresponding best-fit parameters are reported in Tables \ref{table_bestfit} and  \ref{table_single_schech_SF}, respectively\footnote{All the parameters are given after correction for the Eddington bias (cf. Sect. \ref{fitting}).}.

\subsubsection{Quantifying the SMF evolution}
\label{quant_SMF_evol}

\begin{figure*}[!]
\includegraphics[width=0.3525\hsize, trim = 0.3cm 0cm 0.8cm 0cm, clip]{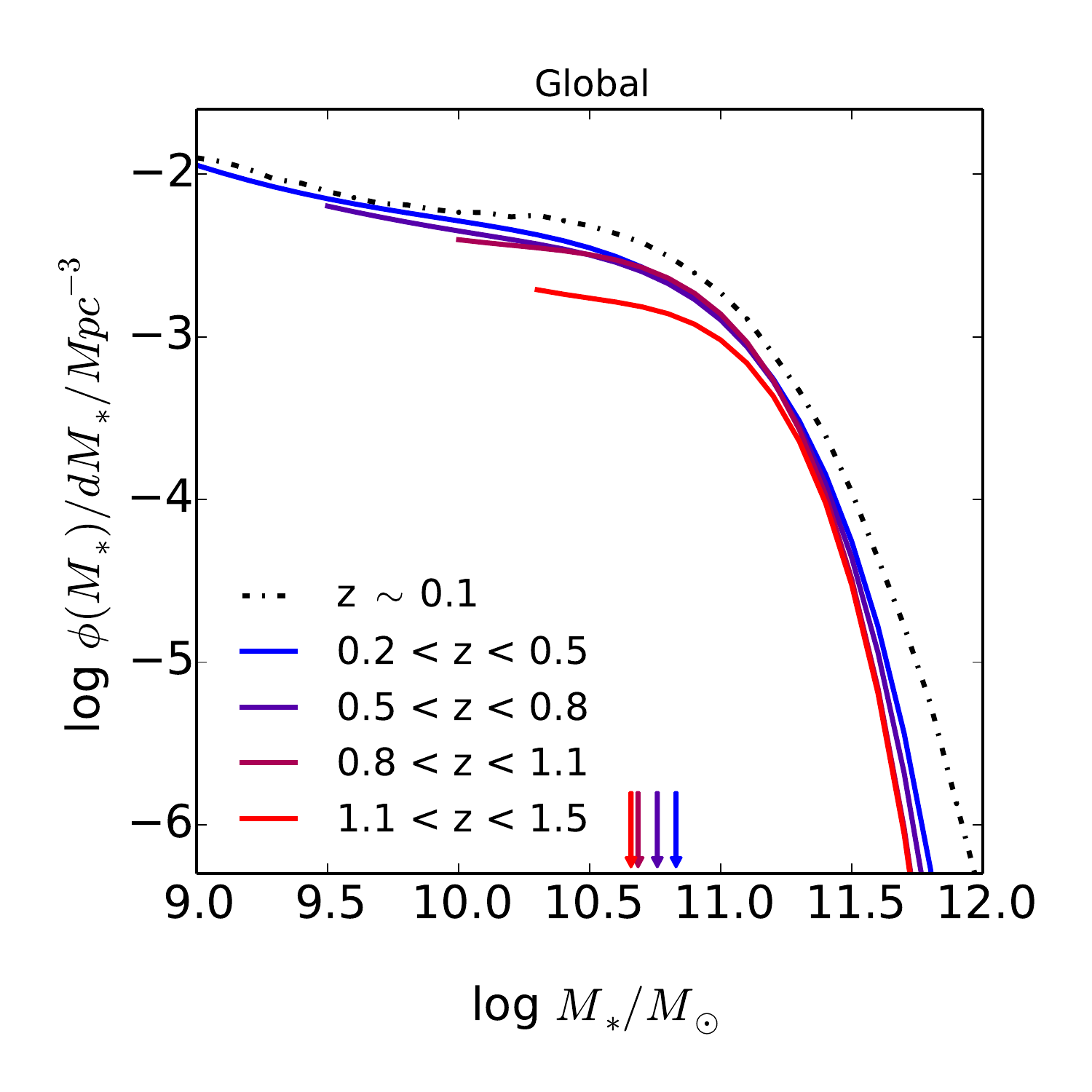}
\includegraphics[width=0.32\hsize, trim = 1.6cm 0cm 0.8cm 0cm, clip]{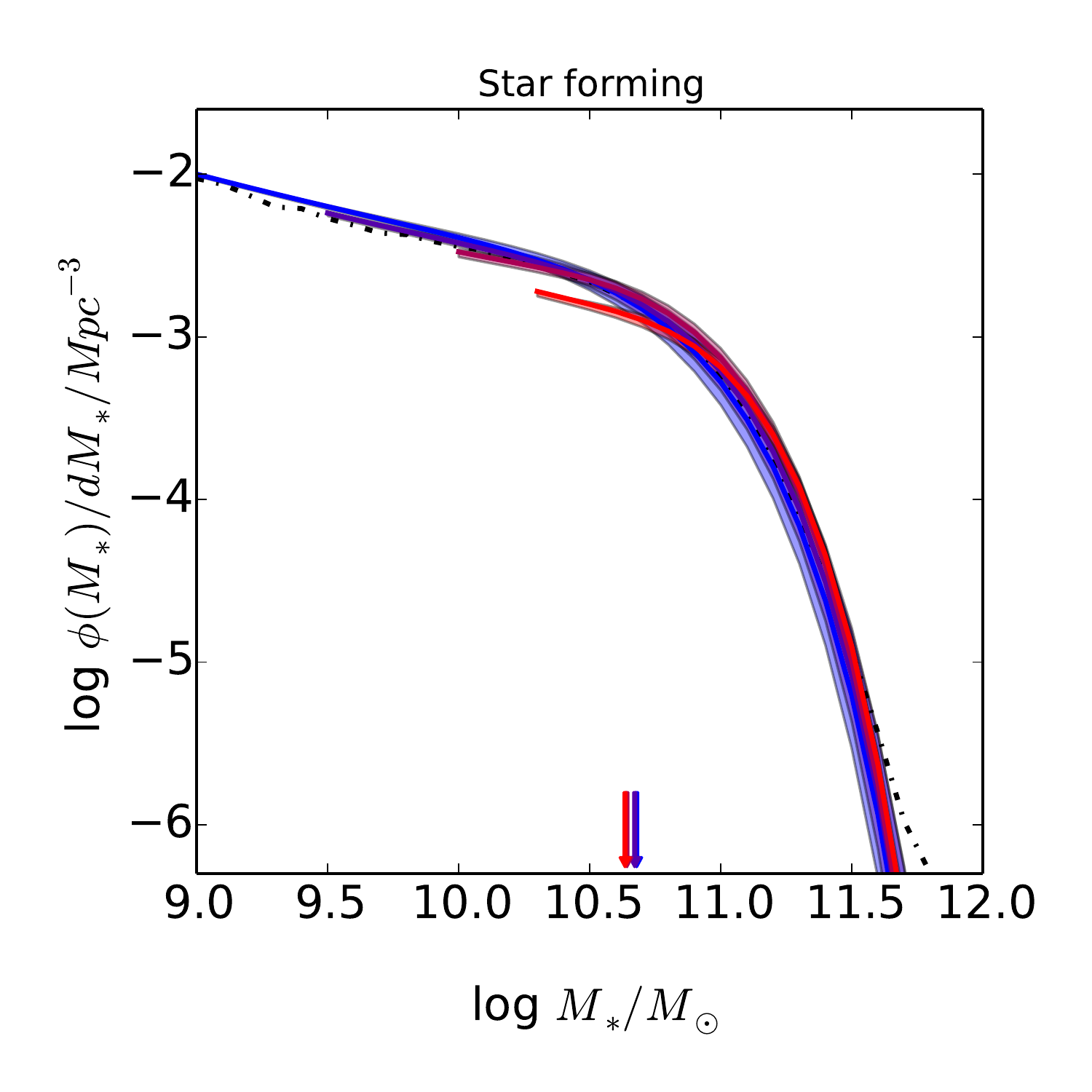}
\includegraphics[width=0.32\hsize, trim = 1.6cm 0cm 0.8cm 0cm, clip]{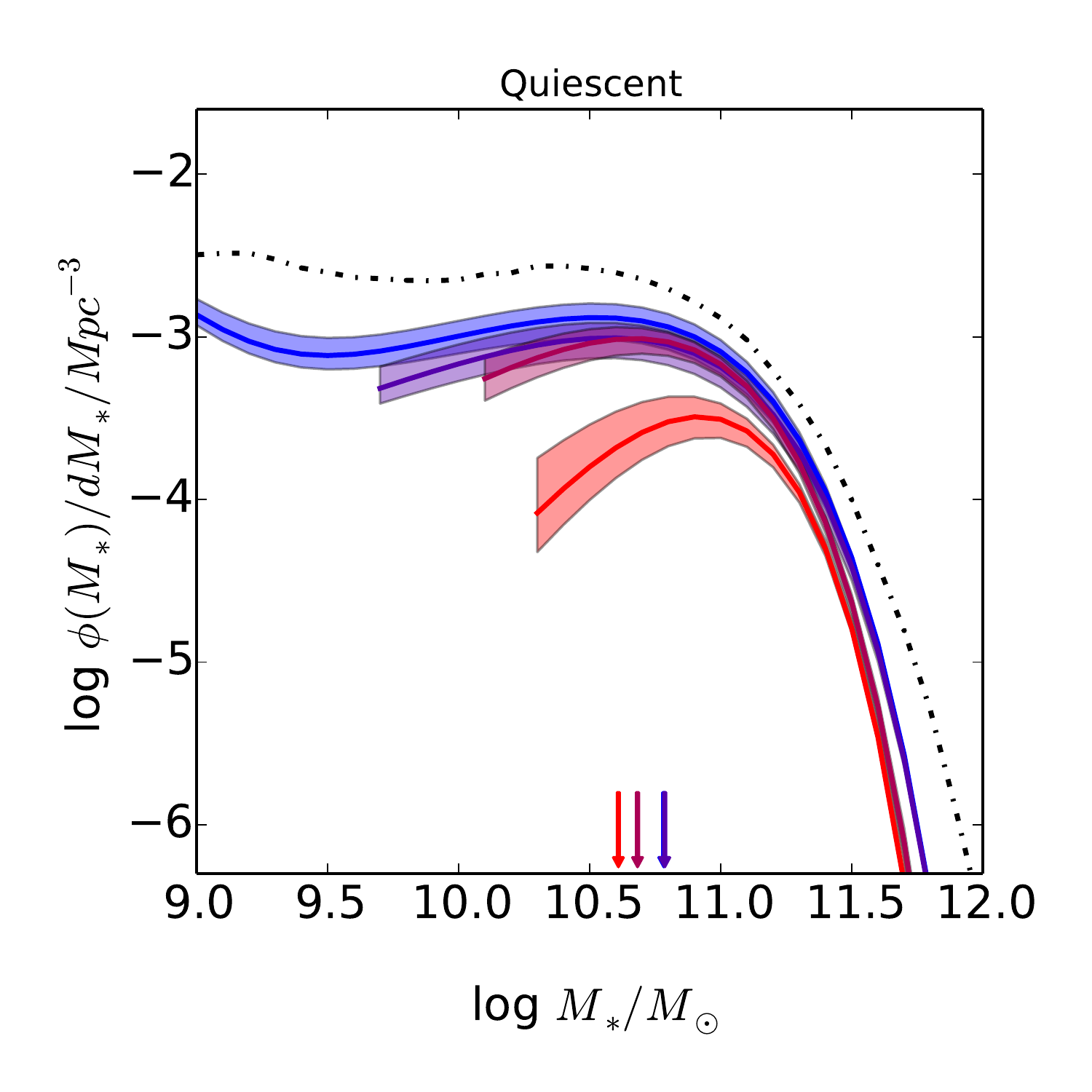}
\caption{Evolution of the SMF for the global (left), star-forming (middle), and quiescent (right) populations. The solid lines represent the best SMF parametric form at each redshift, while the arrows show the corresponding $\mathcal{M}^\star$ parameter positions. \textbf{\textit{Left}} and \textbf{\textit{middle}} panels: The dashed lines show the best fit with a single-Schechter function. \textbf{\textit{Middle}} and \textbf{\textit{right}} panels: The shaded areas represent the systematic uncertainties that are due to the separation into star-forming or quiescent galaxies depending on whether we insert the galaxies in transition (cf. Sect.  \ref{smf_measur}). \label{evolMF}}
\end{figure*}

In Fig. \ref{evolMF} we plot the evolution of the SMF for all (left panel), star-forming (middle panel), and quiescent galaxies (right panel). Each redshift bin is coded with a different colour. As in Fig. \ref{SMF_fitt}, the shaded areas show the systematic uncertainty induced by the star-forming or quiescent classification in the NUVrK diagram, while the solid lines represent the parametric form of reference. The arrows show the position of the corresponding characteristic mass $\mathcal{M}^\star$.

\begin{table*}[!]
\begin{center}
\caption{Best-fit parameters of the SMF parametric form for the total, quiescent, and star-forming populations. \label{table_bestfit}}
\begin{tabular}{l*{9}{c}}
\hline \\
\multicolumn{9}{c}{Quiescent} \\
\noalign{\smallskip}
\hline \\[-2mm]
\hline \\ 
Redshift  & $N_{gal}$ & $\log(M_{lim})$ $^{(a)}$ & $\log(\mathcal{M}^\star)$ $^{(a)}$ & $\log(\Phi^\star_1)$ $^{(b)}$ & $\alpha_1$ & $\log(\Phi^\star_2)$ $^{(b)}$ & $\alpha_2$ & $\log(\rho_*)$ $^{(c)}$\\[1mm] 
\hline \\[-2mm]
$0.2 < z < 0.5$ & 29078 & 8.75 &   10.78$^{{\rm +  0.02}}_{{\rm  -0.02}}$ &    -2.86$^{{\rm +  0.02}}_{{\rm  -0.02}}$ &   -0.44$^{{\rm +  0.05}}_{{\rm  -0.04}}$ &    -5.88$^{{\rm +  0.21}}_{{\rm  -0.42}}$ &   -2.43$^{{\rm +  0.20}}_{{\rm  -0.21}}$ & 7.88$^{{\rm +  0.03}}_{{\rm  -0.03}}$ \\[1mm] 
$0.5 < z < 0.8$ & 38708 & 9.50 &   10.79$^{{\rm +  0.01}}_{{\rm  -0.01}}$ &    -2.97$^{{\rm +  0.01}}_{{\rm  -0.01}}$ &   -0.38$^{{\rm +  0.03}}_{{\rm  -0.03}}$ &    ~~~~~~~~ &    ~~~~~~~~ & 7.76$^{{\rm +  0.02}}_{{\rm  -0.02}}$ \\[1mm] 
$0.8 < z < 1.1$ & 43421 & 9.97 &   10.68$^{{\rm +  0.02}}_{{\rm  -0.02}}$ &    -2.94$^{{\rm +  0.02}}_{{\rm  -0.02}}$ &   -0.03$^{{\rm +  0.10}}_{{\rm  -0.10}}$ &    ~~~~~~~~ &    ~~~~~~~~ & 7.73$^{{\rm +  0.03}}_{{\rm  -0.03}}$ \\[1mm] 
$1.1 < z < 1.5$ & 15567 & 10.28 &   10.61$^{{\rm +  0.02}}_{{\rm  -0.02}}$ &    -3.60$^{{\rm +  0.03}}_{{\rm  -0.04}}$ &    1.04$^{{\rm +  0.15}}_{{\rm  -0.14}}$ &    ~~~~~~~~ &    ~~~~~~~~ & 7.31$^{{\rm +  0.03}}_{{\rm  -0.03}}$ \\[1mm] 
\hline \\  
\end{tabular}   
\begin{tabular}{l*{9}{c}}
\multicolumn{9}{c}{Star-forming} \\
\noalign{\smallskip}
\hline \\[-2mm]
\hline \\
Redshift  & $N_{gal}$ & $\log(M_{lim})$ $^{(a)}$ & $\log(\mathcal{M}^\star)$ $^{(a)}$ & $\log(\Phi^\star_1)$ $^{(b)}$ & $\alpha_1$ & $\log(\Phi^\star_2)$ $^{(b)}$ & $\alpha_2$ & $\log(\rho_*)$ $^{(c)}$\\[1mm] 
\hline \\[-2mm]
$0.2 < z < 0.5$  &   143500 & 8.75 &  10.68$^{{\rm +  0.04}}_{{\rm  -0.04}}$ &    -2.89$^{{\rm +  0.09}}_{{\rm  -0.11}}$ &   -0.82$^{{\rm +  0.30}}_{{\rm  -0.23}}$ &    -3.24$^{{\rm +  0.22}}_{{\rm  -0.48}}$ &   -1.49$^{{\rm +  0.09}}_{{\rm  -0.18}}$ & 7.98$^{{\rm +  0.03}}_{{\rm  -0.03}}$ \\[1mm] 
$0.5 < z < 0.8$ & 155173 & 9.50 &  10.67$^{{\rm +  0.01}}_{{\rm  -0.01}}$ &    -2.85$^{{\rm +  0.02}}_{{\rm  -0.03}}$ &   -0.64$^{{\rm +  0.03}}_{{\rm  -0.03}}$ &    -3.24~~~~~~ &   -1.49~~~~~~ & 8.00$^{{\rm +  0.01}}_{{\rm  -0.01}}$ \\[1mm] 
$0.8 < z < 1.1$ & 114331 & 9.97 &  10.64$^{{\rm +  0.01}}_{{\rm  -0.01}}$ &    -2.78$^{{\rm +  0.02}}_{{\rm  -0.02}}$ &   -0.36$^{{\rm +  0.05}}_{{\rm  -0.05}}$ &    -3.24~~~~~~ &   -1.49~~~~~~ & 8.01$^{{\rm +  0.01}}_{{\rm  -0.01}}$ \\[1mm] 
$1.1 < z < 1.5$ & 73600 & 10.28 &  10.63$^{{\rm +  0.02}}_{{\rm  -0.02}}$ &    -2.97$^{{\rm +  0.02}}_{{\rm  -0.02}}$ &    0.02$^{{\rm +  0.06}}_{{\rm  -0.06}}$ &    -3.24~~~~~~ &   -1.49~~~~~~ & 7.92$^{{\rm +  0.01}}_{{\rm  -0.01}}$ \\[1mm] 
\hline \\
\end{tabular}
%
\begin{tabular}{l*{9}{c}}
\multicolumn{9}{c}{Total} \\
\noalign{\smallskip}
\hline \\[-2mm]
\hline \\
Redshift  & $N_{gal}$ & $\log(M_{lim})$ $^{(a)}$ & $\log(\mathcal{M}^\star)$ $^{(a)}$ & $\log(\Phi^\star_1)$ $^{(b)}$ & $\alpha_1$ & $\log(\Phi^\star_2)$ $^{(b)}$ & $\alpha_2$ & $\log(\rho_*)$ $^{(c)}$\\[1mm] 
\hline \\[-2mm]
$0.2 < z < 0.5$  &  166658 & 8.75 & 10.83$^{{\rm +  0.02}}_{{\rm  -0.03}}$ &    -2.63$^{{\rm +  0.03}}_{{\rm  -0.03}}$ &   -0.95$^{{\rm +  0.10}}_{{\rm  -0.08}}$ &    -4.01$^{{\rm +  0.28}}_{{\rm  -1.14}}$ &   -1.82$^{{\rm +  0.18}}_{{\rm  -0.22}}$ & 8.23$^{{\rm +  0.02}}_{{\rm  -0.02}}$ \\[1mm] 
$0.5 < z < 0.8$ & 185245 & 9.50 & 10.76$^{{\rm +  0.01}}_{{\rm  -0.01}}$ &    -2.66$^{{\rm +  0.02}}_{{\rm  -0.02}}$ &   -0.57$^{{\rm +  0.03}}_{{\rm  -0.03}}$ &    -3.24~~~~~~  &   -1.49~~~~~~ & 8.20$^{{\rm +  0.01}}_{{\rm  -0.01}}$ \\[1mm] 
$0.8 < z < 1.1$ & 153881 & 9.97 & 10.68$^{{\rm +  0.02}}_{{\rm  -0.02}}$ &    -2.57$^{{\rm +  0.03}}_{{\rm  -0.03}}$ &   -0.33$^{{\rm +  0.08}}_{{\rm  -0.08}}$ &    -3.24~~~~~~  &     -1.49~~~~~~ & 8.19$^{{\rm +  0.02}}_{{\rm  -0.03}}$ \\[1mm] 
$1.1 < z < 1.5$ & 85722 & 10.28 & 10.66$^{{\rm +  0.02}}_{{\rm  -0.02}}$ &    -2.88$^{{\rm +  0.01}}_{{\rm  -0.01}}$ &    0.19$^{{\rm +  0.07}}_{{\rm  -0.07}}$ &    -3.24~~~~~~  &   -1.49~~~~~~ & 8.01$^{{\rm +  0.02}}_{{\rm  -0.02}}$ \\[1mm] 
\hline \\[-2mm] 
\multicolumn{9}{l}{\begin{footnotesize} $^{(a)}$ $M_{\odot}$ \end{footnotesize}} \\
\multicolumn{9}{l}{\begin{footnotesize} $^{(b)}$ $dM_*^{-1}$ $Mpc^{-3}$ \end{footnotesize}} \\
\multicolumn{9}{l}{\begin{footnotesize} $^{(c)}$ $M_{\odot}$ $Mpc^{-3}$ \end{footnotesize}} \\
\end{tabular}
\end{center}
\end{table*}

\begin{table*}[!]
\begin{center}
\caption{Best-fit parameters of the SMF parametric form for the total and star-forming populations if a single-Schechter function
is assumed to fit the SMF of  star-forming galaxies. \label{table_single_schech_SF}}
\begin{tabular}{l*{9}{c}}
\hline \\
\multicolumn{9}{c}{Star-forming} \\
\noalign{\smallskip}
\hline \\[-2mm]
\hline \\
Redshift  & $N_{gal}$ & $\log(M_{lim})$ $^{(a)}$ & $\log(\mathcal{M}^\star)$ $^{(a)}$ & $\log(\Phi^\star_1)$ $^{(b)}$ & $\alpha_1$ & $\log(\Phi^\star_2)$ $^{(b)}$ & $\alpha_2$ & $\log(\rho_*)$ $^{(c)}$\\[1mm] 
\hline \\[-2mm]
$0.2 < z < 0.5$  &   143500 & 8.75 &  10.79$^{{\rm +  0.01}}_{{\rm  -0.01}}$ &    -2.89$^{{\rm +  0.02}}_{{\rm  -0.02}}$ &   -1.29$^{{\rm +  0.01}}_{{\rm  -0.01}}$ &    ~~~~~~~~~~~~~~~ &   ~~~~~~~~~~~~~~~ & 7.98$^{{\rm +  0.02}}_{{\rm  -0.02}}$ \\[1mm] 
$0.5 < z < 0.8$ & 155173 & 9.50 &  10.78$^{{\rm +  0.01}}_{{\rm  -0.01}}$ &    -2.83$^{{\rm +  0.02}}_{{\rm  -0.02}}$ &   -1.18$^{{\rm +  0.02}}_{{\rm  -0.02}}$ &    ~~~~~~~~~~~~~~~ &   ~~~~~~~~~~~~~~~ & 7.99$^{{\rm +  0.01}}_{{\rm  -0.01}}$ \\[1mm] 
$0.8 < z < 1.1$ & 114331 & 9.97 &  10.72$^{{\rm +  0.01}}_{{\rm  -0.01}}$ &    -2.70$^{{\rm +  0.02}}_{{\rm  -0.02}}$ &   -0.88$^{{\rm +  0.04}}_{{\rm  -0.04}}$ &    ~~~~~~~~~~~~~~~ &   ~~~~~~~~~~~~~~~ & 7.99$^{{\rm +  0.02}}_{{\rm  -0.02}}$ \\[1mm] 
$1.1 < z < 1.5$ & 73600 & 10.28 &  10.73$^{{\rm +  0.02}}_{{\rm  -0.02}}$ &    -2.83$^{{\rm +  0.02}}_{{\rm  -0.02}}$ &    -0.71$^{{\rm +  0.07}}_{{\rm  -0.04}}$ &    ~~~~~~~~~~~~~~~ &   ~~~~~~~~~~~~~~~ & 7.85$^{{\rm +  0.02}}_{{\rm  -0.02}}$ \\[1mm] 
\hline \\
\end{tabular}
%
\begin{tabular}{l*{9}{c}}
\multicolumn{9}{c}{Total} \\
\noalign{\smallskip}
\hline \\[-2mm]
\hline \\
Redshift  & $N_{gal}$ & $\log(M_{lim})$ $^{(a)}$ & $\log(\mathcal{M}^\star)$ $^{(a)}$ & $\log(\Phi^\star_1)$ $^{(b)}$ & $\alpha_1$ & $\log(\Phi^\star_2)$ $^{(b)}$ & $\alpha_2$ & $\log(\rho_*)$ $^{(c)}$\\[1mm] 
\hline \\[-2mm]
$0.2 < z < 0.5$  &  166658 & 8.75 & 10.83$^{{\rm +  0.02}}_{{\rm  -0.03}}$ &    -2.63$^{{\rm +  0.03}}_{{\rm  -0.03}}$ &   -0.95$^{{\rm +  0.10}}_{{\rm  -0.08}}$ &    -4.01$^{{\rm +  0.28}}_{{\rm  -1.14}}$ &   -1.82$^{{\rm +  0.18}}_{{\rm  -0.22}}$ & 8.23$^{{\rm +  0.02}}_{{\rm  -0.02}}$ \\[1mm] 
$0.5 < z < 0.8$ & 185245 & 9.50 & 10.79$^{{\rm +  0.02}}_{{\rm  -0.02}}$ &    -2.99$^{{\rm +  0.05}}_{{\rm  -0.06}}$ &   -0.40$^{{\rm +  0.07}}_{{\rm  -0.07}}$ &    -2.83~~~~~~  &   -1.18~~~~~~ & 8.19$^{{\rm +  0.01}}_{{\rm  -0.01}}$ \\[1mm] 
$0.8 < z < 1.1$ & 153881 & 9.97 & 10.73$^{{\rm +  0.03}}_{{\rm  -0.04}}$ &    -2.99$^{{\rm +  0.09}}_{{\rm  -0.11}}$ &   -0.33$^{{\rm +  0.08}}_{{\rm  -0.08}}$ &    -2.70~~~~~~  &     -0.88~~~~~~ & 8.17$^{{\rm +  0.02}}_{{\rm  -0.02}}$ \\[1mm] 
$1.1 < z < 1.5$ & 85722 & 10.28 & 10.68$^{{\rm +  0.10}}_{{\rm  -0.05}}$ &    -3.40$^{{\rm +  0.08}}_{{\rm  -0.32}}$ &    0.64$^{{\rm +  0.27}}_{{\rm  -0.73}}$ &    -2.83~~~~~~  &   -0.71~~~~~~ & 7.96$^{{\rm +  0.02}}_{{\rm  -0.02}}$ \\[1mm] 
\hline \\[-2mm] 
\multicolumn{9}{l}{\begin{footnotesize} $^{(a)}$ $M_{\odot}$ \end{footnotesize}} \\
\multicolumn{9}{l}{\begin{footnotesize} $^{(b)}$ $dM_*^{-1}$ $Mpc^{-3}$ \end{footnotesize}} \\
\multicolumn{9}{l}{\begin{footnotesize} $^{(c)}$ $M_{\odot}$ $Mpc^{-3}$ \end{footnotesize}} \\
\end{tabular}
\end{center}
\end{table*}

As mentioned above, the galaxy population at low masses is strongly dominated by its star-forming component, and the global SMF evolution is then mainly driven by the star-forming population. We note
that the evolution of the global SMF that is characterised by a $\sim 0.2$ dex increase of the $\mathcal{M}^\star$ (see the arrows in Fig. \ref{evolMF} left panel). However, there is almost no evolution of the star-forming population (Fig. \ref{evolMF} middle panel): the characteristic mass is nearly constant, with $\log( \mathcal{M}^\star _{\textsc{sf}} / M_{\odot}) = 10.66^{+ 0.02}_{- 0.03}$ in the redshift range $0.2 < z < 1.5$, while the evolution of the low-mass slope remains very stable, as discussed previously. This confirms that the probability of finding a star-forming galaxy declines exponentially above a certain stellar mass $M_* > \mathcal{M}^\star _{\textsc{sf}}$, which is constant with time. This stresses that the star formation seems to be impeded beyond this stellar mass independent of the redshift up to $z = 1.5$.  This is one of the cornerstones of the empirical description proposed by \citet{Peng2010}, in which the evolution of high-mass galaxy is dominated by internal quenching mechanisms (named \textit{mass quenching} by the authors).  \citet{Peng2010} suggested that the efficiency of mass quenching  is proportional to SFR/$\mathcal{M}^\star$ to keep the SMF of star-forming galaxies constant with redshift.  

The right panel of Fig. \ref{evolMF} shows that the main contribution to the evolution of the total SMF is due to the quiescent population build-up. In addition to galaxies that are quenched by mass quenching (around $\mathcal{M}^\star_{\textsc{sf}}$), the SMF evolution of quiescent galaxies reveals an increase of low-mass galaxies with time, as shown in \citet{Ilbert2010}. In particular, the SMF upturn built at $z < 0.5$ suggests that the star formation of $M_* < 10^{9 - 9.5} M_{\odot}$ galaxies is efficiently quenched, at least at low redshift. Ascribed by \citet{Peng2010} to \textit{environmental quenching}, the build-up of the low-mass quiescent population is discussed in Sect. \ref{qu_channels}. The increase of the very high-mass population that we observe in the quiescent sample (and consequently also in the total SMF) is discussed in Sect. \ref{high_mass_evol}.

\subsection{Evolution of the number densities  and stellar mass densities}
\label{dens_evol}

We derived the galaxy number and stellar mass densities, $n_*$ and $\rho_*$ , respectively, by integrating the stellar mass function \begin{equation}
n_* = \int_{M_1}^{M_2} \Phi(M_*) \ dM_*\\
\end{equation}
and 
\begin{equation}
\rho_* = \int_{M_1}^{M_2} \Phi(M_*) \ M_* \ dM_*
.\end{equation}
We adopted the parametric form of the SMF corrected for the Eddington bias. We derived the number densities above the stellar mass completeness limit. The stellar mass density was calculated by integrating the SMF over the stellar mass range $9 < \log( M_* / M_{\odot}) < 13$, as in \citet{Tomczak2014}. We recall that at $z > 0.5$, the stellar mass density relies partially on the extrapolation of the SMF to the lower stellar mass limit.

\begin{figure*}[t!]
\includegraphics[width=0.3525\hsize, trim = 0.3cm 0cm 0.8cm 0cm, clip]{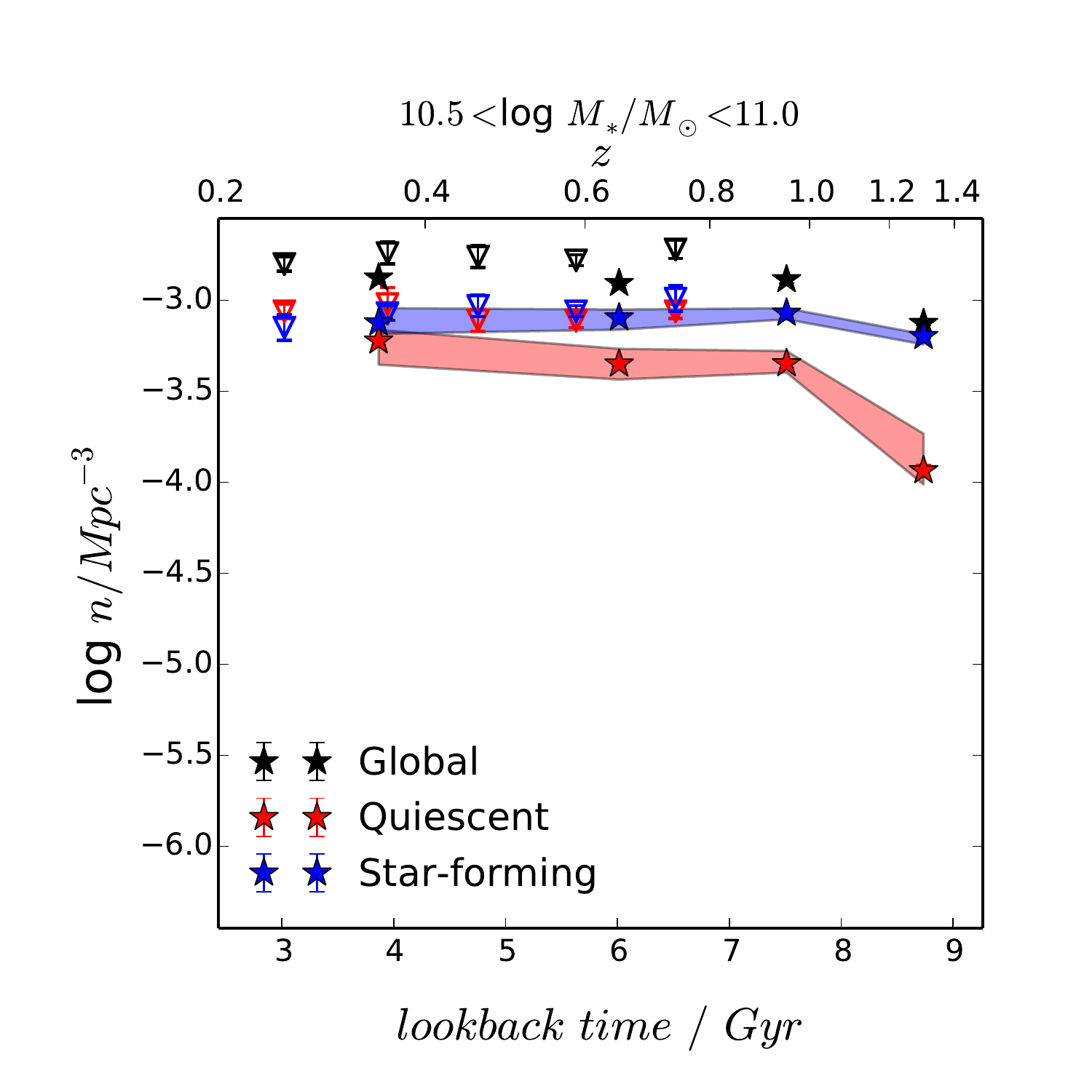}
\includegraphics[width=0.32\hsize, trim = 1.6cm 0cm 0.8cm 0cm, clip]{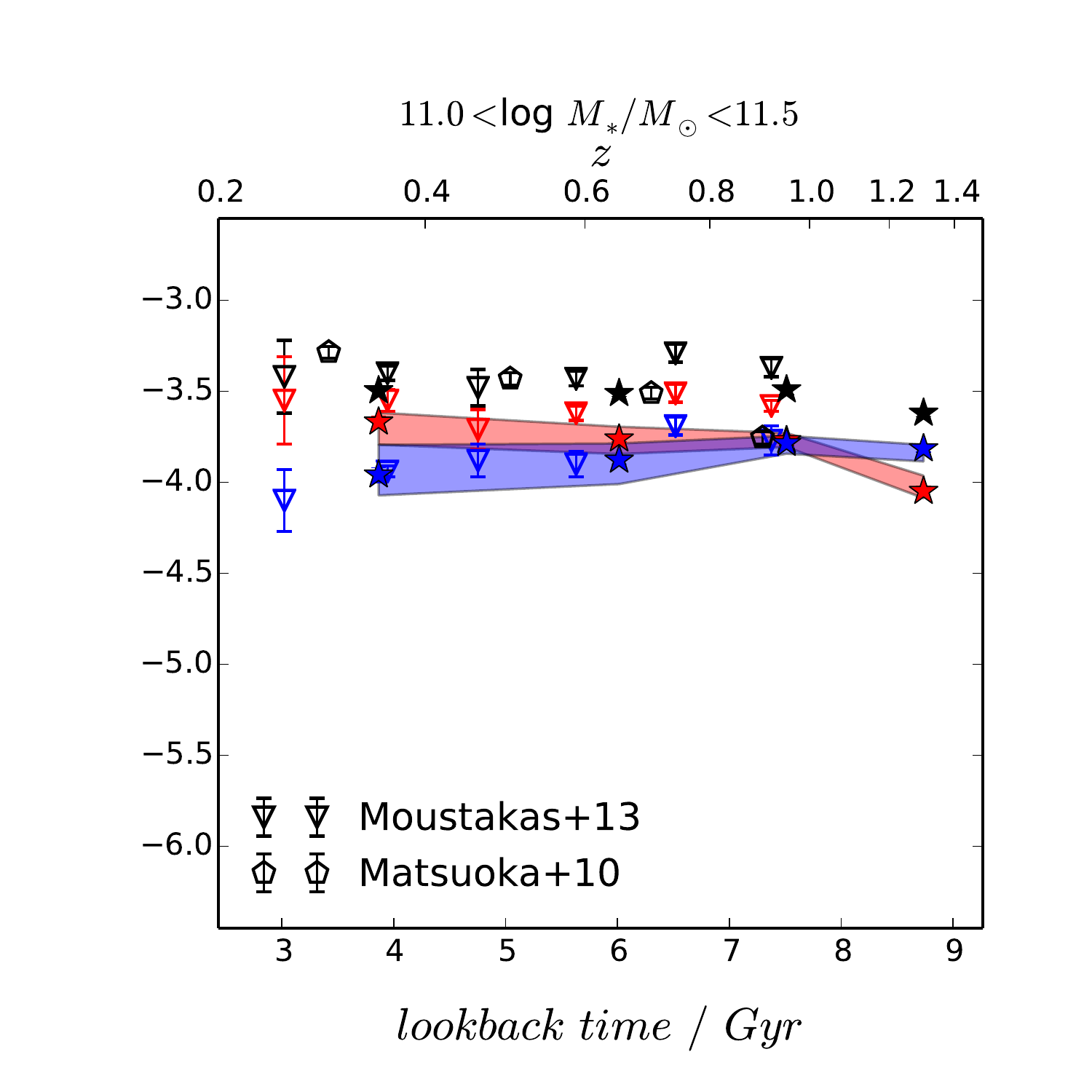}
\includegraphics[width=0.32\hsize, trim = 1.6cm 0cm 0.8cm 0cm, clip]{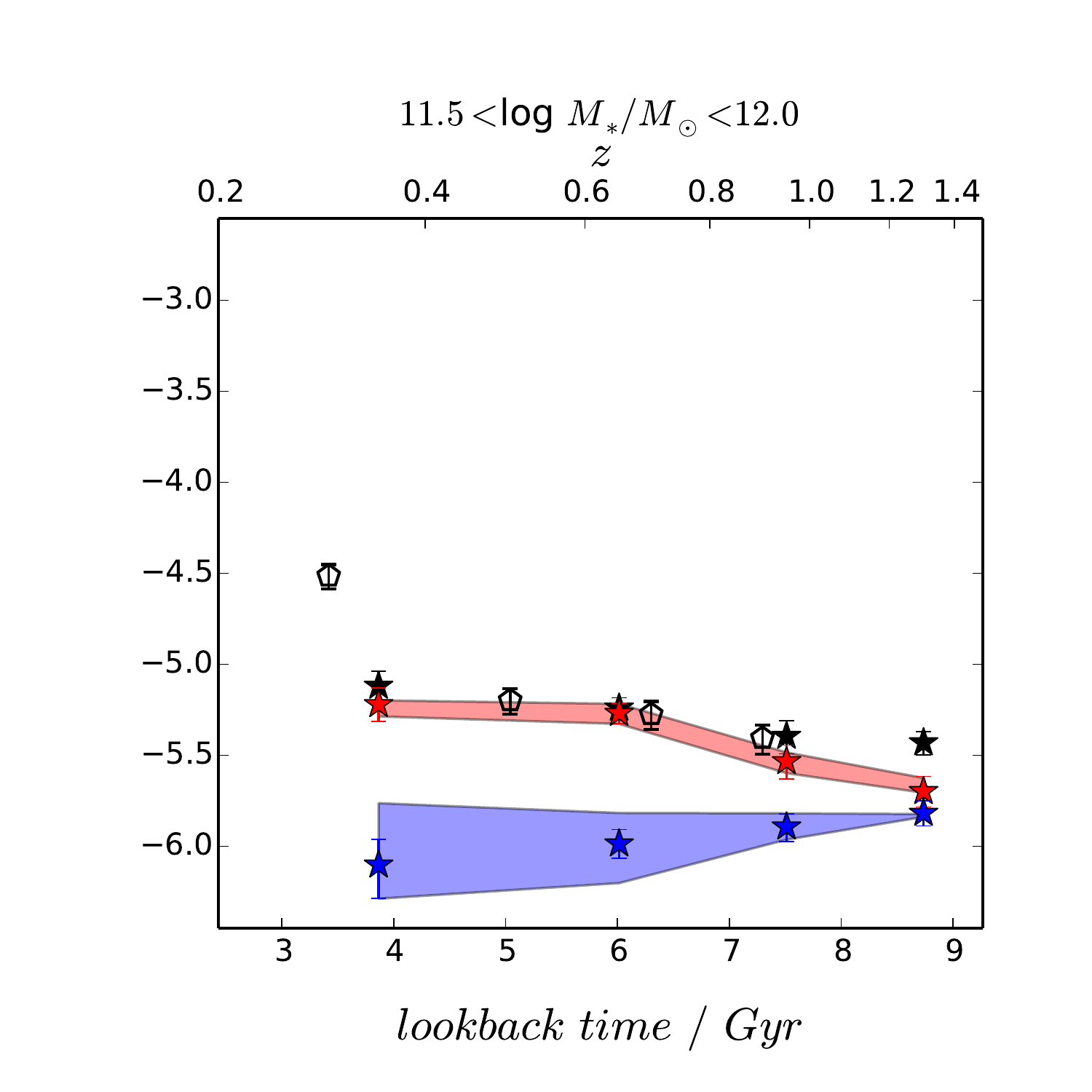}
\caption{Evolution of the number densities in three bins of $M_*$, for the global (black), SF (blue), and Q (red) populations. The corresponding shaded area shows the systematic uncertainty that is due to the SF/Q selection around our reference measurement (stars). The measurements of \citet[][triangles]{Moustakas2013} and \citet[][pentagons]{Matsuoka2010} are plotted for comparison. \label{densities}}
\end{figure*}

In Fig. \ref{densities} we plot the cosmic evolution of the number densities, $n_*$, in the stellar mass bins $10.5 < \log( M_* / M_{\odot}) < 11$ (left), $11 < \log( M_* / M_{\odot}) < 11.5$ (middle), and $11.5 < \log( M_* / M_{\odot}) < 12$ (right), between redshifts $z = 0.2$ and $z = 1.5$. For every mass bin, we  show the densities for the global, star-forming, and quiescent galaxy populations that we compare with the measurements from \citet[][triangles]{Moustakas2013} and \citet[][pentagons]{Matsuoka2010} when available.
For the global population in our sample, we distinguish two types of evolution. In the two lowest stellar mass bins ($10^{10.5} < M_* / M_{\odot} < 10^{11.5}$), we observe a two-phase evolution, with an increase of $\sim25 - 50 \%$  from $z \sim 1.3$ down to $z \sim 1$, followed by a plateau down to $z\sim 0.2$. For the most massive population ($M_* > 10^{11.5} M_{\odot}$), we observe a continuous increase by slightly less than a factor two from $z\sim 1.5$ to $z\sim 0.2$.
A similar, but weaker, trend is seen in VIPERS because of the narrower redshift range. Our results are directly comparable with \citet{Matsuoka2010} for $M_*>10^{11}M_{\odot}$. These authors
also took the Eddington bias in their density estimates into
account (the estimates are based on simulations).
 They also emphasised that their measurements at $z < 0.5$ are strongly biased  because of their less reliable photo-zs. Within these limits, our $n_*$ evolution measurements for the entire population agree well with their results.   
The trend observed with PRIMUS is also similar for the lowest mass bins, $M_* < 10^{11.5} M_{\odot}$, although they have systematically higher densities ($\sim 40\%$ and $\sim 25\%$ for $M_* \sim 10^{10.75} M_{\odot}$ and $M_* \sim 10^{11.25} M_{\odot}$ , respectively), as expected from the higher normalisation of their SMFs (cf. Fig. \ref{SMF_litt}).  In addition, it is important to recall that they did not take Eddington bias into account, which can enhance the differences, especially at $M_*> 10^{11} M_{\odot}$.

For the evolution by galaxy type, we observe a two-phase evolution of $M_* < 10^{11.5} M_{\odot}$ quiescent galaxies, while star-forming galaxies experience a constant evolution, if not a decreasing evolution.  At low mass, $M_* < 10^{11} M_{\odot}$ (left panel), the density of quiescent galaxies increases with redshift and equals the star-forming density in the lowest redshift bin, at $z \sim 0.3$.
For the intermediate masses, $10^{11}< M_*/M_{\odot} < 10^{11.5} $ (middle panel), the quiescent population becomes dominant at higher redshift, $z\sim 0.9$. 
In the highest stellar mass bin ($M_* > 10^{11.5} M_{\odot}$, right panel), the quiescent population always outnumbers the star-forming one by representing already $50-60\%$ of the global population at $z \sim 1.3$ and more than $80\%$ at $z \sim 0.3$ (i.e. $n_*$ multiplied by 2.5). From $z\sim 1$ to $z\sim0.2$, the number density of the massive star-forming galaxies has diminished by a factor of 1.5 and 2 in the two highest mass bins, respectively.   

The number densities computed in VIPERS are not plotted since the stellar mass bins used by \citet{Davidzon2013} are different from ours. However, the authors observed the same general trends, though their uncertainties prevent them from distinguishing the two-phase evolutions observed in our survey \citep[][Fig. 6]{Davidzon2013}.
We also generally agree with \citet{Moustakas2013} for star-forming galaxies, as our studies observe a decreasing $n_*$ between $z= 1$ and $z=0.3$ for $M_* > 10^{10.5} M_{\odot}$\footnote{Our highest stellar mass bin is not explored in \citet{Moustakas2013}, who limited their analysis to $M_*<10^{11.5}\,M_\odot$).}. The continuous increase of the corresponding quiescent population is also detected by \citet{Moustakas2013} between $z=1$ and $z=0.1$ when they measured the weighted linear fits of  $n_*(z)$.

\begin{figure}[!]
\includegraphics[width=\hsize, trim = 0.5cm 0cm 1cm 0cm, clip]{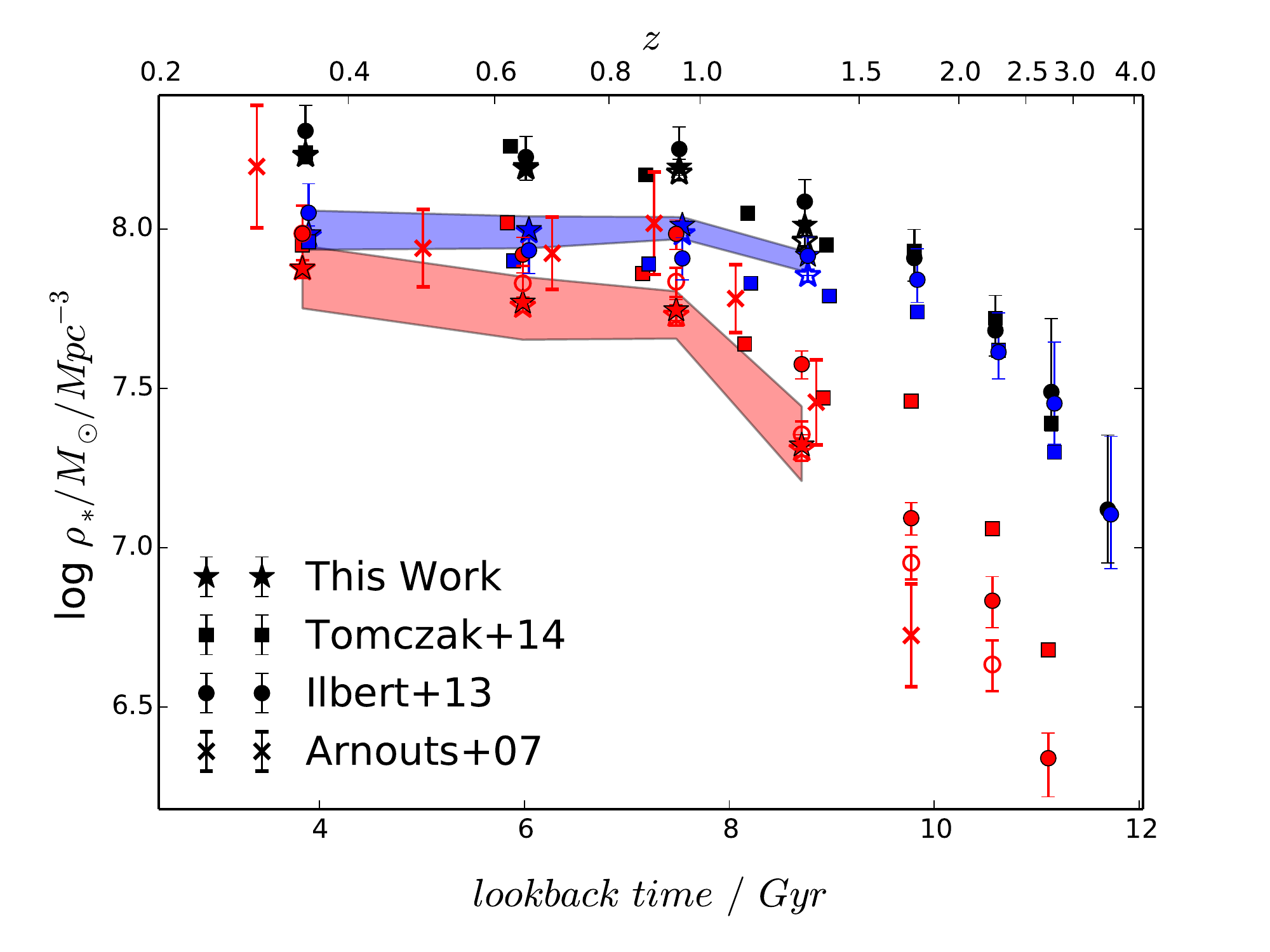}
\caption{Evolution of the cosmic stellar mass density for all (black), star-forming (blue), and quiescent (red) galaxies. The shaded areas show the corresponding systematic uncertainties that are due to the SF/Q selection. The open stars represent the measurement that we obtain by assuming a single-Schechter function to fit the star-forming galaxies. Measurements of  \citet[][squares]{Tomczak2014}, \citet[][circles]{Ilbert2013}, and \citet[][quiescent only, red crosses]{Arnouts2007} are shown for comparison. The filled and open red circles represent the quiescent measurements of \citet{Ilbert2013}, using a selection of quiescent galaxies based on the NUV-r/r-J plan and the sSFR respectively. The quiescent measurement of \citet{Arnouts2007} is based on the $K$-band luminosity density, and the selection uses the SED-fitting. For the sake of clarity, the star-forming
or quiescent measurements are plotted with of shift of +0.03
or /-0.03 Gyr.\label{sm_density}}
\end{figure}

Figure \ref{sm_density} presents the cosmic evolution of the stellar mass density $\rho_*$ for all (black), star-forming (blue), and quiescent (red) galaxies. We compare our results (filled stars) with previous studies. We also plot the stellar mass density obtained by assuming a different slope of the star-forming SMF low-mass end (open stars; cf. Sect. \ref{SMF_evol}), but it does not change the results significantly. In good agreement with \citet[][circles]{Ilbert2013}\footnote{With respect to our results, the slightly higher values measured in COSMOS are expected, given the $8 < \log( M_* / M_{\odot}) < 13$ integration range adopted by \citet{Ilbert2013}.} and \citet[][squares]{Tomczak2014}, our measurement of the global evolution of $\rho_*$ reveals two phases: a $>50 \%$ increase from $z \sim 1.3$ down to $z \sim 1$, and a continuous 12--20\% increase from $z \sim 1$ down to $z \sim 0.3$. 

As mentioned in Sect. \ref{SMF_evol}, our selection of quiescent galaxies is more compatible with the selections of \citet{Ilbert2013} and \citet{Tomczak2014} when we consider that galaxies lying in the green valley are classified as quiescent. This corresponds to the upper red and lower blue envelopes of $\rho_*$ in Fig. \ref{sm_density}. Still, our measurement for quiescent galaxies is smaller than previous measurements by up to 25\%. We do not find this difference when we consider the global stellar mass density. The importance of the Q/SF selection is reinforced by the fact that the agreement is better with \citet{Ilbert2013}, when they use the log $sSFR= -11$ selection\footnote{See \citet{Ilbert2010} concerning this threshold.} (Fig. \ref{sm_density}, open red circles). Our measurement is also consistent with the $\rho_*$ measured by \citet[][ red crosses]{Arnouts2007} for quiescent galaxies, which are selected thanks to SED-fitting (we do not plot their star-forming $\rho_*$\footnote{The $\rho_*$ measurement of \citet{Arnouts2007} is based on the $K$-band luminosity. \citet[][Appendix D]{Ilbert2010} showed that the mass-to-light ratio derived by \citet{Arnouts2007} for star-forming galaxies is not appropriate at low and intermediate masses.}).

As previously suggested, the evolution of the stellar mass density of star-forming galaxies seems to be quite stable at $z < 1.5$. At the same time, a rapid increase of the stellar mass contained in quiescent galaxies is observed, increased by a factor $>2.5$ from $z \sim 1.3$ down to $z \sim 1$. At lower redshift, we detect a small and continuous $\gtrsim30\%$ increase of $\rho_*$ from $z \sim 1$ down to $z \sim 0.3$, which reflects the progressive quenching of less massive galaxies.

\section{Discussion}
\label{discut}

\subsection{High-mass end evolution}
\label{high_mass_evol}

As highlighted above, our sample can be used to investigate the evolution of massive ($M_*> 10^{10.5} M_{\odot}$) and  rare ($M_*> 10^{11.5} M_{\odot}$) galaxies, thanks to the large volume of our survey. Most importantly, we are interested in the evolution of these objects across cosmic time, in particular to understand which mechanisms determine their evolution from star-forming to quiescent galaxies. Several studies \citep[e.g.][]{Kauffmann2003,Bundy2006,Davidzon2013} have characterised galaxy quenching with the so-called  \textit{transition mass}, which is~the stellar mass at which the quiescent and star-forming populations are equal in a given redshift bin.
In the same spirit, we define the transition redshift, $z_{tr}$, at which the quiescent population becomes dominant. As shown in Fig. \ref{densities}, the transition redshift is found to be $z_{tr} \gtrsim 1.4$, $z_{tr} \sim 0.9$ and $z_{tr} \sim 0.2$, for $M_* \sim10^{11.75} M_{\odot}$, $M_* \sim10^{11.25} M_{\odot}$ , and $M_* \sim10^{10.75} M_{\odot}$ galaxies, respectively: globally, the more massive a galaxy, the earlier its star formation is stopped. This is qualitatively consistent with the redshift evolution of the transition mass \citep[e.g. see][]{Davidzon2013}.
As already mentioned, several physical mechanisms could explain this trend within a hierarchical context \citep[e.g.][]{DeLucia2007,Neistein2008,Weinmann2012}. 
For instance, based on the  stellar-halo mass relation from \citet[][]{Coupon2015}, the star-forming galaxies with stellar masses of $M^*_{SF} (\sim 10^{10.64} M_{\odot}$) should reside in dark matter halos of masses of around $M_h \sim 10^{12.4} M_{\odot}$. This value agrees
well with the halo mass threshold invoked by \citet{Cattaneo2006}, corresponding to halo' quenching, but we cannot exclude that some radio-AGN quenching could also explain why massive galaxies cease forming stars and/or are not fuelled anymore by fresh infalling gas \citep{Croton2006}. 

We find that the number density of the most massive ($M_* > 10^{11.5} M_{\odot}$) galaxies almost doubled from $z \sim 1$ to $z \sim 0.3$ (Fig. \ref{densities}). This corresponds to the $< 0.25$ dex increase of the SMF high-mass end that is seen between $z \sim 1$ and $z \sim 0.3$ (Fig. \ref{evolMF}). Because the high-mass end is dominated by quiescent galaxies at $z < 1$, the increase of the $M_* > 10^{11.5} M_{\odot}$ population cannot be explained by incidental star formation \citep{Arnouts2007}. If we assume that, in general, these very high-mass galaxies do not experience significant star formation, they can still assemble stellar mass through mergers at $z<1$, in particular through dry merging.

\subsection{Taming of galaxies}
\label{qu_channels}

\begin{figure}
\includegraphics[width=\hsize, trim = -0.5cm 0cm -1cm 0cm, clip]{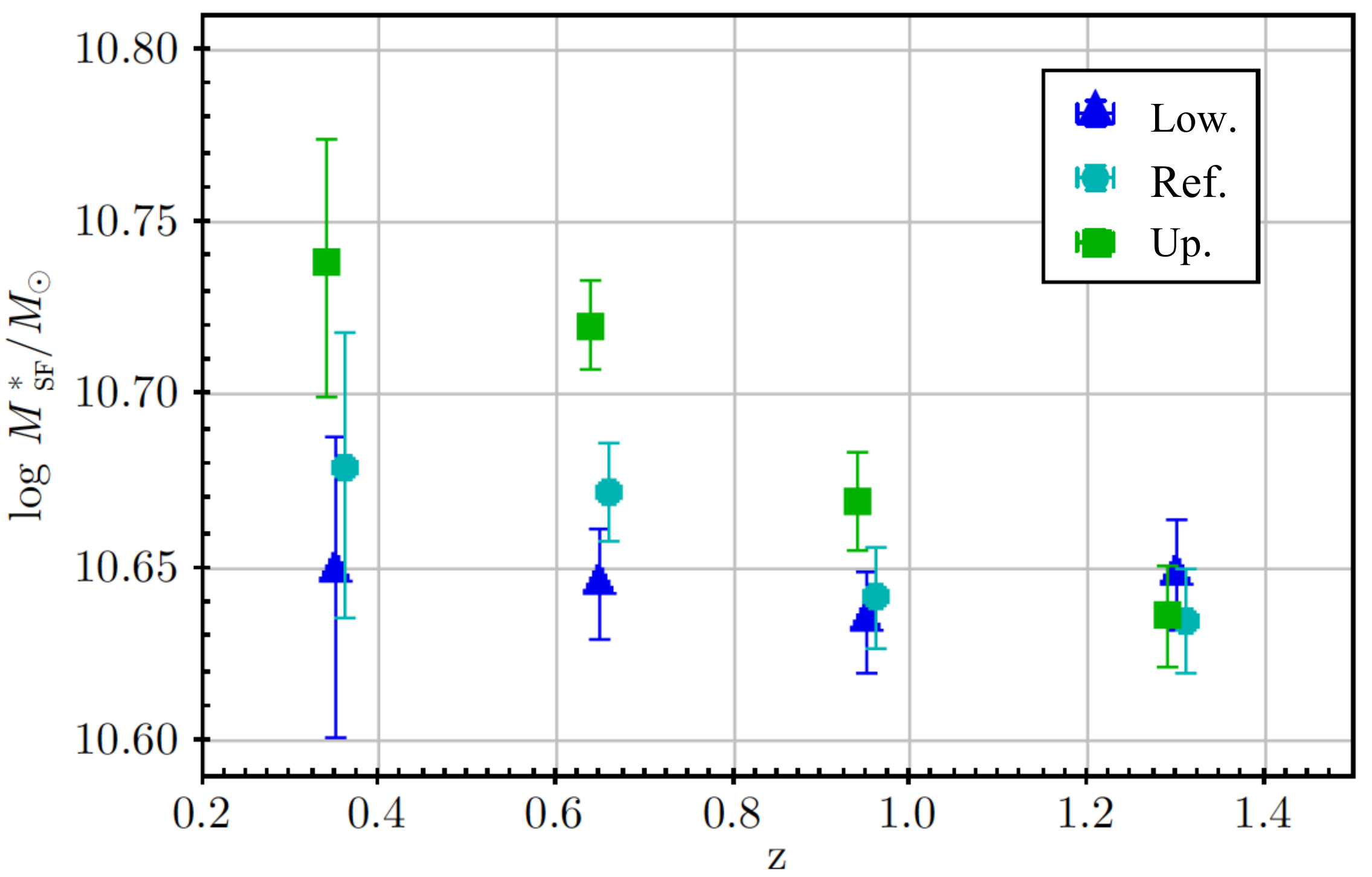}
\caption{Redshift evolution of $\mathcal{M}^\star _{\textsc{sf}}$, corresponding to the three selections of SF galaxies in the NUVrK diagram defined in Sect. \ref{smf_measur}:  the reference selection (for which the limit lies in the middle of the green valley; cyan circles), its \textit{lower} limit (when galaxies in transition are excluded; blue triangles), and the \textit{upper} limit (if the green valley is included in the SF \textit{locus}; green squares).\label{Mstar_z}}
\end{figure}

In Sect. \ref{SMF_evol} we have shown that the characteristic stellar mass of the star-forming SMF does not vary significantly between redshifts $z = 0.2$ and $z = 1.5$. As described in Sect. \ref{smf_measur}, we performed three selections of the SF galaxies, and the values of $\mathcal{M}^\star _{\textsc{sf}}$ differed slightly from one selection to another. In Fig. \ref{Mstar_z} we plot $\mathcal{M}^\star _{\textsc{sf}}$ as a function of the redshift and the SF galaxy selection in the NUVrK diagram. First, we find that $\mathcal{M}^\star _{\textsc{sf}} $ is between $10^{10.6}$ and $10^{10.8} M_{\odot}$ at $0.2 < z < 1.5$, regardless of the SF selection in the NUVrK diagram. More precisely, we find
\begin{itemize}
\item log $\mathcal{M}^\star _{\textsc{sf}} / M_{\odot} = 10.69^{+ 0.04}_{- 0.05}$ if the galaxies in transition are included in the selection of SF galaxies (upper dotted lines in Fig \ref{NUVrK_z}),
\item log $\mathcal{M}^\star _{\textsc{sf}} / M_{\odot} = 10.66^{+ 0.02}_{- 0.03}$ for our intermediate selection and
\item log $\mathcal{M}^\star _{\textsc{sf}} / M_{\odot} = 10.64^{+ 0.01}_{- 0.01}$ for the most conservative selection. 
\end{itemize}
Therefore, the evolution of $\mathcal{M}^\star _{\textsc{sf}}$ is consistent with being constant if the galaxies transitioning in the green valley are excluded from the selection of SF galaxies. The invariance with respect to redshift of $\mathcal{M}^\star _{\textsc{sf}}$ for the most conservative selection strongly supports a mass-quenching process occurring around a constant stellar mass, which makes this selection suitable for investigating the galaxies that are about to quench.

\subsubsection{Tracking galaxies in the green valley}
\label{gal_tracking}

\begin{figure}[!]
\includegraphics[width=0.5\columnwidth, trim = -0.15cm 1.5cm 0cm 0cm, clip]{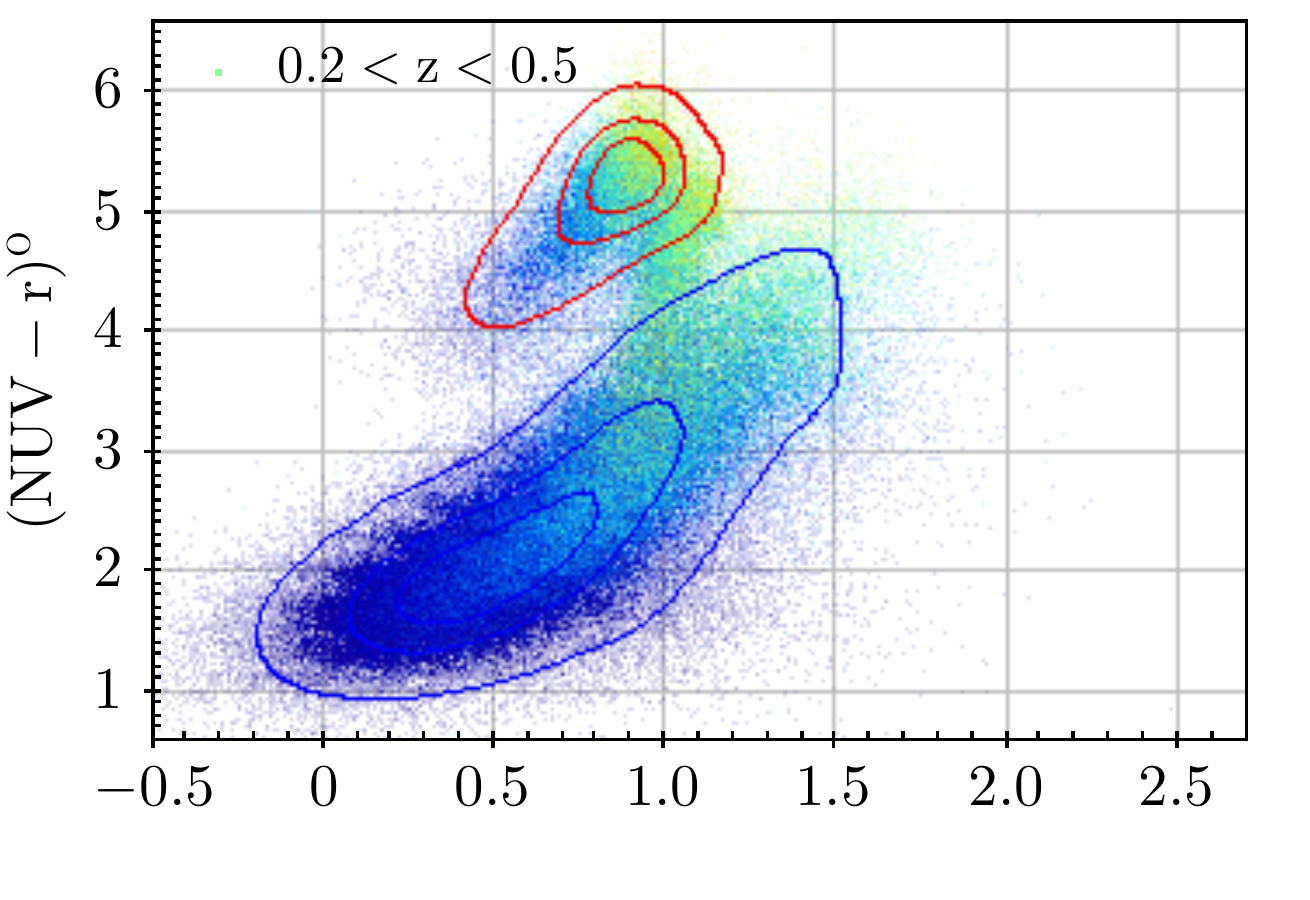}
\hspace{-0.15cm}\includegraphics[width=0.5\columnwidth, trim = -0.15cm 1.5cm 0cm 0cm, clip]{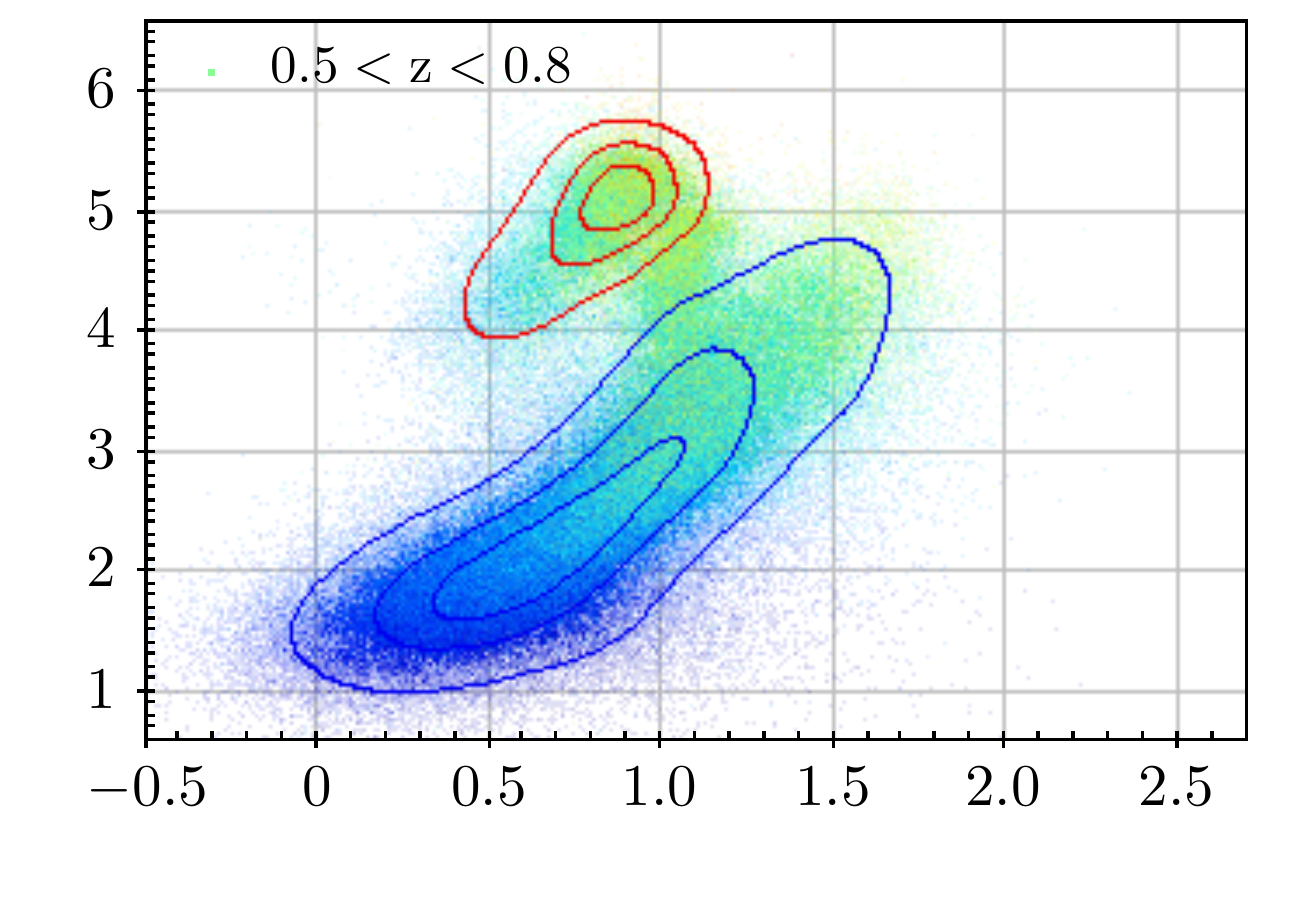}

\includegraphics[width=0.5\columnwidth, trim = -0.05cm 0.9cm 0cm 0cm, clip]{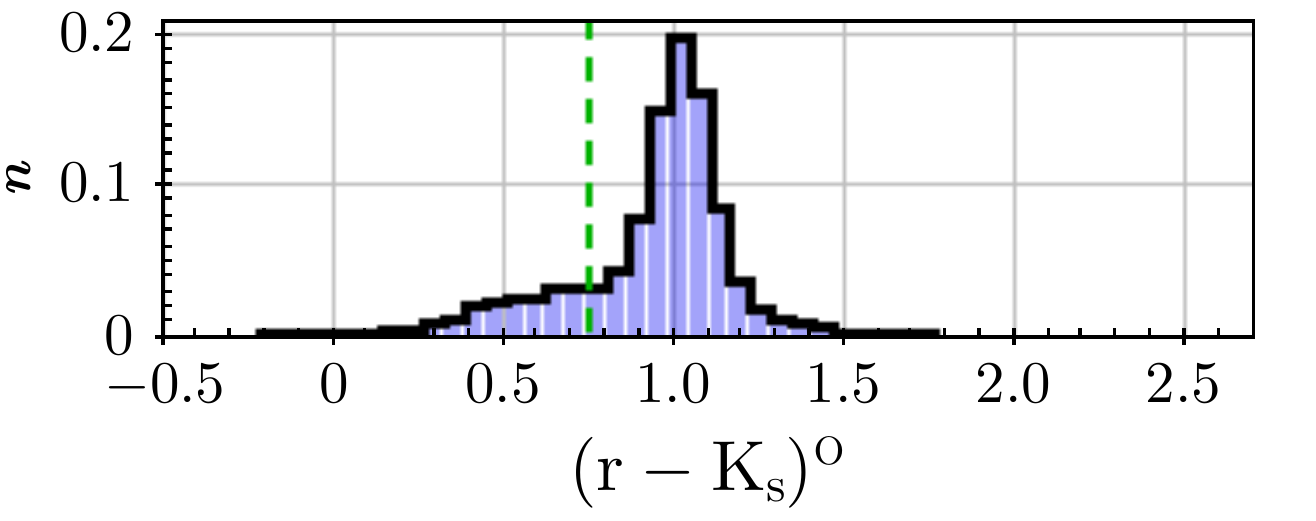}
\includegraphics[width=0.484\columnwidth, trim = 0.5cm 0.9cm 0cm 0cm, clip]{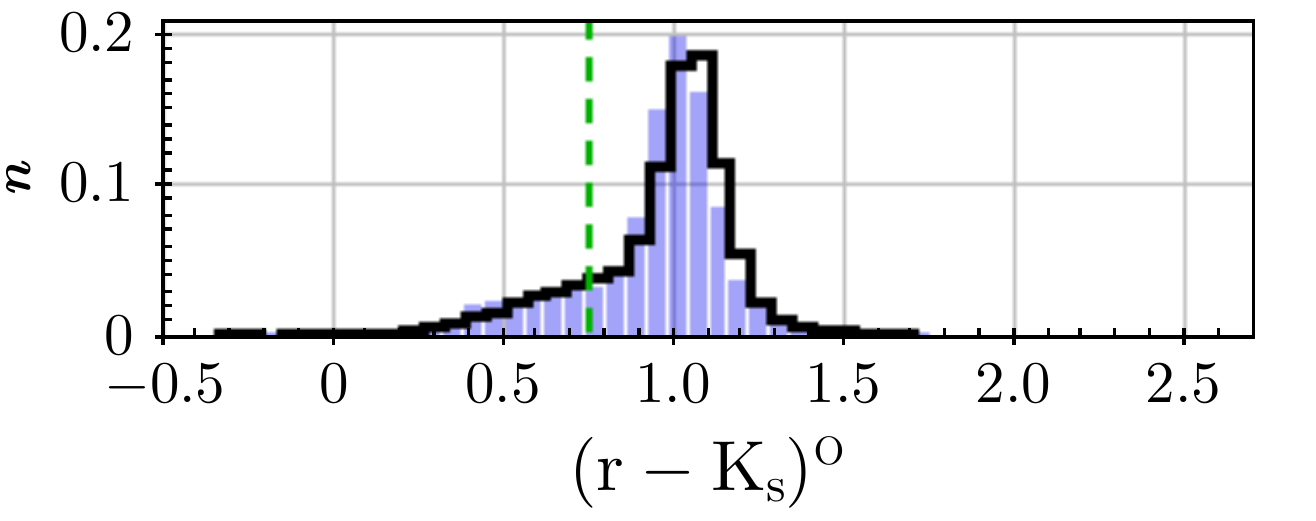}

\includegraphics[width=0.5\columnwidth, trim = -0.15cm 1.5cm 0cm 0cm, clip]{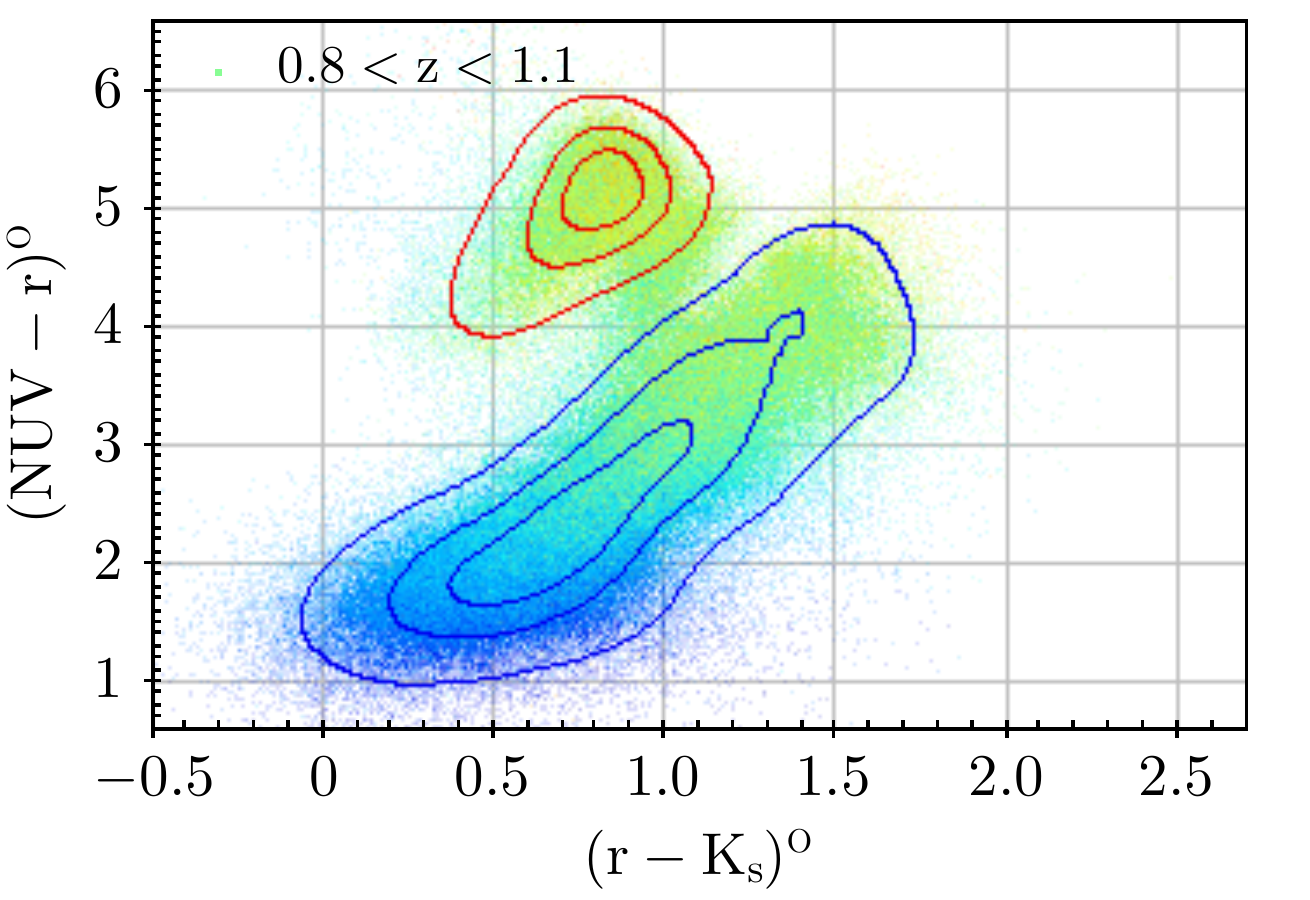}
\hspace{-0.15cm}\includegraphics[width=0.5\columnwidth, trim = -0.15cm 1.5cm 0cm 0cm, clip]{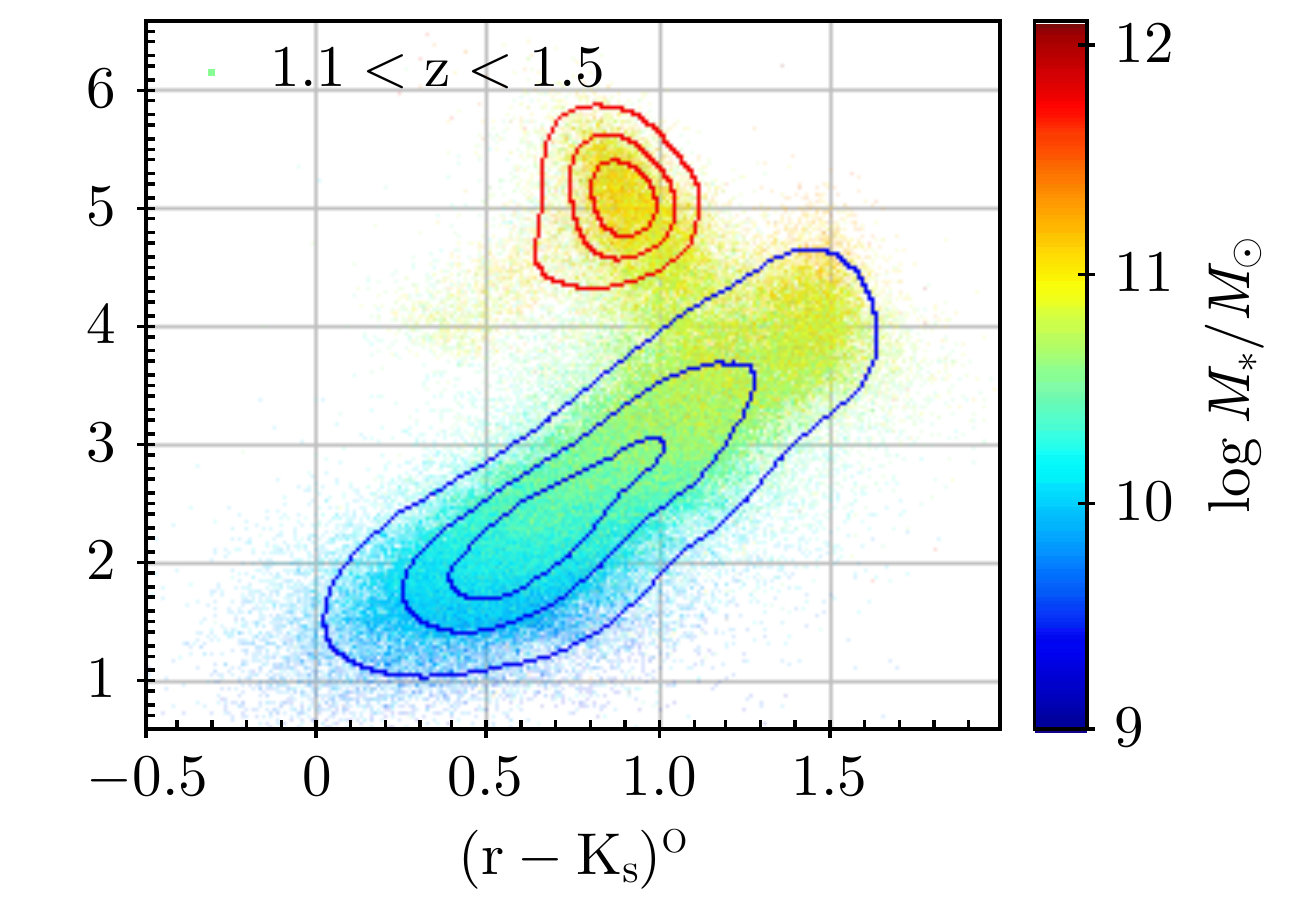}

\includegraphics[width=0.5\columnwidth, trim = -0.05cm 0cm 0cm 0cm, clip]{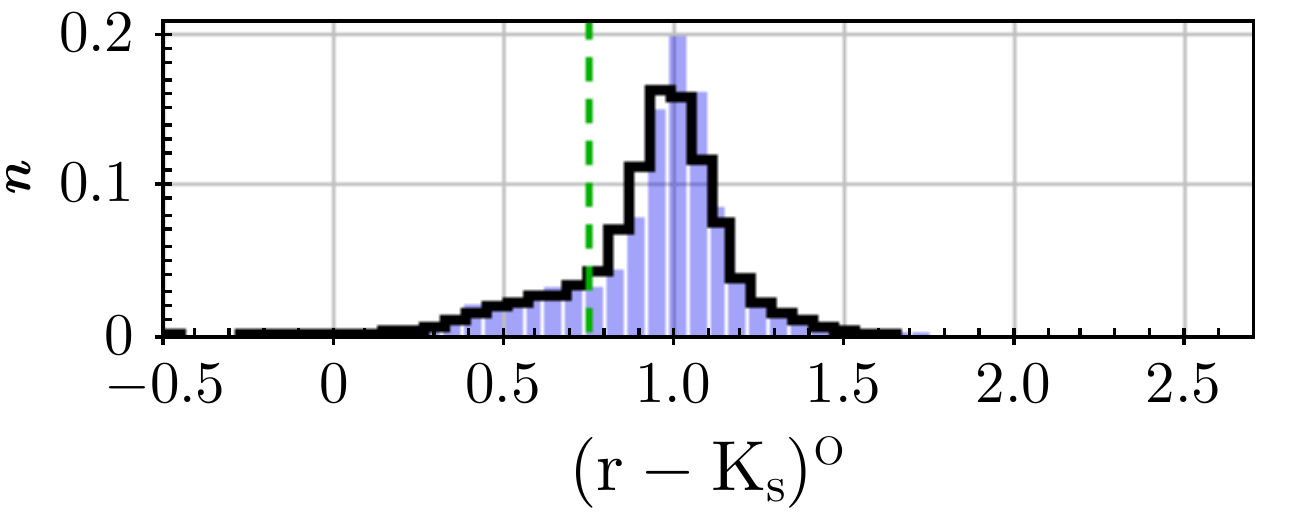}
\includegraphics[width=0.484\columnwidth, trim = 0.5cm 0cm 0cm 0cm, clip]{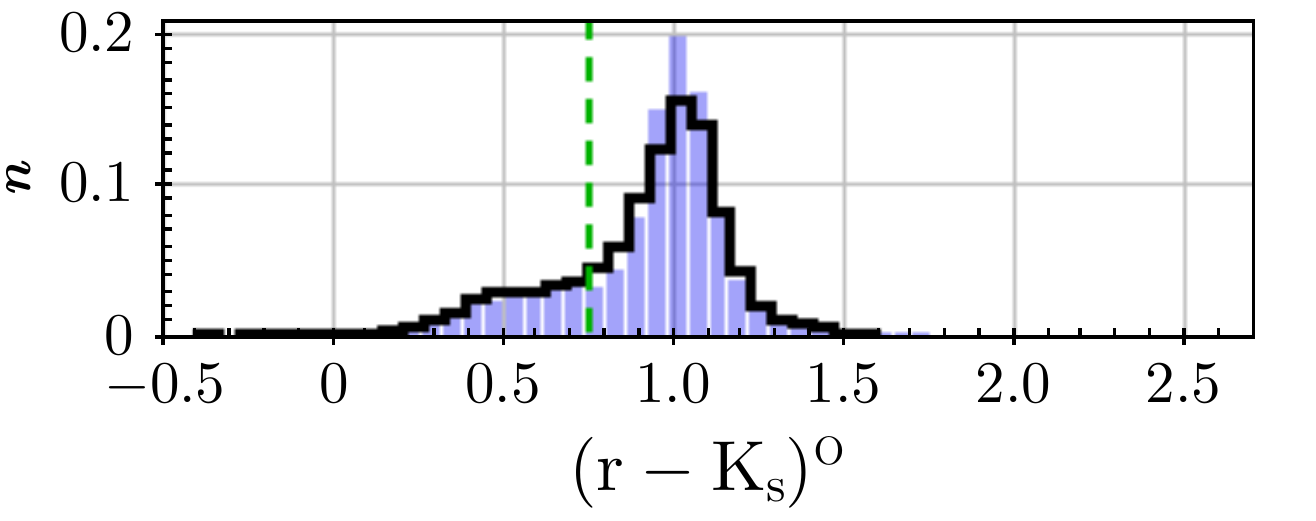}
\caption{NUVrK galaxy distribution ouside and inside the green valley, shown in four redshift bins. \textbf{\textit{Top sub-panels:}} NUVrK diagram as a function of the galaxy stellar mass. The red and blue contours show the equal density of the quiescent and star-forming populations, respectively, after excluding the transitioning galaxies (i.e. the galaxies lying in the green valley defined in Fig. \ref{NUVrK_z}).  \textbf{\textit{Bottom sub-panels:}} Normalised number counts along the $(r-K_s)^\textsc{o}$ colour in the green valley (black solid line). The distribution at $0.2 < z < 0.5$ is repeated in each panel for comparison (blue shaded area). The vertical green dashed line shows the limit of the $\mathcal{M}^\star _{\textsc{sf}}$-quenching channel, as discussed in Sect. \ref{gal_tracking}. \label{NrK_Mass}}
\end{figure}

To identify a potential quenching channel for $\mathcal{M}^\star _{\textsc{sf}}$ galaxies, we isolate and characterise the green valley galaxies in Fig. \ref{NrK_Mass}, where each panel shows a different redshift bin. The contours represent the density of quiescent and star-forming galaxies, when the galaxies in transition are excluded (i.e. using the strictest selection of Q/SF galaxies). The colour code expresses the stellar mass. In the lower panels, we show the rest-frame $(r-K_s)^\textsc{o}$ distribution of the transitioning galaxies (i.e. the galaxies lying in the NUVrK green valley). 
As explained in Sect. \ref{smf_measur}, the NUVrK diagram is very efficient in separating dusty star-forming galaxies from quiescent ones (see Fig.16 of the companion paper), which allows us to properly define transitioning galaxies in the green valley.
We observe that
\begin{itemize}
\item[1] the $(r-K_s)^\textsc{o}$ distribution of galaxies in transition is narrow and does not evolve with redshift ($>80\%$ of these galaxies have $0.76 < (r-K_s)^\textsc{o} < 1.23$), and
that \item[2] the typical stellar mass of galaxies in transition is around $\mathcal{M}^\star _{\textsc{sf}}$ ($>60\%$ of these galaxies have $10^{10.5} < M_* / M_{\odot} < 10^{11}$). 
\end{itemize}
Therefore, we isolated the quenching channel of the $\mathcal{M}^\star _{\textsc{sf}}$-galaxies with the colour criterion $(r-K_s)^\textsc{o} > 0.76$ in the NUVrK green valley (green dashed lines in Fig. \ref{NrK_Mass}, sub-panels).

\begin{figure}[!h]
\includegraphics[width=\hsize]{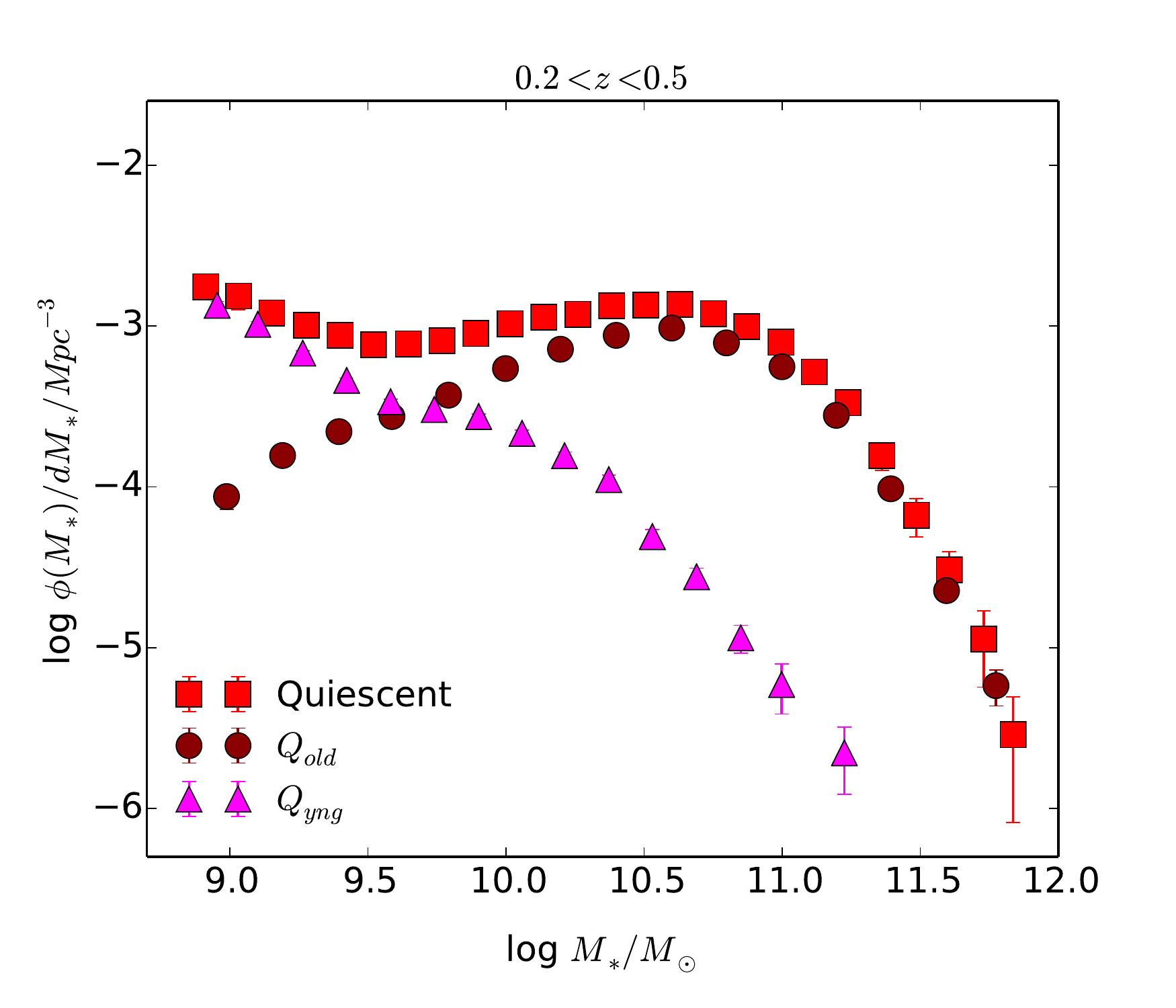}
\caption{Deconstruction of the quiescent SMF at $0.2 < z < 0.5$. The red squares represent the measurement for the whole quiescent population, while the magenta triangles and the darkred circles show the SMF for the \textit{young} ($Q_{yng}$) [ $(r-K_s)^\textsc{o} < 0.76$ ] and \textit{old} ($Q_{old}$) [ $(r-K_s)^\textsc{o} > 0.76$ ] quiescent populations, respectively.\label{SMF_Q}}
\end{figure}

We also detect a clear plume of \textit{young} quiescent galaxies in Fig. \ref{NrK_Mass}, with $(r-K_s)^\textsc{o} < 0.76$ (i.e. bluer than observed for galaxies following the $\mathcal{M}^\star _{\textsc{sf}}$ channel) at $z < 0.5$. It is well established that rest-frame optical-NIR colours are sensitive to both dust attenuation and age of the stellar populations \citep[see e.g.][]{Whitaker2012}.  Under the assumption that, on average, the $(r-K_s)^\textsc{o}$ colour of quiescent galaxies cannot become bluer with time, the \textit{young} part of the quiescent population should have used another quenching channel. According to the limit that we defined to isolate the $\mathcal{M}^\star _{\textsc{sf}}$ quenching channel (green dashed line in Fig. \ref{NrK_Mass}), we separated the \textit{young} quiescent ($Q_{yng}$) and \textit{old} quiescent ($Q_{old}$) galaxies with $(r-K_s)^\textsc{o}=0.76$. Figure \ref{NrK_Mass} also reveals that $Q_{yng}$ galaxies are characterised by relative low masses ($M_* \lesssim 10^{9.5} M_{\odot}$), which seems to match the low-mass upturn of the quiescent SMF (see Fig. \ref{SMF_fitt}) at $z < 0.5$. In Fig. \ref{SMF_Q} we compute the SMF for $Q_{yng}$ (magenta triangles) and $Q_{old}$ (dark red circles) galaxies at $0.2 < z < 0.5$. The $Q_{yng}$ galaxies dominate at low mass, and they are responsible for the low-mass upturn in the quiescent SMF.
At the same time, the SMF of $Q_{old}$ galaxies peaks at $\mathcal{M}^\star_{\textsc{sf}}$, which clearly supports the idea that the building of the quiescent SMF is led through two quenching channels that can be distinguished with a cut in the NUVrK diagram at $(r-K_s)^\textsc{o} = 0.76$. 
The timescale might then be a key element for characterising the mechanisms that are involved in each channel.

\subsubsection{Quenching timescales}
\label{qu_timesc}

\begin{figure*}[!]
\begin{small}
\hspace*{2.3cm}$t_Q$ = 1 Gyr  \hspace*{5.54cm}$\tau$ = 0.1 Gyr  \hspace*{3.51cm}$\tau$ = 1 Gyr
\end{small}

\includegraphics[width=0.35\hsize, trim = 0cm 0cm 0cm 0cm, clip]{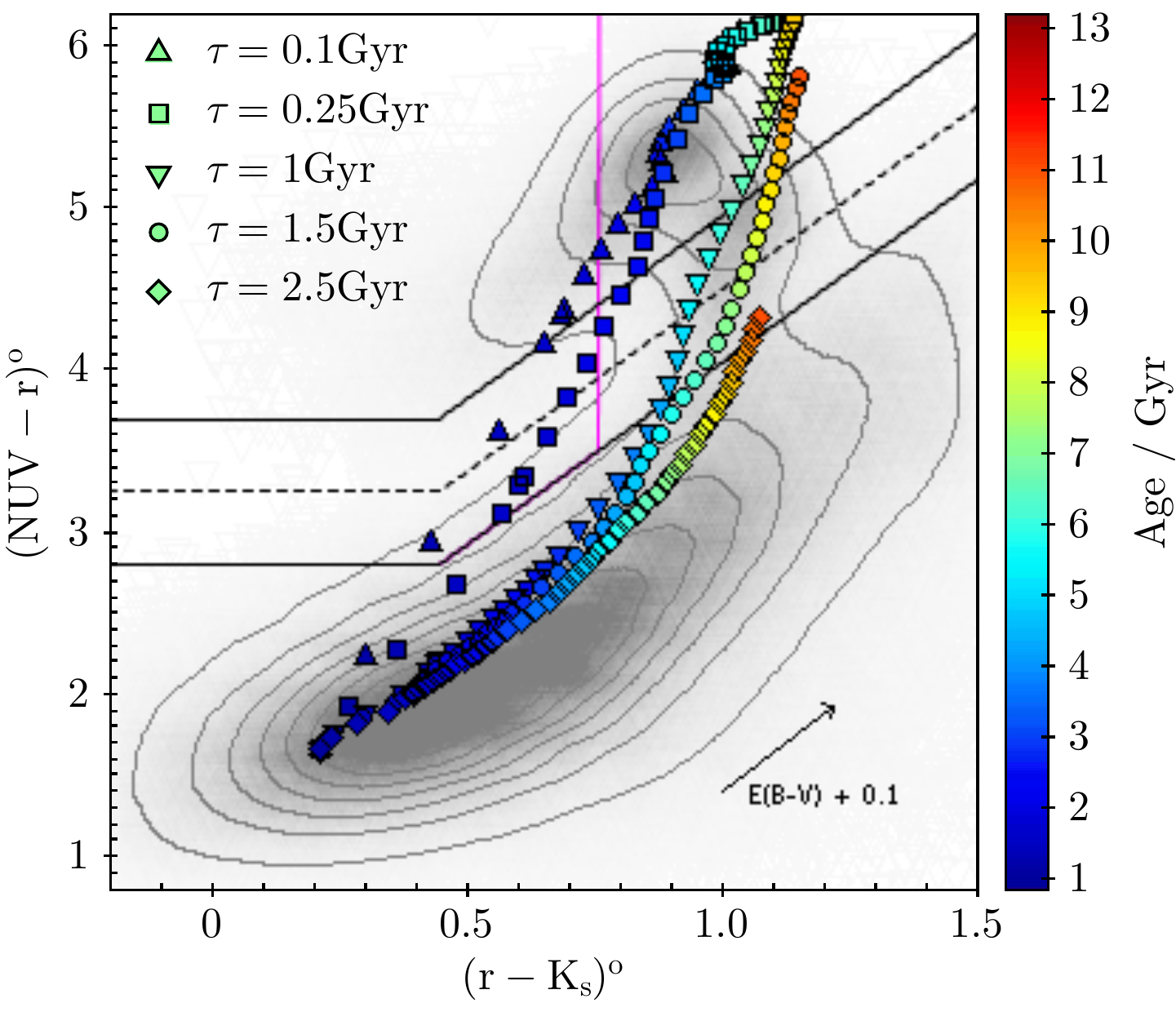}
\includegraphics[width=0.35\hsize, trim = -1.35cm 0cm 1.35cm 0cm, clip]{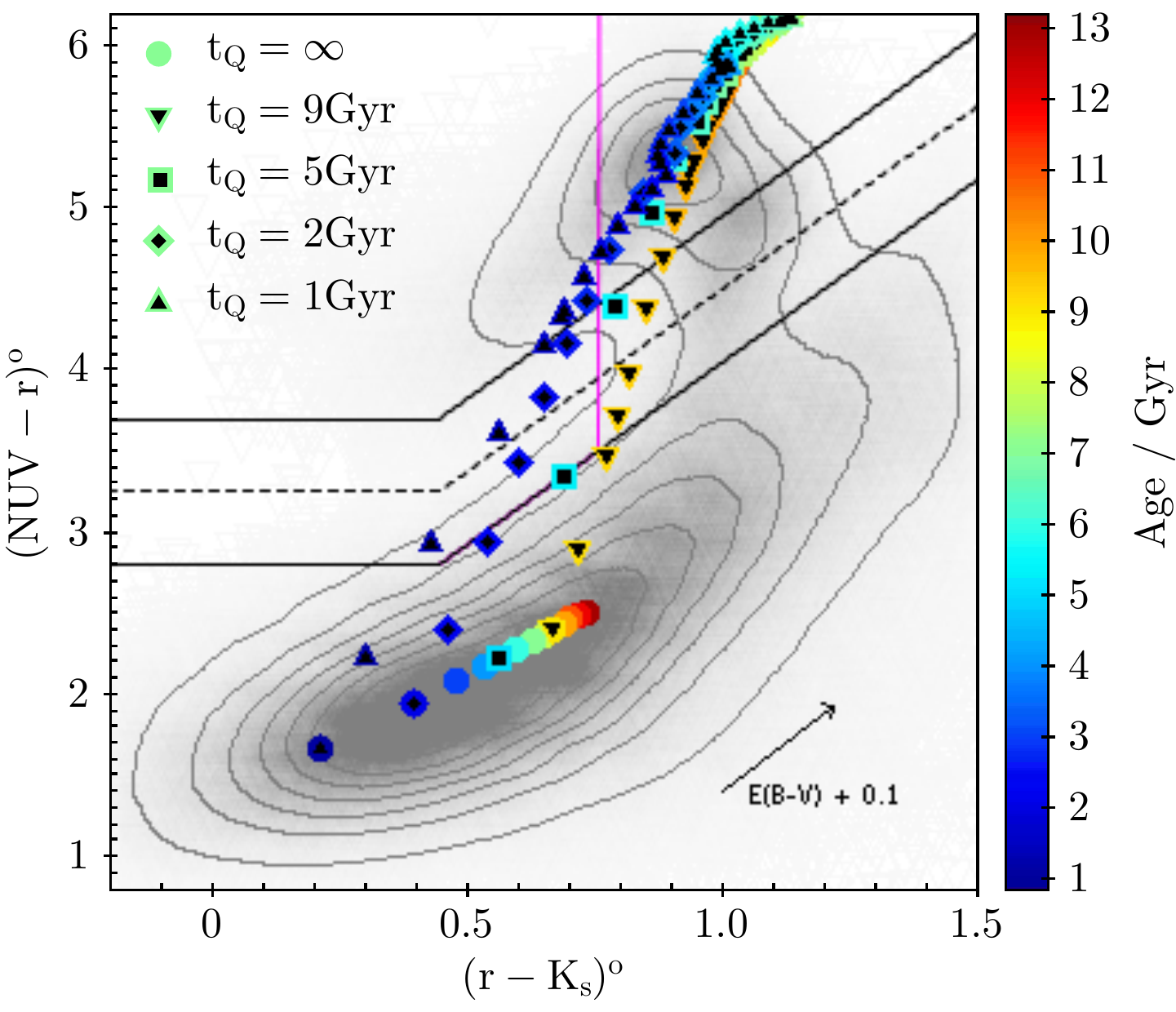}
\hspace*{-0.5cm}\includegraphics[width=0.35\hsize, trim = 1.35cm 0cm -1.35cm 0cm, clip]{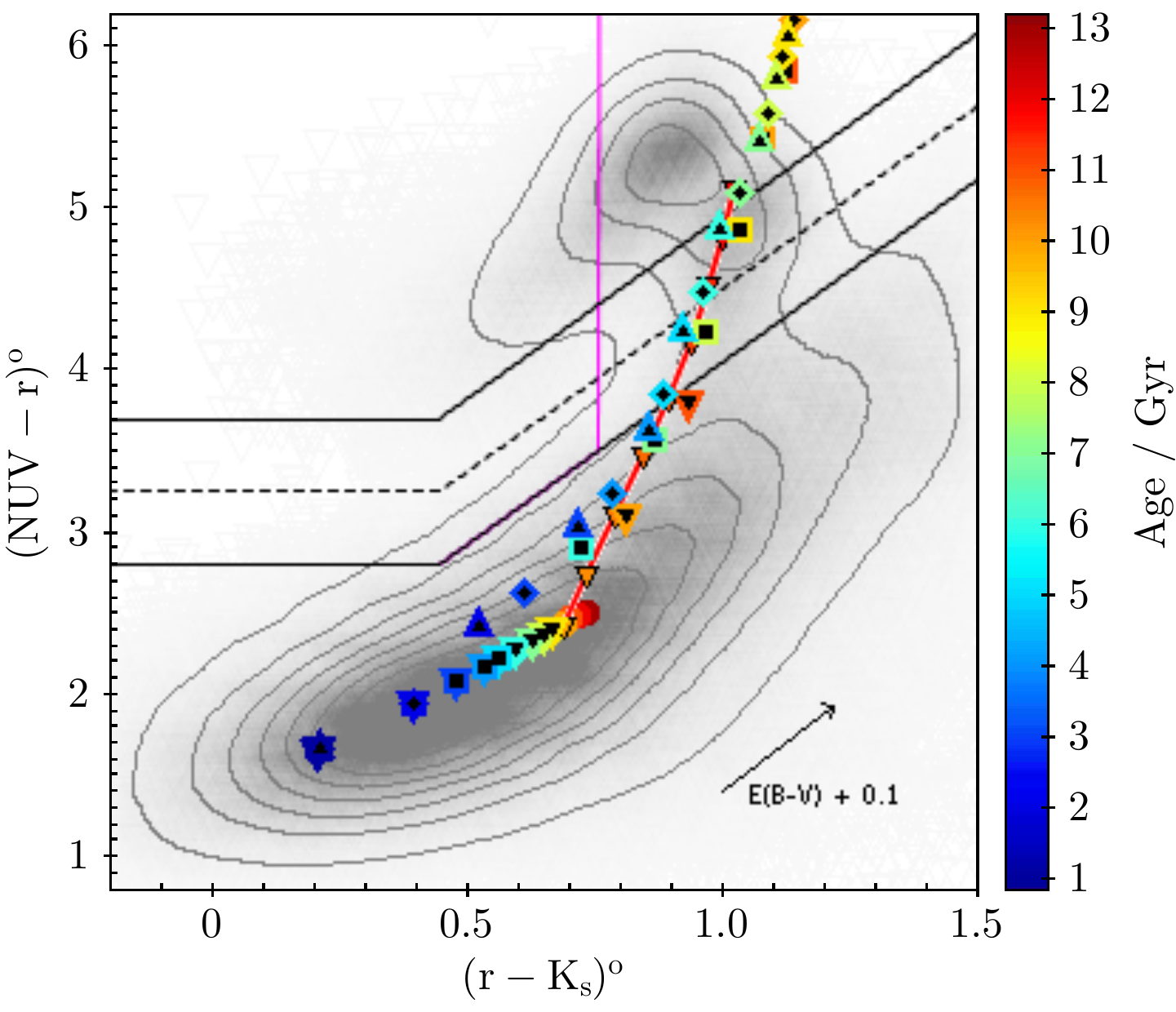}
\caption{Predicted BC03 tracks in the NUVrK diagram at $0.2 < z < 0.5$ for $Z=0.008$ \citep{Calzetti2000} and E(B-V) = 0.2. The arrow shows the shift expected for E(B-V) + 0.1. Analogously to Fig. \ref{NUVrK_z}, the black solid and dashed lines correspond to the limits of the green valley and its middle, respectively, while we report the $(r-K_s)^{o}$-limit of the $\mathcal{M}^\star$-quenching channel with a vertical magenta solid line. The grey contours outline the galaxy density distribution. Each marker is coloured with respect to the  corresponding stellar age (in Gyr). \textit{\textbf{Left panel}}: Only one quenching time is considered: $t_Q$ = 1 Gyr, with $\tau$ = 0.1 Gyr (triangles), $\tau$ = 0.25 Gyr (squares), $\tau$ = 1 Gyr (inverted triangles), $\tau$ = 1.5 Gyr (circles), and $\tau$ = 2.5 Gyr (diamonds). \textbf{\textit{Right panels}}: Two quenching timescales are considered: $\tau$ = 0.1 Gyr (middle panel) and $\tau$ = 1 Gyr (right panel), for $t_Q$ = 1 Gyr (triangles), 2 Gyr (diamonds), 5 Gyr (squares), and 9 Gyr (inverted triangles). The filled circles show the track for a continuous star formation without quenching. The red solid line linking the black edge triangles shows the track for $t_Q$ = 9 Gyr and $\tau$ = 0.5 Gyr.\label{NrK_qu}}
\end{figure*}

In Sect. \ref{gal_tracking} we have identified two possible channels in which galaxies are transitioning to build the quiescent population. We now investigate the nature of these channels through their characteristic timescales.

The restframe UV is sensitive to timescales of $10^{-2} - 10^{-1} Gyr$, and the scarcity of \textit{young}/low-mass galaxies in the green valley allows us to expect that some quenching processes occur on timescales of the same order or shorter. 
To better constrain the timescale of the quenching that affects the star formation of low-mass and $\mathcal{M}^\star_{\textsc{sf}}$ galaxies, we  explored the behaviour of simple scenarios of star formation history (SFHs) within the NUVrK diagram in a similar way as the approach adopted by \cite{Schawinski2014}. We performed this analysis at $0.2 < z < 0.5,$ where both \textit{old} and \textit{young} quiescent galaxies are well identified.
 The use of simple e-folding SFHs implies that we assumed that galaxies can only become redder with time. This is motivated by the fact  that the fraction of quiescent galaxies has continuously increased between $z \sim 3$ and $z \sim 0.2$ \citep[e.g.][]{Ilbert2010,Muzzin2013,Mortlock2015} and by assuming that most green valley galaxies are transitioning for the first time \citep{Martin2007}. 
Doing so, we neglect the green valley galaxies produced by \textit{rejuvenation}
processes, as observed in the local Universe \citep[e.g][]{Salim2010,Thomas2010} and recently predicted at higher redshift in the \textsc{eagle} simulations \citep{Trayford2016}. However, in these simulations, the rejuvenation is responsible for a small fraction of the green valley galaxies.

Figure \ref{NrK_qu} presents the resulting tracks in the NUVrK diagram for SFHs constructed in the same way: a continuous star formation up to the quenching time at $t_{Q}$, followed by an exponentially declining star formation characterised by $\tau$. To mimic the average properties of our $K_s < 22$ sample at $0.2 < z < 0.5,$ the example is plotted for one metallicity ($Z=0.008$), one extinction law \citep{Calzetti2000}, one value of the dust attenuation (E(B-V) = 0.2), and with a stellar age of at least
1 Gyr. The stellar age is colour coded, and only the ages allowed by the given redshift bin are plotted. In the left panel of Fig. \ref{NrK_qu} the SFHs are characterised by $t_{Q}$ = 1 Gyr, with $\tau$ = 0.1, 0.25, 1, 2, and 2.5 Gyr. The tracks are constructed in a very simple way, and the evolution assumes a constant dust attenuation based on its average value for the bluest SF galaxies. The arrows show the shift that is due to a 0.1 increase of E(B-V). It is expected that quiescent galaxies are less affected by dust, which would tend to make the tracks steeper in the NUVrK green valley. Keeping this effect in mind, we see as a first result that the presence of $Q_{yng}$ galaxies is expected if any quenching process occurs early ($t_Q \sim 1$ Gyr) with a typical timescale
of $\tau \lesssim 0.25$ Gyr (triangles and squares in the left
panel of Fig. \ref{NrK_qu} ). As a second result, $\tau$ = 1 Gyr (inverted triangles in Fig. \ref{NrK_qu} left panel) seems to be a lower limit for the quenching timescale that is compatible with the channel drawn by $\mathcal{M}^\star_{\textsc{sf}}$ galaxies. The galaxies with a quenching $\tau >$ 2 Gyr do not reach the quiescent cloud.

In the middle and right panels of Fig. \ref{NrK_qu}, we also investigate the effect of the quenching epoch. We fixed $\tau$ = 1 Gyr and $\tau$ = 0.1 Gyr for several values of $t_Q$ between 1 and 9 Gyr. Any $t_Q>9$ Gyr will produce the same result as $t_Q=9$ Gyr since the NUVrK colours of SF galaxies saturate at ages $> 9$ Gyr, as shown by the predicted track with a continuous star formation (circles). All the models with $\tau$ = 1 Gyr are able to explain the galaxy presence within the $\mathcal{M}^\star$ channel. We could also imagine that a shorter timescale combined with a late quenching time can reproduce the observed $\mathcal{M}^\star$ channel. However, if we consider an SFH with $\tau$ = 0.1 Gyr after 9 Gyr on the SF main sequence (middle panel, inverted  triangles), the track seems to move away from the channel that is drawn by $\mathcal{M}^\star$-galaxies in the NUVrK diagram. To produce a track that is compatible with this channel, we need to consider a quenching timescale $\tau \gtrsim$ 0.5 Gyr (red solid line), regardless of the considered SHF. We recall that we have considered the shortest timescales compatible with the $\mathcal{M}^\star_{\textsc{sf}}$-\textit{quenching channel}, and we could pick out SFHs that agree better. Namely, SFHs characterised by $t_Q$ = 1 Gyr and $\tau$ = 1.5 Gyr, $t_Q$ = 5 Gyr and $\tau$ = 1 Gyr, or $t_Q$ = 9 Gyr and $0.5 < \tau < 1$ Gyr could also explain the presence of this channel. This suggests a quenching timescale range of $0.5 < \tau < 2$ Gyr for $\mathcal{M}^\star$-galaxies, which corresponds to a quenching duration of between $\sim 1$ and 3.5 Gyr\footnote{These values agree with the estimate of \citet{Fritz2014} in VIPERS, who found that massive ($\log(M_*/M_{\odot})>11$) galaxies are expected to turn quiescent in $\sim$1.5 Gyr at $0.7 < z < 1.3$, and more slowly at $z < 0.7$ (i.e. with longer quenching durations).}.
Therefore, the physical mechanism explaining the building of the quiescent SMF around $\mathcal{M}^\star_{\textsc{sf}}$ at $z < 1$ seems to be a slow process. 
Such a \textit{mass dependent} mechanism is compatible with a \textit{strangulation} picture where the star formation quenching occurs on several Gyr, moving slowly away from the SF main sequence in the NUVrK diagram, while the gas supply is progressively halted \citep{Schawinski2014, Peng2015}.

Figure \ref{NrK_qu} shows that the plume formed by $Q_{yng}$ galaxies in the NUVrK plan is explained by a $\sim 0.1$ Gyr-quenching process occurring during the first $\sim 5$ Gyr of the galaxy life (squares, diamonds, and triangles in the middle panel of Fig. \ref{NrK_qu}). The absence of these low-mass galaxies lying in the green valley can be first explained by the rapidity of their quenching. Indeed, a galaxy quenching with $\tau$ = 0.1 Gyr (triangles in the left and middle panels of Fig. \ref{NrK_qu}) is expected to cross the green valley (delimited by the black solid lines) in $\sim 0.4$ Gyr, while a galaxy with $\tau \sim 0.5-2$ Gyr spends $\sim 1-3.5$ Gyr there, on average. Nevertheless, the potential reservoir of SF $M_* < 9.5$ galaxies that can quench is $ \text{about ten}$ times larger than for galaxies around $\mathcal{M}^\star _{\textsc{sf}}$ (cf. Fig. \ref{SMF_fitt}). We could then expect to see more low-mass galaxies in transition. By adopting a conservative approach, we can assume that the ratio between the two quenching timescales is $\sim 10$ (0.1 Gyr for $M_* < 9.5$ galaxies, 1 Gyr around $\mathcal{M}^\star _{\textsc{sf}}$). The corresponding quenching rate should consequently be $ \text{about
ten}$ times lower for the low-mass galaxies that are the progenitors of the $Q_{yng}$ galaxies than for the $\mathcal{M}^\star_{\textsc{sf}}$ galaxies. The resulting flux of quenching  galaxies (i.e. quenching rate $\times$ SF reservoir) that cross the green valley is then expected to be of the same order of magnitude, both at low and high mass, except when only a fraction of the low-mass galaxies is likely to be affected by the quenching. The SF satellite galaxies, which are estimated to be $\gtrsim 3$ times less abundant than field galaxies \citep{Yang2009,Peng2012}, are therefore good candidates for this low-mass quenching mechanism. Moreover, its typical timescale is compatible with the scenario suggested by \citet{Schawinski2014} for the rapid formation of young early-type galaxies. In this picture, the quiescent low-mass galaxies are formed through dramatic events such as major mergers and not through \textit{ram-pressure stripping} or \textit{strangulation}, by explaining both the almost instantaneous star formation shutdown and the morphological transformation.

\section{Summary}

 We analysed the evolution of the stellar mass function in an unprecedentedly large ($>22$ deg$^2$) NIR selected ($K_s < 22$) survey. This allowed us to provide reliable constraints on the evolution of massive galaxies and to investigate quenching processes below redshift $z\sim 1.2$.  Covering the VIPERS spectroscopic survey, we computed highly reliable photometric redshifts, with usual estimates of the precision $\sigma_{\Delta z/(1+z)}< 0.03$ and $\sigma_{\Delta z/(1+z)}< 0.05$ for bright ($i < 22.5$) and faint ($i > 22.5$) galaxies, respectively.

Paying particular attention to several sources of uncertainties (photometry, star-galaxy separation, photometric redshift, dust extinction treatment, and classification into quiescent and star-forming galaxies), we computed the SMF between redshifts $z = 0.2$ and $z = 1.5$. The unique size of our sample enabled us
to drastically reduce the statistical uncertainties affecting the SMFs and stellar mass densities with respect to other current surveys over the stellar mass range we consider: the Poissonian error and cosmic variance are reduced by factors of $\sim 3.3$ and $\sim 2, $ respectively, compared to a 2 deg$^2$-survey.
Combined with a careful treatment of the Eddington bias that
is due to the stellar mass uncertainty, we produced an unprecedentedly precise measurement of the massive end of the SMF at $z<1.5$. In particular, we stress the importance of constraining all sources of systematic uncertainties, which quickly become the dominant sources of error in large-scale surveys such as those planned with Euclid or LSST.

Using the $(NUV-r)$ versus $(r-K)$  rest-frame colour diagram to classify star-forming and quiescent galaxies in our sample, we measured the evolution of the  SMFs of the two populations and investigated the possible quenching processes that could explain the build-up of the quiescent population. Our main conclusions are summarised below.
\begin{itemize}
\item[1)] We provided clear evidence that the number density of the most massive ($M_* > 10^{11.5} M_{\odot}$) galaxies increases by a factor $\sim 2$ from $z \sim 1$ to $z \sim 0.3,$ which was first highlighted by \citet[][]{Matsuoka2010}. This population is largely dominated 
 by the quiescent population since $z\sim 1$, allowing for the possibility of  galaxy mass assembly through dry-mergers in very massive galaxies.    

\item[2)] The characteristic mass of the SF population was found to be very stable in the redshift range $0.2 < z < 1.5$, with $\log(\mathcal{M}^\star_{\textsc{sf}} / M_{\odot}) = 10.64 \pm 0.01$.  
 This confirms that the star formation is impeded above a certain stellar mass \citep[][]{Ilbert2010,Peng2010}. 

\item[3)] Using the NUVrK diagram as a tracer of the galaxy evolution,
 we identified one main \textit{quenching channel} between the star-forming and quiescent sequences at $0.2 < z < 1.5$, which is followed by galaxies with stellar masses around $\mathcal{M}^\star_{\textsc{sf}}$. This channel is characterised by a colour $(r-K_s)^\textsc{o}>0.76$, typical of evolved massive star-forming galaxies, which should feed the majority of the quiescent population.  
We also identified a \textit{young} quiescent population with $(r-K_s)^\textsc{o}<0.76$, whose galaxies likely followed another path to reach the quiescent sequence. We showed that this \textit{blue} quiescent population, dominated by low-mass galaxies, is responsible for the upturn of the quiescent SMF at low redshift.  

\item[4)] Assuming simple e-folding SFHs (galaxies can only become redder with time), we found that the $\mathcal{M}^\star_{\textsc{sf}}$ channel is explained by long quenching timescales, with 0.5 Gyr $< \tau \lesssim$ 2 Gyr. 
 Galaxies in this channel are expected to turn quiescent after $\sim 1-3.5$ Gyr on average. 
 This is compatible with \textit{strangulation} processes occurring when the gas cooling or the cold gas inflows are impeded, allowing the galaxy to progressively consume its remaining gas reservoir \citep{Peng2015}. Conversely, the quenching of low-mass galaxies that is visible at low redshift is characterised by short timescales with $\tau \sim$ 0.1 Gyr.  This quenching that halts star formation in $\sim$ 0.4 Gyr can be consistent with major merging \citep[][]{Schawinski2014} and may preferentially affect satellite galaxies. 
\end{itemize}

\begin{acknowledgements}
We gratefully acknowledge the anonymous referee, whose advice helped much in improving the clarity of the paper.
We wish to thank J.-C. Cuillandre, H. Aussel, S. de la Torre and B. C. Lemaux for helpful discussions. We would also like to thank J. Moustakas for providing the PRIMUS stellar mass estimates. 
This research is in part supported by the Centre National d'Etudes Spatiales (CNES) and the Centre National de la Recherche Scientifique (CNRS) of France, and the ANR Spin(e) project (ANR-13-BS05-0005, http://cosmicorigin.org).
L.G. acknowledges support of the European Research Council through the Darklight ERC Advanced Research Grant (\# 291521).
This paper is based on observations obtained with MegaPrime/MegaCam, a joint project of CFHT and CEA/DAPNIA, and with WIRCam, a joint project of CFHT, Taiwan, Korea, Canada and France, at the Canada-France-Hawaii Telescope (CFHT).The CFHT is operated by the National Research Council (NRC) of Canada, the Institut National des Science de l'Univers of the Centre National de la Recherche Scientifique (CNRS) of France, and the University of Hawaii. 
This work is based in part on data products produced at Terapix available at the Canadian Astronomy Data Centre as part of the Canada-France-Hawaii Telescope Legacy Survey, a collaborative project of NRC and CNRS. 
We thank the Terapix team for the reduction of all the WIRCAM images and the preparation of the catalogues matching with the T0007 CFHTLS data release.
This paper is based on observations made with the Galaxy Evolution Explorer (GALEX). GALEX is a NASA Small Explorer, whose mission was developed in cooperation with the Centre National d'Etudes Spatiales (CNES) of France and the Korean Ministry of Science and Technology. GALEX is operated for NASA by the California Institute of Technology under NASA contract NAS5-98034.
This paper uses data from the VIMOS Public Extragalactic Redshift Survey (VIPERS). VIPERS has been performed using the ESO Very Large Telescope, under the "Large Programme" 182.A-0886. The participating institutions and funding agencies are listed at http://vipers.inaf.it.
This paper uses data from the VIMOS VLT Deep Survey (VVDS) obtained at the ESO Very Large Telescope under programs 070.A-9007 and 177.A-0837, and made available at the CESAM data center,  Laboratoire d'Astrophysique de Marseille, France.
Funding for PRIMUS is provided by NSF (AST-0607701, AST-0908246, AST-0908442, AST-0908354) and NASA (Spitzer-1356708, 08-ADP08-0019, NNX09AC95G). 
Funding for SDSS-III has been provided by the Alfred P. Sloan Foundation, the Participating Institutions, the National Science Foundation, and the U.S. Department of Energy Office of Science. The Participating Institutions of the SDSS-III Collaboration are listed at http://www.sdss3.org/.

\end{acknowledgements}

\bibliographystyle{aa}
\bibliography{bib_VMLS_MF}

\end{document}